\documentclass[a4paper,11pt]{article}
\usepackage{xcolor}
\usepackage{epsfig}
\usepackage{amsmath}
\usepackage{amssymb}
\usepackage{amsthm}
\usepackage{mathtools}
\usepackage[top=1cm, bottom=1.5cm, left=1.5cm, right=1.5cm]{geometry}
\usepackage{microtype}
\usepackage{lmodern}
\usepackage[utf8]{inputenc}
\usepackage{float}
\usepackage{hyperref}
\usepackage{cite}
\usepackage{bm}
\usepackage{verbatim}
\newcommand{\ds}{\displaystyle }

\newcommand{\beq}{\begin{equation}}
\newcommand{\eeq}{\end{equation}}

\title{Stochastic predator-prey compartment model with cyclic and non-cyclic transition pathways and mortality}

\author{
\Large 
Thomas M. Michelitsch$^{a}$\footnote{Corresponding, e-mail: thomas.michelitsch@sorbonne-universite.fr}, Ayten D. Baher$^b$, Michael Bestehorn$^c$ \\ 
\Large Andrzej F. Nowakowski$^d$, Alejandro P. Riascos$^e$, Bernard A. Collet$^a$
\\[2ex]
$^a$~Sorbonne Université, CNRS, Institut Jean Le Rond d’Alembert \\ F-75005 Paris, France
\\[2ex]
$^b$~Department of Physics, Shahid Beheshti University, Evin, Tehran, Iran
\\[2ex]
$^c$~Institut f\"ur Physik, Brandenburgische Technische Universit\"at Cottbus-Senftenberg \\
Cottbus, Germany
\\[2ex]
$^d$~School of Mechanical, Aerospace and Civil Engineering \\  University of Sheffield, 
Sheffield S1 3JD, United Kingdom
\\[2ex]
$^e$~Departamento de Física, Universidad Nacional de Colombia, Bogot\'a, Colombia
}

\begin{document}

\maketitle

\begin{abstract}
We study variants of a stochastic SEIRS--SEID compartmental model incorporating presymptomatic infectiousness and disease-related mortality, with independent compartmental sojourn times drawn from arbitrary distributions, in particular Weibull distributions. We consider a large population of agents that occupy the states (``compartments'') S (susceptible to infection), E (exposed: infected but asymptomatic during the latency period and already infectious, including diseases with presymptomatic infections), I (infected and infectious, with symptoms), R (recovered, immune, and non-infectious), or D (dead). 
We consider {\bf (i)} cyclic transmission pathways, where 
compartmental sojourn times are finite and immunity is temporary, and
{\bf (ii)} non-cyclic pathways, where immunity is permanent. 
In the absence of mortality, cyclic transmission pathways support an endemic equilibrium above the spreading threshold
which is either globally stable or oscillatory (Hopf) unstable, whereas for non-cyclic pathways a stationary disease-free state is emerging. 
From a stochastic perspective, we elucidate the connection between compartmental sojourn-time distributions and the appearance of fractional operators in the evolution equations: fat-tailed compartment waiting-time distributions necessarily lead to fractional derivatives. For cyclic transition pathways we obtain an explicit expression for the expected number of infection cycles that an agent survives.
Our model is validated by a combined Monte Carlo--Runge--Kutta simulation scheme and random-walk simulations on Barabási–Albert networks. It has cross-disciplinary applications beyond epidemiology, including wildfire spreading, certain classes of chemical reactions, population dynamics, finance, economics, and biology.
\\[2ex]
\end{abstract}

\newpage
\tableofcontents
\section*{}
%
%
%
%
{\bf 
Spreading phenomena are ubiquitous in nature and relevant 
in cross-disciplinary contexts including population dynamics, chemistry, economy, epidemiology and others.
In many cases these processes can be reduced to
compartmental models where agents undergo transitions among specific states (`compartments').
The vast field of applications includes disease-spreading with 
the widely-known class of SIR (Susceptible–Infectious–Recovered) models. 
We provide a stochastic framework for a large class of predator-prey compartmental models in which agents have a memory on the transition dynamics. 
We focus on cyclic and non-cyclic transmission pathways with and without mortality.
Agents in `predator' compartments exhibit infectivity and are able to transmit the disease to agents in the `prey' compartment. We highlight the connection with widely considered memoryless compartment models which are contained as special cases.
We demonstrate that broad compartmental sojourn time distributions necessarily result in fractional transition dynamics. 
We investigate the rich dynamics resulting from the complex interplay of presymptomatic infections, disease-related mortality, and 
memory effects originating from features of the
compartmental waiting time distributions. 
}

\section{Introduction}
\label{Intro}
Spontaneous outbreaks of epidemics recurrently affect human populations and represent major challenges for societies and public health systems. Since the emergence of the COVID-19 pandemic, epidemic models have attracted considerable attention, highlighting the need for a deeper understanding of the underlying mechanisms governing epidemic propagation. A popular approach to modeling epidemic spreading is provided by compartmental models, in which the population is divided into categories (``compartments'') that characterize the health states of individuals. The first compartmental models can be traced back to the seminal work of Kermack and McKendrick \cite{KermackMcKendrick1927}, which is widely regarded as the foundation of modern epidemiological modeling. In their SIR model, individuals are susceptible (healthy and susceptible to infection), denoted by S, infected and infectious, denoted by I, or recovered (healthy and immune), denoted by R. The classical SIR model is characterized by a noncyclic infection pathway, S $\to$ I $\to$ R, with a finite, exponentially distributed sojourn time in compartment I and an infinite sojourn time in compartment R, corresponding to everlasting immunity. This class of models contrasts with models exhibiting cyclic pathways, such as SIS and SIRS, in which all compartments except S have finite sojourn times. In the absence of mortality, all individuals in these models eventually return to compartment S. While classical SIR models capture the dynamics of certain common infectious diseases, such as measles, mumps, and rubella, they cannot describe quasi-periodic or persistent oscillatory temporal patterns, which require cyclic infection pathways. The classical work of Soper was among the first to develop an analytical approach to oscillatory epidemic spreading \cite{Soper1929}.

The classical SIR, SIS, and SIRS models have been generalized in various directions \cite{LiuHeathcote1987,Li-etal1999,Anderson1992,Martcheva2015,Harris2023}. Many studies have sought to establish connections between macroscopic compartmental models and microscopic dynamics. In this context, epidemic spreading on complex networks, including random graphs,
or evolving networks has been extensively investigated \cite{Jafari2017,Basnakov2021,Satoras-Vespignani-etal2015,Pastor-SatorrasVespignani2001,OkabeShudo2021,Barabasi2016,BarabasiAlbert1999,Barrat-etal2008,Granger-et-al2023,entropy_paper}. Combinations of compartmental models and random walk approaches have also been explored \cite{Ross1996,fractional_book_MiRia2019,BesMi-etal2021,vanKampen1981,VanMiegem2014}. In contrast to the classical Markovian assumption, in which compartmental waiting times are exponentially distributed, and the resulting evolution equations are memoryless, non-Markovian variants in which waiting times are drawn from arbitrary non-exponential distributions remain relatively rare in the literature. Recent studies addressing various aspects of memory effects and non-Markovian dynamics include \cite{Sherborne_etal2017,BasnakovSandev-etal2020,BesMiRias2022,Granger-et-al2023,entropy_paper}. 
A potential source of memory effects is environmental heterogeneity. Human populations live in spatially heterogeneous and often modular environments, where population density and contact patterns may exhibit broad, scale-free distributions. For example, urban population sizes  follow power-law distributions in some settings. Such heterogeneity can generate nontrivial anomalous transport leading to spreading dynamics with memory on the infection history. In some cases waiting time distributions exhibit fat tails, leading necessarily to fractional evolution equations, a subject which also is considered in the present paper. We refer to \cite{Jafari_etal_2017} for a related SIR model with memory. 

Other works have considered cyclic SEIRS dynamics \cite{Batista_etal2021}, including variants that incorporate birth and death processes \cite{Bornstadt_et_al2020}. Interesting connections also exist between compartmental models with mortality and evanescent (mortal) random walks, which constitute an increasingly broad interdisciplinary field with applications including random search problems and foraging \cite{Ben2014,MRW_miria2025}.

In the present study, our aim is to provide a stochastic framework for non-Markovian spreading dynamics that captures a broad class of predator-prey compartmental models with independent and arbitrary compartmental waiting times (sojourn times in the compartments). We emphasize the cross-disciplinary nature of the model and its applications beyond epidemiology. The model includes both cyclic and non-cyclic infection pathways and incorporates mortality in one compartment, namely compartment I. Furthermore, we allow disease transmission from agents in the ``predator'' compartment E (presymptomatic infections) and from agents in compartment I through contacts with susceptible agents (the prey). The assumption that transmission can occur from agents in compartment E corresponds to diseases that are infectious during the asymptomatic latency phase. This feature is observed in various infectious diseases, including COVID-19 (SARS-CoV-2), measles, and chickenpox. As long as the waiting times in all compartments are finite, cyclic infection pathways occur. Whenever the sojourn time in at least one compartment is infinite, as in the case of everlasting immunity (an infinite sojourn time in R), a non-cyclic transition pathway emerges.

Our paper is organized as follows. In Sect. \ref{mean-field}, we introduce a stochastic formulation of the SEIRS-SEID compartmental model and derive its ``memory representation'' (Sect. \ref{general}). This approach encompasses a wide range of common compartmental models, most of which are memoryless. To account for mortality and other effects, we introduce the distributions governing the compartmental waiting times and the survival of agents. In our numerical evaluations, we use Weibull distributed compartmental waiting times because of their flexibility in describing a variety of shapes observed in real-world situations. In Sect. \ref{fractional_dyn}, we demonstrate that broad (fat-tailed) distributions of compartmental sojourn times correspond to a slow power-law decay of the `associated memory kernel'. This dynamics necessarily involves fractional (Riemann-Liouville) derivatives in the evolution equations. In Sect. \ref{SEIRS_zero_infect}, we consider the dynamics in the absence of infection rates in order to demonstrate the effect of delayed transitions arising from the initial conditions. Sect. \ref{full_SEIRDS_latetime} is devoted to the late-time asymptotics of the full SEIRS-SEID system. We consider both the cyclic case, in which all compartmental waiting times are finite, and the non-cyclic case of everlasting immunity. For the cyclic case, we obtain an asymptotic formula for the expected number of infection cycles an agent undergoes during its lifetime. A linear stability analysis of the disease-free state, corresponding to the absence of agents in the predator compartments E and I, is performed in Sect. \ref{spreading_cond}, leading to the basic and effective reproduction numbers in the presence of mortality and vaccinated agents. Sect. \ref{no_mortality} is devoted to diseases without mortality, for which a cyclic SEIRS pathway emerges when all compartmental waiting times are finite. We demonstrate that, for a basic reproduction number ${\cal R}_0 > 1$, an endemic equilibrium exists that is either globally asymptotically stable or oscillatory (Hopf) unstable. We show, by means of inequalities, that Hopf instabilities are possible for arbitrary compartmental waiting-time distributions. This result extends a previous one \cite{Granger-et-al2023}. We also consider everlasting immunity with a non-cyclic SEIR pathway, for which we find that an endemic equilibrium does not exist. In Sect. \ref{pertinent_special}, we consider several pertinent variants contained within our model, including SIS, SIR, SIR-SID, and SEIR-SEID transmission pathways, and highlight their connections to common models in the literature. The non-cyclic SIR-SID pathway is investigated in Sect. \ref{SIRD}, where we compare the numerical results with those obtained from simulations of multiple random walkers on a Barabási-Albert graph. In Sect. \ref{RW_simulations_SEIRDS}, we explore the complex interplay between presymptomatic infections, mortality, and finite immunity times for the non-cyclic SEIR-SEID pathway. Finally, Sect. \ref{RW_simulations_SEIRDS_dynamics} compares the non-cyclic SEIR-SEID dynamics with those of the fully cyclic SEIRS-SEID pathway, where we explore the effects of multiple infection cycles and confinement measures on overall mortality.

\section{Stochastic SEIRS - SEID compartment model}
\label{mean-field}
\subsection{Model formulation}
\label{general}

We consider a compartment model in which the agents are in one of the states (compartments) S (susceptible to infection), E 
(exposed: infected and already infectious corresponding but without symptoms corresponding to the latency period), I (infected and infectious), R (recovered, immune), and D (dead). 
The dead compartment has obviously infinite sojourn time. Presymptomatic transmissions, i.e. transmissions from agents which are still asymptomatic has been observed in several diseases, including
COVID-19, Influenza, Measles, Chickenpox (Varicella), Hepatitis A, Ebola, to quote but a few \cite{Heymann2022}. Works addressing asymptomatic transmission (susceptible S exposed E, asymptomatic A, infected and infectious I, recovered R) are sparse. Consult \cite{Basnakov2021} for such a model 
of COVID-19.

\begin{figure}[H]
\centerline{
\includegraphics[width=0.35\textwidth]{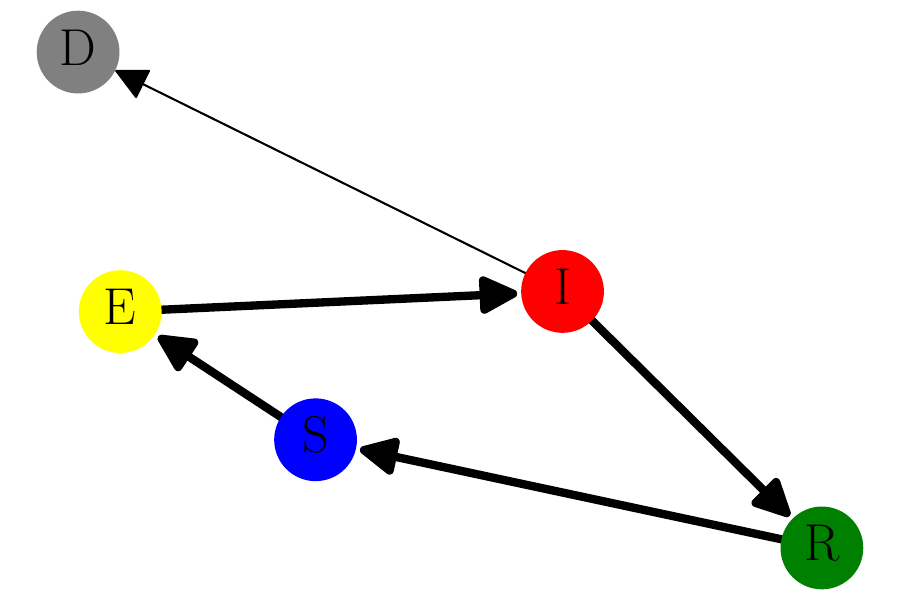} 
}
\caption{Schematic representation of the SEIRS-SEID infection pathway for finite $t_{E,I,R,M}$.} 
\label{pathway}
\end{figure}

The compartmental sojourn times $t_E,t_I,t_R >0$ (sojourn times in the compartments E, I, R) are assumed to be independent random variables drawn from specific probability density functions (PDFs). 
The sojourn time $t_E$ is the latency period (asymptomatic period), i.e. the time between the instant of infection (transition S to E) and outbreak of the disease (transition E to I).
We consider a large class of infectious diseases with `direct' infection pathway among agents: Agents being in `predator compartments' (E or I) may infect agents in the prey compartment (S) while only
ill agents (agents in I) exhibit mortality.
In order to capture this kind of mortality, we introduce an independent random variable $t_M$ which defines a random clock starting from the time instant of 
outbreak of the disease (instant of transition from E into compartment I): An agent in I survives the disease if $t_M>t_I$ with a transition from I to R alive 
(having spent the time $t_I$ in I)and dies otherwise, i.e. for $t_M<t_I$ (having 
spent the time $t_M$ in I)
undergoing a transition from I to D. The `real' sojourn time in compartment I is hence $t_{IM} = min(t_I,t_M)$. It equals $t_I$ in the limit of immortality (i.e. when $t_M=\infty$).
An agent that survives the disease runs through the cyclic SEIRS pathway (given all sojourn times $t_{E,I,R}$ are finite), whereas 
an agent which is dying while being in compartment I, passes through the non-cyclic SEID pathway. 
This behavior is represented schematically in the Fig. \ref{pathway}: 
The compartments are represented as nodes of a directed graph. The arrows indicate allowed transitions.
We call hereafter the SEIRS-SEID (SEIRS with mortality) `cyclic' whenever $t_{E,I,R} < \infty$ although the SEID pathway is non-cyclic.
We allow transmission of the disease if an agent in E or I (the predator compartments) gets in contact with an agent in S (the prey compartment). This assumption suggests an infection law of the form
\beq
\label{infection_law}
{\cal A}(t) = {\cal A}[S(t);E(t),J(t)] =\beta S(t) \Lambda(t) ,  \hspace{0.5cm} \Lambda(t)= \mu_E E(t)  + J(t)  ,\hspace{0.5cm} \beta >0, \hspace{0.1cm} \mu_E \geq 0
\eeq
in which $\beta$ and $\beta\mu_{E} $ are phenomenological rate constants capturing (though not being identical with) the probabilities of transmission of the disease.
The quantity $\mu_E$ in (\ref{infection_law}) is a dimensionless tuning parameter characterizing the contagiosity of presymptomatic agents. We point out that by writing
${\cal A}(t)$ we mean that ${\cal A}$ is an implicit function of $S(t)$, 
$E(t)$ and $J(t)$ describing
the interaction of the predator agents in E and I with prey agents in S.
The evolution equations of the compartmental fractions
have the following stochastic formulation 
\beq
 \label{SCEIRS-D-model}
 \begin{array}{clr}
 \displaystyle \frac{d S(t)}{dt} = & \ds  - {\cal A}(t)  + \Big\langle {\cal A}(t-t_E-t_I-t_R)\Theta(t_M-t_I)\Big\rangle &  \\[3ex]
 & \ds  + \Big\langle \left\{e_0\delta(t-t_E-t_I-t_R) \Big\rangle  +j_0\delta(t-t_I-t_R) \right\}\Theta(t_M-t_I) \Big\rangle + r_0\Big\langle \delta(t-t_R) \Big\rangle &\\[3ex]
 \displaystyle \frac{d E(t)}{dt}  = & \ds {\cal A}(t) - \Big\langle {\cal A}(t-t_E)\Big\rangle \,  -e_0\Big\langle \delta(t-t_E)\Big\rangle   &\\[3ex] 
 \displaystyle  \frac{d J(t)}{dt}   = & \ds  \Big\langle {\cal A}(t-t_E) \Big\rangle 
  - \Big\langle  {\cal A}(t-t_E-t_I) \Theta(t_M-t_I) \Big\rangle  \, -  j_0  \Big\langle \delta(t-t_I) \Theta(t_M-t_I) \Big\rangle  - \frac{d D(t)}{dt} & \\[3ex]
& \ds + e_0\Big\langle \left\{\delta(t-t_E) - \Theta(t_M-t_I)\delta(t-t_E-t_I) \right\} \Big\rangle & \\[3ex]
 \displaystyle  \frac{d R(t)}{dt} = & \ds \Big\langle \left\{{\cal A}(t-t_E-t_I) - {\cal A}(t-t_E-t_I-t_R)\right\} \Theta(t_M-t_I) \Big\rangle 
  &\\[3ex] & \ds +  j_0 \Big\langle \left\{ \delta(t-t_I) -\delta(t-t_I-t_R) \right\} \Theta(t_M-t_I) \Big\rangle  - r_0 
 \Big\langle \delta(t-t_R)\Big\rangle & \\[3ex]
 & \ds +  e_0\Big\langle \left\{\delta(t-t_E-t_I) -\delta(t-t_E-t_I-t_R) \right\} \Theta(t_M-t_I) \Big\rangle
 \end{array}
\eeq
containing the mortality rate (which is due to $S(t)+E(t)+J(t)+ R(t) +D(t)=1$ redundant)
\beq
 \label{SCEIRS-D-model-mortality}
 \frac{d D(t)}{dt}  = \ds \Big\langle \left\{e_0\delta(t-t_E-t_M) +j_0\delta(t-t_M)\right\}\Theta(t_I-t_M)\Big\rangle 
 + \Big\langle {\cal A}(t-t_E-t_M)\Theta(t_I-t_M) \Big\rangle  . 
\eeq
$S(t), E(t), J(t), R(t), D(t)$ quantify the fractions of the susceptible, exposed, infected and infectious, recovered (immune), and 
dead population with $S(t)+E(t)+J(t)+R(t)+D(t)=1$. The dead compartment D has infinite sojourn time and hence $D(t)$ is non-decreasing.
We denote with
$\Theta(..)$ the Heaviside unit step function, and with $\delta(..)$ the Dirac's $\delta$-function. Moreover, $\Big\langle \ldots\Big\rangle$ stands for averaging with respect to the contained (independent) random variables $t_E,t_I,t_R, t_M >0$.
The waiting time (sojourn time) an agent spends in compartment I (alive) is $t_{IM}=min(t_I,t_M)$, i.e. it is $t_I$ if the individual survives the disease (when $t_M >t_I$, i.e. $\Theta(t_M-t_I)=1$), and is $t_M$ if the individual dies during being in compartment I (when $t_I>t_M$, i.e. $\Theta(t_I-t_M)=1$). More strictly speaking, $t_I$ is actually the sojourn time in I for zero mortality which is retrieved for $t_M=\infty$. We consider general initial conditions 
$S(0)=s_0$, $J(0)=j_0$, $E(0)=e_0$, $R(0)=r_0$, but $D(0)=0$, where the time instant $0$ marks the beginning of the observation. 
It has to be pointed out that (\ref{SCEIRS-D-model}) is causal, i.e. the infection rate ${\cal A}$ is a causal function\footnote{We call a function $A$ causal if $A(t)=0$ for $t<0$.}, meaning that
all contributions for which ${\cal A}$ has negative time-arguments are null.
One can see in above evolution equations that agents have a memory of the infection history which is wiped out
after one SEIRS cycle, i.e. once an agent re-enters compartment S alive.

Indeed, all variants of transition pathways considered hereafter are contained as special cases in system (\ref{SCEIRS-D-model}).
In our demonstration, we will extensively use Laplace transforms.
We define the Laplace transform (LT) of a suitable function $f(t)$ as
\beq
\label{LT_def}
{\widehat f}(\lambda) = \int_0^{\infty} f(t) e^{-\lambda t} {\rm d}t 
\eeq
with suitably chosen Laplace variable $\lambda$.
The waiting times (sojourn times) $t_{E,I,R,M}$ in the compartments are drawn from the probability density functions (PDFs)
\beq
\label{compartment_PDFs}
Prob(t_{E,I,R,M}\in [\tau,\tau+{\rm d}\tau]) =  K_{E,I,R,M}(\tau){\rm d}\tau .
\eeq
Generally applies the averaging rule 
\beq
\label{averag_rule}
\Big\langle f(t_{E,I,R,M}) \Big\rangle = \int_0^{\infty}f(\tau) K_{E,I,R,M}(\tau){\rm d}\tau 
\eeq
from which we observe that $ K_{E,I,R,M}(t) = \Big\langle \delta(t-t_{E,I,R,M}) \Big\rangle$ retrieving the waiting time PDFs.
For convolutions of two causal functions $a(t), b(t)$, 
we will use the notation $(a \star b)(t) = \int_0^t a(\tau)b(t-\tau){\rm d}\tau$.
Applying (\ref{averag_rule}) on causal functions $A$ such as the infection rate (\ref {infection_law}) takes us to the convolutional relation
\beq
\label{convol_mean}
\Big\langle A(t-t_{E,I,R,M}) \Big\rangle = (A \star K_{E,I,R,M})(t)  = \int_0^tA(t-\tau)K_{E,I,R,M}(\tau){\rm d}\tau .
\eeq
Of interest later on is the persistence probability 
\beq
\label{peristent_probas}
\Phi_{E,I,R,M}(t)= Prob[t_{E,I,R,M} >t]= \Big\langle\Theta(t_{E,I,R,M}-t)\Big\rangle = 1-\int_0^t K_{E,I,R,M}(\tau){\rm d}\tau 
\eeq
with $\Phi_{E,I,R,M}(0) =1$ (telling that $t_{E,I,R,M} >0$) and
$\Phi_{E,I,R,M}(\infty) = 0$ reflecting proper normalization of the PDFs $K_{E,I,R,M}(t)$, 
and telling us that $t_{E,I,R,M} < \infty$.
The complementary probabilities of $\Phi_{E,I,R,M}(t)$ are the cumulative distribution functions for which we use the equivalent notations
\beq
\label{cumulative_dist}
\begin{array}{clr}
\ds 1-\Phi_{E,I,R,M}(t) & = \ds  Prob[t_{E,I,R,M} < t]= \Big\langle\Theta(t-t_{E,I,R,M})\Big\rangle = 
\left(1 \star K_{E,I,R,M}\right)(t) &\\[2ex]
&= \ds \int_0^t\Theta(t-\tau)K_{E,I,R,M}(\tau){\rm d}\tau =\int_0^t K_{E,I,R,M}(\tau){\rm d}\tau , &
\end{array}
\eeq
i.e., we identify throughout our paper "$1$" in convolutions with the (causal)
Heaviside function $\Theta(t)$.
We also will need the PDF governing the transitions (dead or alive) out from I, namely
\beq
\label{end_of_inf}
K_{IM}(t) = - \frac{d}{dt} \left\{\Big\langle \Theta(t_M-t)\Big\rangle\Big\langle \Theta(t_I-t) \Big\rangle\right\} = - \frac{d}{dt}[\Phi_M(t)\Phi_I(t)] = b_d(t)+b_r(t)
\eeq
where we use the independence of the waiting times $t_M, t_I$. In this relation we have for the persistence probability
that an agent is in I (and alive)
\beq
\label{Phi_IM_def}
\Phi_{IM}(t) = Prob[min(t_I,t_M) >t] =  \Big\langle [\Theta(t_M-t)\Big\rangle\Big\langle \Theta(t_I-t)] \Big\rangle = \Phi_M(t)\Phi_I(t).
\eeq
(\ref{end_of_inf}) contains the `defective' PDFs
\beq
\label{b_r_def}
b_r(t) = \Big\langle \delta(t-t_I)\Big\rangle\Big\langle \Theta(t_M-t) \Big\rangle = \Big\langle \delta(t-t_I)\Theta(t_M-t_I) \Big\rangle = K_I(t) \Phi_M(t)
\eeq
governing the transitions of agents out of compartment I alive, i.e. I $\to$ R. 
Further
\beq
\label{b_d_def}
b_d(t) = \Big\langle \delta(t-t_M)\Big\rangle\Big\langle \Theta(t_I-t) \Big\rangle = \Big\langle \delta(t-t_M)\Theta(t_I-t_M) \Big\rangle = K_M(t) \Phi_I(t)
\eeq
describes the transitions I $\to$ D.
We call a PDF `defective' when it lacks proper normalization \cite{Dono_Mi_et_al2024}, i.e.
$\int_0^{\infty} b_r(t){\rm d}t =B_r < \int_0^{\infty}K_I(t){\rm d}t = 1$ (where the r.h.s. is an upper bound representing the limit without mortality corresponding to $t_M=\infty$, $\Phi_M(t)=1$). In contrast, the above PDFs $K_{E,I,R,M}$ and $K_{IM}$
of (\ref{compartment_PDFs}) and (\ref{end_of_inf}) are properly normalized
since $t_{E,I,R,M}$ and $t_{IM}$ are finite. We call such PDFs `complete' or non-defective.
The quantity $B_d=\int_0^{\infty} b_d(t){\rm d}t \in [0,1)$ quantifies the overall probability to 
die from the disease and $B_r = \int_0^{\infty} b_r(t){\rm d}t =1-B_d$
indicates the complementary probability, i.e. the probability to survive the disease.
Since $t_{E,I,R,M}$ are independent random variables, one has with (\ref{averag_rule})
\beq
\label{LTs}
\begin{array}{clr}
\ds \Big\langle e^{-\lambda(t_E+t_I+t_R)}\Big\rangle & = \ds  \Big\langle e^{-\lambda t_E}\Big\rangle  \Big\langle e^{-\lambda t_I}\Big\rangle  \Big\langle e^{-\lambda t_R}\Big\rangle , \hspace{1cm} \Re\{\lambda\} > 0
& \\[2ex]
& = \ds  {\widehat K}_E(\lambda) {\widehat K}_I(\lambda) {\widehat K_R}(\lambda) &\\[2ex]
& = \ds \Big\langle \int_0^{\infty} e^{-\lambda t}\delta(t-(t_E+t_I+t_R)) {\rm d}t \Big\rangle
 &
\end{array} 
\eeq
involving the LTs of the convolution sequence of PDFs $\Big\langle \delta(t-(t_E+t_I+t_R)) \Big\rangle = (K_E\star K_I\star K_R)(t)$ which itself is
the PDF from which is drawn the random time $T=t_E+t_I+t_R$.
\begin{figure}[H]
\centerline{
\includegraphics[width=0.55\textwidth]{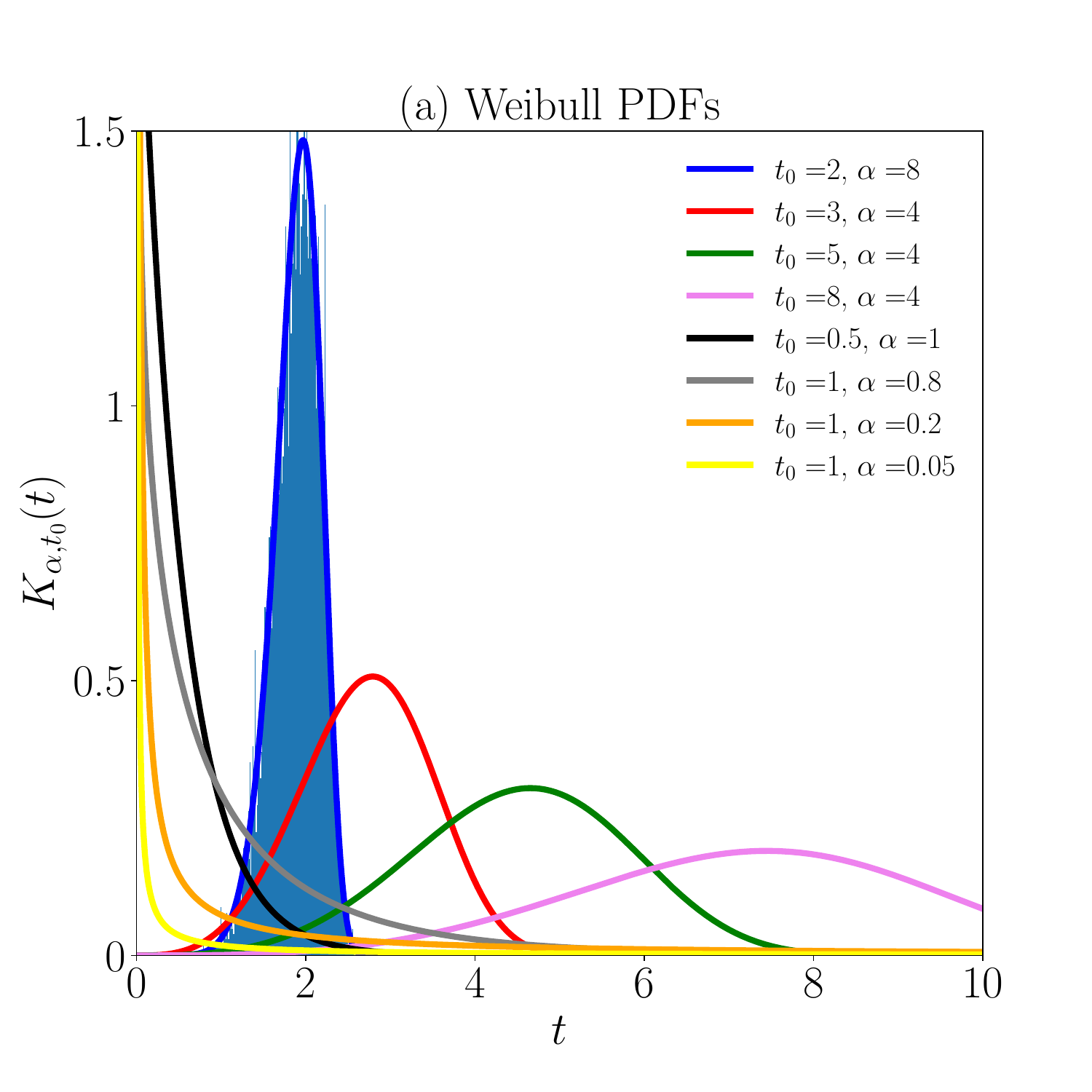} 
\includegraphics[width=0.55\textwidth]{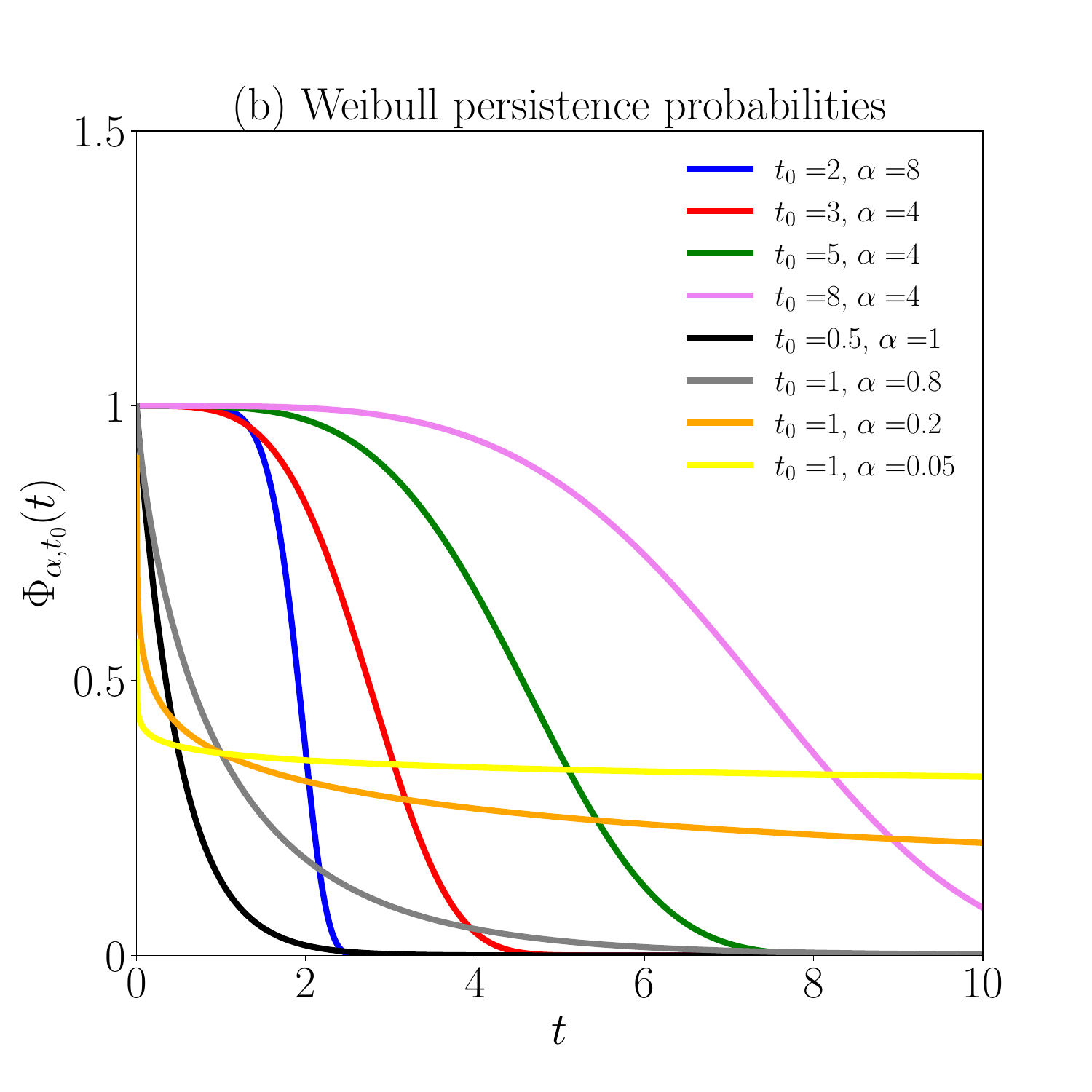}
}
\caption{
(a) Weibull PDF (\ref{Weibull_proba_funct}) for selected values of the Weibull parameters $\alpha$ and $t_0$. For $t_0=2$ and $\alpha=8$, we also show a normalized histogram (blue zone)
 obtained from $10^4$ realizations of Weibull waiting times, which closely reproduces the analytical Weibull PDF.
%
%
(b) Weibull persistence probability (\ref{persistence_weib}) for the same parameter values and color codes as in frame (a).
}
\label{weibull_pdf}
\end{figure}
In the numerical evaluations of the present paper we mainly focus on compartmental sojourn times $t_{E,I,R}$ and $t_M$ drawn from Weibull distributions \cite{Weibull_1951}. The Weibull PDF is given by
\beq
\label{Weibull_proba_funct}
K_{\alpha,t_0}(t) = \frac{\alpha}{t_0^{\alpha}} t^{\alpha-1} \exp{\Big\{-\frac{t^\alpha}{t_0^\alpha}\Big\}} , \hspace{1cm} \alpha, t_0 >0
\eeq
containing two parameters, the shape parameter $\alpha$ and the scale parameter $t_0$. It has the persistence distribution (`survival function)'
which is a stretched exponential
\beq
\label{persistence_weib}
\Phi_{\alpha,t_0}(t) = \int_t^{\infty} K_{\alpha,t_0}(t){\rm d}t' =  \exp{\Big\{-\frac{t^\alpha}{t_0^\alpha}\Big\}} .
\eeq
For $\alpha=1$ the exponential distribution is retrieved and for $0< \alpha \leq 1 $ the Weibull PDF and its survival function are completely monotonic \cite{SchillingVondracek2012},
i.e. they fulfil $$(-)^m \frac{d^m}{dt^m}(K(t), \Phi(t)) >0 , \hspace{1cm} m =0,1,2,\ldots \in \mathbb{N}.$$
For $0<\alpha <1$ the Weibull PDF is weakly singular and
exhibits for $\alpha>1$ unimodal shape with a limit to the Dirac's $\delta$-distribution
\beq
\label{delta_limit}
\lim_{\alpha \to \infty} K_{\alpha,t_0}(t) = \delta(t-t_0)
\eeq
which corresponds to sharp waiting times.
This can easily be seen from the feature of (\ref{persistence_weib}): Since $\ds \lim_{\alpha \to \infty}\exp{\Big\{-\frac{t^\alpha}{t_0^\alpha}\Big\}} 
= \Theta(t_0-t)$, one has $K_{\alpha,t_0}(t) = -\frac{d}{dt} \Phi_{\alpha,t_0}(t) \to \delta(t-t_0)$.
We will also need the Weibull mean
\beq
\label{Weibull_mean}
\Big\langle t_{\alpha,t_0} \Big\rangle = \int_0^{\infty} 
t K_{\alpha,t_0}(t){\rm d}t 
= t_0 \int_0^{\infty} u^{\frac{1}{\alpha}} e^{-u}{\rm d}u = t_0 \,\Gamma\left(1+\frac{1}{\alpha}\right) .
\eeq
For $\alpha \to \infty$ one has $\Big\langle t_{\alpha,t_0} \Big\rangle \to t_0$, 
consistent with the limit (\ref{delta_limit}).
We depict some characteristic Weibull PDFs and their persistence probabilities in Fig. \ref{weibull_pdf} for some values of the parameters.

Above machinery allows averaging of the stochastic system (\ref{SCEIRS-D-model}) 
leading to its `memory representation' (see Appendix \ref{explicit_evol_eqs})
\beq
\label{convol_memory_SEIRDS}
\begin{array}{clr}
\ds \frac{dS(t)}{dt} = & \ds - {\cal A}(t) + (M_R \star R)(t) &\\[3ex]
\ds \frac{dE(t)}{dt}  = & \ds - (M_E \star E)(t) + {\cal A}(t)  & \\[3ex]
\ds \frac{dJ(t)}{dt} = & \ds -(M_{IM} \star J)(t) + (M_E \star E)(t) &\\[3ex]
\ds \frac{dR(t)}{dt} = & \ds  -(M_R \star R)(t) + (m_r \star J)(t)  & \\[3ex]
\ds \frac{dD(t)}{dt} = & \ds (m_d \star J)(t) &
\end{array}
\eeq
containing the `memory kernels' $M_{E,I,R}, M_{IM}(t),\, m_{r,d}(t)$ defined by
\beq
\label{memory_kernels}
\begin{array}{lcr}
\ds K_{E,I,R}(t) &= \ds (M_{E,I,R} \star \Phi_{E,I,R})(t)  & \\[2ex]
\ds K_{IM}(t) & = \ds  (M_{IM} \star \Phi_{IM})(t) & \\[2ex]
\ds b_{r,d}(t) & = \ds  (m_{r,d} \star \Phi_{IM})(t) , &
\end{array}
\eeq
and recall (\ref{end_of_inf})-(\ref{b_d_def}). In the last equation, we observe that $m_r(t)+m_d(t) = M_{IM}(t)$ reflecting $b_r(t)+b_d(t) = K_{IM}(t)$.
The immortal limit can be readily retrieved by setting $m_d(t)=0$ and 
$m_r(t) = M_{IM}(t) = M_I(t)$ corresponding to $t_M=\infty$.
We define a memory kernel $M(t)$ such that it connects a PDF $K(t)$ with its persistence probability $\Phi(t)=1 - \int_0^tK(\tau){\rm d}\tau$ by a convolutional relation, namely
$K(t) = (M \star \Phi)(t)$. This uniquely determines the memory kernel
which reads in Laplace space
\beq
\label{memor_kernel_general_der}
{\widehat M}(\lambda) = \frac{{\widehat K}(\lambda)}{{\widehat \Phi}(\lambda)} = \lambda  \frac{{\widehat K}(\lambda)}{1-{\widehat K}(\lambda)} = 
\lambda {\widehat {\cal L}}(\lambda) .
\eeq
In the subsequent discussion we refer to the kernel ${\cal L}(t)$ as `associated memory kernel' (AMK).
The AMK has an evocative interpretation: By expanding its LT into a geometrical series and inverting it back to time leads to an infinite series of successive convolutions of
the PDF $K(t)$ with itself, namely
\beq
\label{exit_rate}
{\cal L}(t) = \sum_{n=1}^{\infty} K^{(n)}(t) , \hspace{1cm} K^{(n)}(t) =
(K^{(n-1)}\star K)(t)
\eeq
where $K^{(1)}(t)=K(t)$. Since all PDFs $K^{(n)}(t) \geq 0$, it follows that 
${\cal L}(t)$ is a non-negative function (with units $sec^{-1}$), 
and can be conceived as the mean exit rate from a respective compartment 
when a unit fraction of agents is considered. From (\ref{memor_kernel_general_der})
one can see (see (\ref{limit_value})) that the AMK converges for late times to the inverse of the mean of the respective 
compartmental sojourn time ${\cal L}_{E,I,R}(\infty) = 1/\big\langle t_{E,I,R} \big\rangle$.
The convolution $(E\star {\cal L}_E)(t)$ is the fraction of agents having left the compartment E up to time $t$.
We will see this a little later more closely (system (\ref{convol_memory_SEIRDS_gen_der_integral})). 
There is also a connection with renewal processes: In this context the AMK (\ref{exit_rate}) plays the role of the mean renewal rate in a continuous time renewal process in which the renewal times are drawn from PDF $K(t)$ \cite{Kochubei2011}, and see \cite{mi_dono_et_al_resetting2025}
for a corresponding derivation of renewal rates in the context of stochastic resetting.

It has to be emphasized that in a specific case local integrability of the respective memory kernels $M(t)$ has to be verified, consult \cite{Diethelm2020} for related issues concerning singularities. For non-integrable memory kernels representation (\ref{convol_memory_SEIRDS}) does not exist.
Instead, one has to employ
the connection with general (fractional) derivatives, see \cite{Kochubei2011} for this theory. The general (fractional) derivative of a sufficiently good function $y(t)$
in the Riemann-Liouville (R-L) sense is defined by 
\beq
\label{RL_sense}
D_t y(t) = \frac{d}{dt}\int_0^t k(t-\tau)y(\tau){\rm d}\tau =\frac{d}{dt}(k \star y)(t)
\eeq
where the kernel $k(t)$ is a non-negative locally integrable function. For $k(t) =\delta(t)$ (\ref{RL_sense}) recovers the standard first order derivative.
(\ref{RL_sense}) has the LT 
$\lambda {\widehat k}(\lambda) {\widehat y}(\lambda)$ 
(given that $\lim_{\tau \to 0}(k \star y)(\tau) =0$). 
This allows us to rewrite (\ref{convol_memory_SEIRDS}) in terms of general 
(fractional) derivatives in the R-L sense, namely
\beq
\label{convol_memory_SEIRDS_gen_der}
\begin{array}{clr}
\ds \frac{dS(t)}{dt} = & \ds - {\cal A}(t) + \frac{d}{dt}\int_0^t{\cal L}_R(t-\tau)R(\tau){\rm d}\tau &\\[3ex]
\ds \frac{dE(t)}{dt}  = & \ds - \frac{d}{dt} \int_0^t{\cal L}_E(t-\tau)E(\tau){\rm d}\tau
 + {\cal A}(t)  & \\[3ex]
\ds \frac{dJ(t)}{dt} = & \ds - \frac{d}{dt} \int_0^t{\cal L}_{IM}(t-\tau)J(\tau){\rm d}\tau
 + \frac{d}{dt} \int_0^t{\cal L}_E(t-\tau)E(\tau){\rm d}\tau &\\[3ex]
\ds \frac{dR(t)}{dt} = & \ds  -\frac{d}{dt}\int_0^t{\cal L}_R(t-\tau)R(\tau){\rm d}\tau
 + \frac{d}{dt}\int_0^t{\cal L}_r(t-\tau)J(\tau){\rm d}\tau
 & \\[3ex]
\ds \frac{dD(t)}{dt} = & \ds \frac{d}{dt}\int_0^t{\cal L}_d(t-\tau)J(\tau){\rm d}\tau . &
\end{array}
\eeq
We point out that these equations are (slightly) different from the ones obtained in 
\cite{Granger-et-al2023} since here
the transitions of the initial conditions are taken into account (due to the terms $ \Big\langle \delta(t-t_{E,I,R}) \Big\rangle$). 
It is straightforward to integrate
(\ref{convol_memory_SEIRDS_gen_der}) to obtain a self-consistent memory representation equivalent to (\ref{convol_integraytion_appendix}) in terms of convolutions with AMKs, namely
\beq
\label{convol_memory_SEIRDS_gen_der_integral}
\begin{array}{clr}
\ds S(t) = & \ds s_0 - (1 \star {\cal A})(t) +({\cal L}_R \star R)(t) &\\[3ex] 
\ds E(t)  = & \ds e_0  + (1 \star {\cal A})(t) - ({\cal L}_E \star E)(t)  & \\[3ex]
\ds J(t) = & \ds j_0 +({\cal L}_E \star E)(t)- ({\cal L}_{IM} \star J)(t) &\\[3ex]
\ds R(t)= & \ds r_0 
 + ({\cal L}_r\star J)(t) - ({\cal L}_R \star R)(t) & \\[3ex]
\ds D(t) = & \ds ({\cal L}_d\star J)(t) . &
\end{array}
\eeq

\subsection{Connection with fractional dynamics}
\label{fractional_dyn}
Especially interesting is the question on the connection with fractional dynamics. In order to shed light on this
widely discussed issue, we consider the case of long memory for which the AMK exhibits a slow power-law decay ${\cal L}_{\alpha}(t) = \xi \frac{t^{\alpha-1}}{\Gamma(\alpha)}$ ($\alpha \in (0,1)$)
where $\xi >0$ is a dimensional constant with units ${\rm sec}^{-\alpha}$.
Importantly this kernel 
is weakly singular at $t=0$ and locally integrable 
having LT ${\widehat {\cal L}}_{\alpha}(\lambda) = \xi \lambda^{-\alpha}$ with
${\widehat M}_{\alpha}(\lambda) = \xi \lambda^{1-\alpha}$ 
whereas the memory kernel $M_{\alpha}(t)$
is fat-tailed and non-integrable due to a strong $t^{\alpha-2}$--singularity, and hence
representation (\ref{convol_memory_SEIRDS}) does not exist. 
Therefore, convolutional relations in Laplace space 
$\lambda {\widehat {\cal L}}_{\alpha}(\lambda){\widehat y}(\lambda)$ (with $\alpha \in (0,1)$) have to be inverted using (\ref{RL_sense}) involving the integrable AMK, taking us to a Riemann-Liouville fractional derivative of order $\nu=1-\alpha$, namely
\beq
\label{RL_der}
{\cal D}_t^{\nu} y(t) = \frac{d}{dt} D^{-(1-\nu)} y(t) = \frac{d}{dt} 
\int_0^t \frac{(t-\tau)^{-\nu}}{\Gamma(1-\nu)}y(\tau){\rm d}\tau , 
\hspace{1cm} \nu =1-\alpha \in (0,1)   
\eeq
in which $D^{-(1-\nu)} y(t)$ is the R-L fractional integral of order $1-\nu$.
Recall that for $\nu \to 1_{-}$ (\ref{RL_der}) recovers the standard first order derivative and
$\nu \to 0_{+}$ identity.
Among the myriad of existing literature on the subject, we refer to \cite{Samko_Kilbas1993,Metzler_Klafter2000,Gorenflo-et-al2008,Mainardi_Gorenflo2007} (and see the references therein) for the related theory. 
Relation (\ref{memor_kernel_general_der}) determines uniquely
the LT of waiting time PDF as
\beq
\label{ML_PDF}
{\widehat K}_{\alpha}(\lambda) = 
\frac{{\widehat {\cal L}}_{\alpha}(\lambda)}{1+ {\widehat {\cal L}}_{\alpha}(\lambda)} = \frac{\xi}{\xi + \lambda^{\alpha}} , \hspace{1cm} \alpha \in (0,1).
\eeq
Its inversion yields the Mittag-Leffler (M-L) PDF 
\cite{Gorenflo-et-al2008,Mainardi_Gorenflo2007}, exhibiting a fat-tailed $t^{-\alpha-1}$--late time scaling, a feature which is well-known in fractional dynamics.
Assuming now fat-tailed distributed $t_{E,R}$ (scaling respectively 
with $t^{-\alpha_{E,R}-1}$--laws) yields for (\ref{convol_memory_SEIRDS_gen_der}) the system
\beq
\label{convol_memory_SEIRDS_fractional RL_der}
\begin{array}{clr}
\ds \frac{dS(t)}{dt} = & \ds - {\cal A}(t) + \xi_R {\cal D}_t^{1-\alpha_R} R(t)
 &\\[3ex]
\ds \frac{dE(t)}{dt}  = & \ds - \xi_E {\cal D}_t^{1-\alpha_E} E(t)
 + {\cal A}(t)  & \\[3ex]
\ds \frac{dJ(t)}{dt} = & \ds - \frac{d}{dt} ({\cal L}_{IM} \star J)(t)
 + \xi_E D_t^{1-\alpha_E} E(t) &\\[3ex]
\ds \frac{dR(t)}{dt} = & \ds  - \xi_R {\cal D}_t^{1-\alpha_R} R(t) 
 + \frac{d}{dt} ({\cal L}_r \star J)(t) 
 & \\[3ex]
\ds \frac{dD(t)}{dt} = & \ds \frac{d}{dt} ({\cal L}_d \star J)(t)  &
\end{array} ,   \hspace{1cm} \alpha_{E,R} \in (0,1)
\eeq
involving on the r.h.s. R-L-fractional derivatives defined in (\ref{RL_der}). We left the distributions of $t_M$ and $t_{IM}=min(t_M,t_I)$ unspecified.
Moreover, we observe that letting $\alpha_{E,R} \to 1$ represents the memoryless limit ${\cal D}_t^{1-\alpha} y(t) \to y(t)$ equivalent to
$M_{\alpha\to 1}(t)= \xi \delta(t)$ corresponding to exponentially distributed waiting times 
widely considered in the literature.

\subsection{SEIRS-SEID transition pathway with vanishing infection rate}
\label{SEIRS_zero_infect}
It has to be pointed out that accounting for the transition rates $\Big\langle \delta(t-t_{E,I,R,M}) \Big\rangle$ in system (\ref{SCEIRS-D-model}) describing the transition dynamics of the initial state is crucial 
in order to obtain fixed points which are independent of the initial conditions. Without these terms a given initial state $s_0, e_0, j_0, r_0$ would persist forever which is in contradiction to our assumption that agents stay in the compartments E, I, R for finite sojourn times. 
To demonstrate their effect, it is instructive to consider zero infection rate $\beta=0$, 
i.e. ${\cal A}(t)=0$ in presence of mortality. 
We consider both, finite waiting times $t_{E,I,R,M} < \infty$  
(case {\bf (i)}), and thereafter {\bf (ii)} of everlasting immunity 
$t_R=\infty$ and $t_{E,I,M} < \infty$).
System (\ref{SCEIRS-D-model}) with initial condition $[S(0), E(0), J(0), R(0)] = [s_0, e_0,j_0,r_0]$ 
then boils down to
\beq
 \label{SCEIRS-D-model_null_infection}
 \begin{array}{clr}
 \displaystyle \frac{d S(t)}{dt} = & \ds  \Big\langle \left\{e_0\delta(t-t_E-t_I-t_R) +j_0\delta(t-t_I-t_R) \right\}\Theta(t_M-t_I) \Big\rangle + r_0\Big\langle \delta(t-t_R) \Big\rangle &\\[3ex]
 \displaystyle \frac{d E(t)}{dt}  = & \ds  -e_0\Big\langle \delta(t-t_E)\Big\rangle   &\\[3ex] 
 \displaystyle  \frac{d J(t)}{dt}   = & \ds  e_0\Big\langle \delta(t-t_E)\Big\rangle  +
 j_0 \frac{d}{dt} \Big\langle \Theta(t_M-t)\Theta(t_I-t)\Big\rangle   & \\[3ex]
& \ds  - e_0\Big\langle\Theta(t_M-t_I)\delta(t-t_E-t_I) \Big\rangle -e_0\Big\langle\Theta(t_I-t_M)\delta(t-t_E-t_M) \Big\rangle & \\[3ex]
 \displaystyle  \frac{d R(t)}{dt} = & \ds   j_0 \Big\langle \left\{ \delta(t-t_I) -\delta(t-t_I-t_R) \right\} \Theta(t_M-t_I) \Big\rangle  
 - r_0 \Big\langle \delta(t-t_R)\Big\rangle & \\[3ex]
 & \ds +  e_0\Big\langle \left\{\delta(t-t_E-t_I) -\delta(t-t_E-t_I-t_R) \right\} \Theta(t_M-t_I) \Big\rangle
 \end{array}
\eeq
and the redundant equation for the mortality rate becomes
\beq
\label{mortality_rate_without_infection}
\frac{d D(t)}{dt}  = \ds \Big\langle \left\{e_0\delta(t-t_E-t_M) +j_0\delta(t-t_M)\right\}\Theta(t_I-t_M)\Big\rangle .
\eeq
These equations describe the relaxation of the initial conditions to an asymptotically disease-free state.
We can see this by integrating (\ref{SCEIRS-D-model_null_infection}) directly to arrive at the explicit solution 
\beq
\label{SCEIRS-D-model_null_infection_integtral}
\begin{array}{clr}
\ds S(t) = & \ds s_0 + e_0(1\star b_r\star K_E\star K_R)(t) + j_0(1 \star b_r\star K_R)(t)+r_0 (1\star K_R)(t)   &\\[3ex]
\ds E(t)  = & \ds e_0\Phi_E(t)   & \\[3ex]
\ds J(t) = & \ds j_0 \Phi_{IM}(t) + e_0 \left(\Phi_{IM} \star K_E\right)(t)   &\\[3ex]
\ds R(t) = & \ds  r_0\Phi_R(t) + j_0 \left(\Phi_R \star b_r\right)(t)+e_0 \left(\Phi_R \star K_E \star b_r\right)(t)   &\\[3ex]
\ds D(t) = & \ds 1- S(t)-E(t)-J(t)-R(t) = j_0 \left(1 \star b_d\right)(t)+e_0 \left(1 \star K_E \star b_d\right)(t)  &
\end{array}
\eeq
This solution contains the dynamics without mortality (limit $t_M\to \infty$) for $b_r(t)=K_{IM}(t)=K_I(t)$, $b_d(t)=0$ and $\Phi_{IM}(t) =\Phi_I(t)$ (thus $D(t)=0$). Accounting for $\Phi_{E,I,R,IM}(0)=1$ and
$\Phi_{E,I,R,IM}(\infty)=0$ takes us to the disease-free late time limit of 
(\ref{SCEIRS-D-model_null_infection_integtral}), namely
\beq
\label{SCEIRS-D-model_null_infection_integtral_infinite_time}
\begin{array}{clr}
\ds S(\infty) = & \ds s_0 + r_0 + (e_0+j_0) B_r = 1- B_d(e_0+j_0)   &\\[3ex]
\ds E(\infty)  = & \ds J(\infty) =  R(\infty) = 0   &\\[3ex]
\ds D(\infty) = & \ds  B_d(e_0+j_0) &
\end{array}
\eeq
i.e. the compartments with finite sojourn times E, I and R become empty, whereas $S(t)$ and $D(t)$ converge to finite values. Without infections (in case {\bf (i)}) only non-cyclic pathways 
occur which either end in S or in D without exits from S. The fraction $r_0$ of immune agents are not exposed to mortality and all end up in S, whereas agents $e_0,j_0$ are exposed to mortality during their sojourn in I.
System (\ref{SCEIRS-D-model_null_infection_integtral_infinite_time}) contains
the probabilities that an agent survives ($B_r$) or dies ($B_d$) from the disease
\beq
\label{BRD_relations}
B_{r,d} = \int_0^{\infty}b_{r,d}(t){\rm d}t 
\eeq
with $B_r+B_d=1$, recall (\ref{b_r_def}), (\ref{b_d_def}).
For zero mortality we have $B_d=0$ where the disease-free globally stable fixed point $S(\infty)=1$ is approached.

\paragraph{Everlasting immunity without infections $\beta=0$}
We now consider case {\bf (ii)} for {\it everlasting immunity} ($t_R=\infty$ with 
$\Phi_R(t) =1$ and $K_R(t)=0$ whereas $t_{E,I} <\infty$).
Then (\ref{SCEIRS-D-model_null_infection_integtral}) reads
\beq
\label{SCEIRS-D-model_null_infection_integtral_immunity_ever}
\begin{array}{clr}
\ds S(t) = & \ds s_0   &\\[3ex]
\ds E(t)  = & \ds e_0\Phi_E(t) \to 0  & \\[3ex]
\ds J(t) = & \ds j_0 \Phi_{IM}(t) + e_0 \left(\Phi_{IM} \star K_E\right)(t) \to 0  &\\[3ex]
\ds R(t) = & \ds  r_0 + j_0 \left(1 \star b_r\right)(t)+e_0 \left(1 \star K_E \star b_r\right)(t) \to r_0 + B_r(e_0+j_0)  &\\[3ex]
\ds D(t) = & \ds 1- S(t)-E(t)-J(t)-R(t) =   j_0 \left(1 \star b_d\right)(t)+e_0 \left(1 \star K_E \star b_d\right)(t) \to B_d(e_0+j_0) . &
\end{array}
\eeq
The governing memory equations can be easily retrieved from (\ref{convol_memory_SEIRDS}) by setting $M_R(t) = 0$ ($K_R(t)=0$),
reflecting absence of transitions from R to S.
Here, the fraction of agents that survive the disease accumulate in the everlasting compartment R 
saturating to the value $R(\infty)$ instead of returning to S.
In both cases {\bf (i)} and {\bf (ii)} the fraction of dead agents is the same, namely $D(\infty) =B_d(e_0+j_0)$ where
compartments with finite sojourn times become empty at infinite times.
For zero mortality ($B_d=0$) the disease-free state $S(\infty)=s_0$, $R(\infty) =1-s_0$ is approached
lying on the one-dimensional manifold of fixed points $S+R=1$ of case {\bf (ii)}.

\subsection{Late time asymptotics of the full SEIRS-SEID system}
\label{full_SEIRDS_latetime}
Of interest is the late time asymptotics of the full system 
(\ref{convol_integraytion_appendix}) with non-zero infection rates. We consider first case {\bf (i)}.
Turning into the Laplace space leads to
\beq
\label{convol_integration_laplace}
\begin{array}{clr}
\ds \lambda {\widehat S}(\lambda) = & \ds s_0 -\left\{1 - {\widehat b}_r(\lambda) {\widehat K}_E(\lambda) {\widehat K}_R(\lambda \right\} {\widehat {\cal A}}(\lambda)+ e_0 {\widehat b}_r(\lambda){\widehat K}_E(\lambda) {\widehat K}_R(\lambda)+ j_0{\widehat b}_r(\lambda){\widehat K}_R(\lambda)+r_0 {\widehat K}_R(\lambda)   &\\[3ex]
\ds \lambda {\widehat E}(\lambda)  = & \ds 
[1-{\widehat K_E}(\lambda)] \left\{e_0 +{\widehat {\cal A}}(\lambda)\right\}  & \\[3ex]
\ds \lambda{\widehat J}(\lambda) = & \ds [1-{\widehat K}_{IM}(\lambda)]\left\{j_0 + 
 {\widehat K}_E(\lambda) \left(e_0 + {\widehat {\cal A}}(\lambda)\right) \right\}   &\\[3ex]
\ds \lambda {\widehat R}(\lambda) = & \ds  [1-{\widehat K}_R(\lambda)] \left\{r_0 + {\widehat b}_r(\lambda)\left[j_0+
{\widehat K}_E(\lambda) \left(e_0 + {\widehat {\cal A}}(\lambda)\right)\,\right]\,
\right\} 
&\\[3ex]
\ds \lambda {\widehat D}(\lambda) = &
{\widehat b}_d(\lambda) \left\{ j_0 +  
{\widehat K}_E(\lambda) \left[e_0 + {\widehat {\cal A}}(\lambda) \, \right] \, \right\}
  &
\end{array}
\eeq
which per construction fulfills $\lambda\left({\widehat S}(\lambda)+{\widehat E}(\lambda)+{\widehat J}(\lambda)+{\widehat R}(\lambda)+{\widehat D}(\lambda)\right)=1$. Recall (\ref{end_of_inf}) and consult Appendix \ref{explicit_evol_eqs} for complementary details. 
In order to explore the late time behavior we make use of the limit value theorem of LTs
\beq
\label{limit_value}
f(\infty) = \lim_{t\to \infty} f(t) = \lim_{\lambda \to 0} \lambda {\widehat f}(\lambda)
\eeq
together with (recall (\ref{BRD_relations}))
\beq
\label{defective_LT}
B_r = {\widehat b}_r(0)= {\widehat b}_r(\lambda)\Big|_{\lambda = 0} = \int_0^{\infty} \Phi_M(t)K_I(t){\rm d}t = 1-B_d \leq 1
\eeq
and $B_d={\widehat b}_d(0) = \int_0^{\infty} K_M(t)\Phi_I(t){\rm d}t$ with $B_r+B_d=1$. 
$B_r$ indicates the overall probability that an agent survives the disease (probability to recover), and the $B_d$ the `mortality risk', i.e. the overall probability to die from the disease. 
In the limit of no mortality one has $t_M\to \infty$ thus $ b_r(t) \to K_I(t)$ and $B_r \to 1$ (and mortality risk $B_d \to 0$). 
We observe that $${\widehat K}_{E,I,R, IM}(\lambda)\Big|_{\lambda=0} = \Big\langle \int_0^{\infty}\delta(t-t_{E,I,R,M, IM}) \,{\rm d}t  \Big\rangle = 1 ,$$ i.e. $t_{E,I,R,M}, t_{IM} < \infty$ corresponds to proper normalization of their PDFs. The effect of mortality is such that
\beq
\label{infect_hist}
{\widehat {\cal A}}(0) = (1 \star {\cal A})(\infty)  = {\widehat {\cal A}}(\lambda)\Big|_{\lambda=0} = 
\beta \int_0^{\infty} S(t)\Lambda(t) {\rm d}t < \infty
\eeq
implying that
${\cal A}(t) = \beta S(t) \Lambda(t) \to 0$ as $t \to \infty$, especially 
$\Lambda(\infty)=0$ as a result of the interplay of mortality and the finiteness of $t_{E,I}$.
Letting $\lambda \to 0$ in (\ref{convol_integration_laplace}) yields the late time limit
\beq
\label{infinite_time_limit_with_mort}
\begin{array}{clr}
\ds S(\infty) & = \ds s_0+r_0 +B_r(e_0+j_0) -B_d {\widehat {\cal A}}(0) = 1 - D(\infty)&\\[3ex]
\ds E(\infty) & =\ds  J(\infty) = R(\infty) =0 &\\[3ex] 
\ds D(\infty) & = \ds B_d\left(e_0+j_0+{\widehat {\cal A}}(0)\right) &
\end{array}
\eeq
i.e., compartments with a finite sojourn times become empty. For $\beta=0$ (\ref{infinite_time_limit_with_mort}) retrieves our previous result for no infections
(\ref{SCEIRS-D-model_null_infection_integtral_infinite_time}).

How many infection cycles an agent is expected to survive? 
To shed light on this issue, we can establish the probability $\Phi_n(t)$ that an agent
at time $t$ has suffered and survived $n$ infections. 
With conditional arguments one has in the Laplace space
the representation
\beq
\label{proba_suffer_n}
{\widehat \Phi}_n(\lambda) = \frac{1- {\widehat b}_r(\lambda)}{\lambda} ({\widehat b}_r(\lambda))^n .
\eeq
In this consideration we can ignore the compartments E and R
with zero mortality as they dot affect the number of survived infection cycles.
If we accounted for them, we would have to replace in the computation to follow 
${\widehat b}_r(\lambda)$
with ${\widehat K}_E(\lambda){\widehat K}_R(\lambda){\widehat b}_r(\lambda)$
which is not changing the final result as we confirm below.
The LT of the expected number of infections 
$\big\langle {\cal N}(t) \big\rangle$ an agent survives is 
\beq
\label{expecte_number_surved}
\Big\langle {\widehat {\cal N}}(\lambda) \Big\rangle = \sum_{n=0}^{\infty} n {\widehat \Phi}_n(\lambda) = \frac{d}{d \zeta} P(\zeta,\lambda) \Big|_{\zeta=1} = \frac{1}{\lambda}\frac{{\widehat b}_r(\lambda)}{1-{\widehat b}_r(\lambda)} =  
\frac{1}{\lambda} \frac{{\widehat b}_r(\lambda)}{1-{\widehat K}_{IM}(\lambda) + {\widehat b}_d(\lambda)} 
\eeq
where we introduced
$$P(\zeta,\lambda)= \sum_{n=0}^{\infty}\zeta^n{\widehat \Phi}_n(\lambda) = 
\frac{1- {\widehat b}_r(\lambda)}{\lambda}\frac{1}{1-\zeta {\widehat b}_r(\lambda)}. $$
From (\ref{expecte_number_surved}) by applying (\ref{limit_value}) we can determine the infinite time limit of the expected number of infections an agent survives as
\beq
\label{late_time}
\big\langle {\cal N}(\infty) \big\rangle = \lim_{\lambda \to 0} \lambda \, \Big\langle {\widehat {\cal N}}(\lambda) \Big\rangle = 
\frac{{\widehat b}_r(0)}{{\widehat b}_d(0)} = \frac{B_r}{B_d} .
\eeq
This number is finite since $b_r(t), b_d(t)$ are defective PDFs (i.e. ${\widehat b}_r(0)=B_r <1$). The cyclic SEIRS infection pathway in presence of mortality is a transient stochastic process.
Clearly (\ref{late_time}) remains invariant if we replace
${\widehat b}_r(\lambda)$
with ${\widehat K}_E(\lambda){\widehat K}_R(\lambda){\widehat b}_r(\lambda)$
since ${\widehat K}_E(0),{\widehat K}_R(0) =1$ justifying as stated above that we ignore compartments without mortality E and R.
The higher the mortality probability $B_d$ to die from an infection, the smaller the expected number of survived infections ${\cal N}(\infty)$ is. In the limit of zero mortality ($B_d\to 0$, $B_r\to 1$), one has
$\big\langle {\cal N}(\infty) \big\rangle \to \infty$ corresponding to the recurrent nature of
the cyclic SEIRS pathway without mortality.

\paragraph{Everlasting immunity: non-cyclic SEIR-SEID pathway}

Consider now the late time behavior for everlasting immunity in case {\bf (ii)}.
Then one has $K_R(t)=0$ (no transitions from R to S) from (\ref{convol_integration_laplace})
\beq
\label{infinite_time_limit_everlasting_imm}
\begin{array}{clr}
\ds S(\infty) & = \ds s_0 - {\widehat {\cal A}}(0) &\\[3ex]
\ds E(\infty) & =\ds  J(\infty) = 0 &\\[3ex] 
\ds R(\infty) & = \ds r_0+ B_r\left(e_0+j_0+{\widehat {\cal A}}(0)\right) &\\[3ex]
\ds D(\infty) & = \ds B_d\left(j_0+ e_0+{\widehat {\cal A}}(0)\right) .&
\end{array}
\eeq
Interestingly, the ratio $\frac{R(\infty) -r_0}{D(\infty)} = \frac{B_r}{B_d}$ depends only on
characteristics of the defective distributions $b_r(t),b_d(t)$ being independent of the infection history and even independent of the specific choice of infection law ${\cal A}(t)$ 
(recall (\ref{infection_law})). We corroborate this result numerically later on (Sects. \ref{SIRD} and \ref{RW_simulations_SEIRDS})).

\section{Spreading condition: Linear stability analysis of the disease-free equilibrium in presence of mortality}
\label{spreading_cond}

Here our goal is to derive the effective reproduction number and the spreading condition
of the full system by investigating the stability of the disease-free fixed points for both cases, namely case {\bf (i)} with $t_{E,I,R,M}< \infty$ where the disease-free fixed point is $S=1$, and case 
{\bf (ii)} $t_R=\infty$, $t_{E,I,M}< \infty$ with disease-free fixed point manifold $S+R=1$.
In case {\bf (i)} it makes sense to perform a linear stability of initial condition $s_0=1$, in case {\bf (ii)}
of the disease-free manifold $s_0+r_0=1$ with presence of `vaccinated' agents $r_0>0$.
To this end we set (recall (\ref{infection_law}))

\beq
\label{ansatz}
S(t) =s_0 +S_1 e^{\xi t} , \hspace{0.5cm} \Lambda(t)= \Lambda_1 e^{\xi t},\hspace{1cm} D(t) = D_1 e^{\xi t}, \hspace{0.5cm} {\cal A}(t)= 
\beta s_0 \Lambda_1 e^{\xi t}
\eeq
with $\Lambda(t)\Big|_{t=0} = \Lambda_0 =\mu_E e_0 +j_0 =0 $ and $S_1, \Lambda_1, D_1 \ll 1$ are small constants. Read $s_0=1$ in case {\bf (i)} and $s_0=1-r_0$ 
in case {\bf (ii)} representing the respective disease-free equilibria.
The limit of zero mortality is contained (by setting $D_1=0$ and $b_d(t)=0$, $K_{IM}(t)=K_I(t)$, 
$\Phi_M(t)=1$) in the subsequent computations. The disease-free state is unstable if there is a $\xi>0$ such that
(\ref{ansatz}) solves (\ref{SCEIRS-D-model}) in linear approximation.
Plugging ansatz (\ref{ansatz}) into (\ref{SCEIRS-D-model}) leads after some manipulations to the homogeneous system 
\beq
\label{linearised_dynamics}
\left(\begin{array}{lll}\ds  1 ;\hspace{0.1cm} & \ds  \beta s_0
\frac{1- {\widehat b}_r(\xi){\widehat K}_E(\xi) {\widehat K}_R(\xi)}{\xi} ;\hspace{0.1cm}& \ds 0 \\[3ex] 
\ds   0 \hspace{0.1cm};  & \ds G_M(\xi) -1 ; \hspace{0.1cm}& \ds 0 \\[3ex] 
 \ds 0   \hspace{0.1cm}  ;& \ds \beta s_0 \frac{{\widehat b}_d(\xi){\widehat K}_E(\xi)}{\xi} \hspace{0.1cm} & \ds -1
  \end{array}\right)\cdot 
\left(\begin{array}{l} S_1 \\[3ex] \Lambda_1 \\[3ex] D_1 \end{array}\right) = \left(\begin{array}{l} 0 \\[3ex] 0 \\[3ex] 0 \end{array}\right) .
\eeq
where the LTs of some PDFs come into play (recall (\ref{LTs})). In case {\bf (ii)}
read ${\widehat K}_R(\xi) = 0$ in (\ref{linearised_dynamics}).
Setting the determinant of above matrix to zero yields
\beq
\label{sovability_cond}
G_M(\xi)-1= 0 ,\hspace{1cm} G_M(\xi) = G_M(\xi,\mu_E) = \beta s_0 {\widehat f}_{IM}(\xi;\mu_E) , \hspace{1cm} \xi >0
\eeq
which is independent of ${\widehat K}_R(\xi)$, i.e. is the same equation in cases {\bf (i)} and {\bf (ii)}.
(\ref{sovability_cond}) contains
\beq
\label{def_f_et_g_with_mortality}
\begin{array}{clr}
\ds {\widehat f}_{IM}(\xi; \mu_E) &= \ds  \mu_E \frac{1 -{\widehat K}_E(\xi)}{\xi} +  {\widehat K}_E(\xi) \frac{1 -{\widehat K}_{IM}(\xi)}{\xi}   &\\[3ex]
& = \ds 
\mu_E {\widehat \Phi}_{E}(\xi) + {\widehat K}_E(\xi) {\widehat \Phi}_{IM}(\xi) . &
\end{array}
\eeq
This expression contains the LT of the persistence probability 
$\Phi_{IM}(t) =\Phi_M(t)\Phi_I(t)$ (\ref{Phi_IM_def}) that an agent is in I and alive
\beq
\label{LT_complete}
{\widehat \Phi}_{IM}(\xi) =\frac{1-{\widehat K}_{IM}(\xi)}{\xi}  = \int_0^{\infty} \Phi_M(t)\Phi_I(t) e^{-\xi t}{\rm d}t
\eeq
with the non-defective (end-of-infection) PDF $K_ {IM}(t) = -\frac{d}{dt}\left[\Phi_M(t)\Phi_I(t)\right]$.
Since $\lim_{\xi \to \infty}G_M(\xi) =0$, Eq.
(\ref{sovability_cond}) has a solution $\xi=\xi_*>0$ if $G_M(0)=G_M(\xi)\Big|_{\xi=0} >1$ 
which therefore is the spreading condition, i.e. the condition that the disease-free equilibrium
($s_0=1$ in case {\bf (i)} and $s_0=1-r_0$, $r_0 \in [0,1)$ in case {\bf (ii)}) is unstable.
Therefore, we interpret $G_M(0)$ as {\it basic reproduction number} in case {\bf (i)} when $s_0=1$, and as
{\it effective reproduction number} (which we sometimes denote subsequently with ${\cal R}_{eff}$) in case {\bf (ii)} when $s_0=1-r_0$ corresponding to the presence of vaccinated agents $r_0$.
It is obtained as
\beq
\label{repro_num}
{\cal R}_M = G_M(0)= \beta s_0  {\widehat f}_{IM}(\xi;\mu_E)\big|_{\xi=0} = 
\beta s_0 
\left(\mu_E \Big\langle t_{E} \Big\rangle + \Big\langle t_{IM} \Big\rangle\right) = \beta s_0\Big\langle \tau_{eff,M} \Big\rangle 
\eeq
where $\Big\langle \tau_{eff,M} \Big\rangle$ can be conceived as the effective contact time
of predator and prey agents.
(\ref{repro_num}) contains the mean waiting times
\beq
\label{mean_comp_ER}
 \Big\langle t_{E,R, IM} \Big\rangle = \lim_{\xi \to 0} {\widehat \Phi}_{E, IM}(\xi)  = \int_0^{\infty} \Phi_{E,IM}(t){\rm d}t = -\frac{d}{d\xi}{\widehat K}_{E,IM}(\xi)\Big|_{\xi=0} 
\eeq
holding for the complete (properly normalized) involved PDFs $K_{E}, K_{IM}(t)$.
We observe that for fat-tailed waiting times (recall Sect. \ref{fractional_dyn}) 
which have infinite means, the 
basic or effective reproduction numbers diverge, reflecting infinite effective contact time $\Big\langle \tau_{eff,M} \Big\rangle$.
Especially we observe that
\beq
\label{mean_comIM}
\Big\langle t_{IM} \Big\rangle = \Big\langle min(t_I,t_M)\Big\rangle =\lim_{\xi \to 0} {\widehat \Phi}_{IM}(\xi) 
= \int_0^{\infty} \Phi_M(t)\Phi_I(t){\rm d}t \leq \int_0^{\infty} \Phi_I(t){\rm d}t =\Big\langle t_I \Big\rangle 
\eeq
reflecting the obvious feature $min(t_M,t_I) \leq t_I$.
Therefore, 
\beq
\label{Rbasic_eff_no_mort}
{\cal R}_M \leq {\cal R}_{eff} = \beta s_0\left(\mu_E \Big\langle t_{E} \Big\rangle + \Big\langle t_{I} \Big\rangle \right)
\eeq 
(equality only in the limit of zero mortality) where the r.h.s. refers to zero mortality.
Mortality reduces the basic or effective reproduction number and prevents the disease to spread
if ${\cal R}_M < 1$.

\begin{figure}[H]
\centerline{
\includegraphics[width=0.75\textwidth]{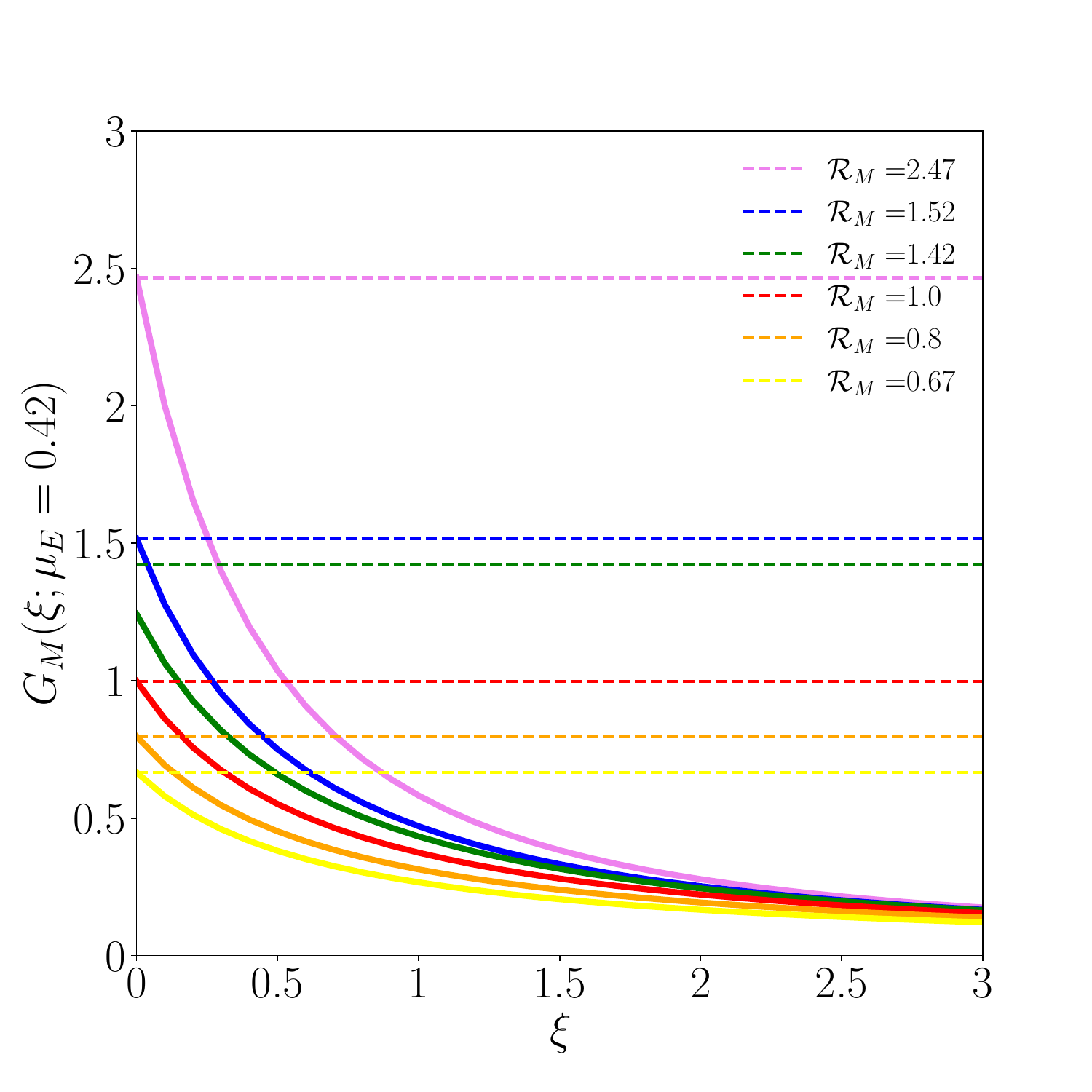} 
}
\caption{
Effect of mortality on ${\cal R}_M$ and on the stability of the disease-free state. We depict the $\xi$-dependence of $G_M(\xi)$, see (\ref{sovability_cond}), for $s_0\beta=1$ and increasing mortality, corresponding to decreasing ${\cal R}_M$. For ${\cal R}_M>1$, the disease-free initial state is unstable, i.e. (\ref{sovability_cond}) is fulfilled for $\xi_* > 0$, whereas it is stable for ${\cal R}_M<1$.
}
\label{f_IM_hat}
\end{figure}

We plot in Fig. \ref{f_IM_hat} the function $G_M(\xi)$ of (\ref{sovability_cond}) for a couple of Weibull parameters 
and for $s_0=1$, $\beta=1$ generating cases of different mortality, expressed by ${\cal R}_M$. Increasing mortality (decrease of ${\cal R}_M$) may prevent spreading of the disease as soon as ${\cal R}_M< 1$. This can be understood by the fact that if mortality is increased, the effective contact time of the predator with the prey is reduced.

\section{Compartment model without mortality}  
\label{no_mortality}
\subsection{Cyclic SEIRS pathway-- emergence of an endemic equilibrium}
\label{endemic_equilibrium}

Diseases without mortality can be captured from (\ref{SCEIRS-D-model}) by setting $t_M=\infty$ 
thus $\Theta(t_M-t_I)=1$ and $\Theta(t_I-t_M)=0$ and hence $\frac{d}{dt}D(t) =0$ for any finite $t$.
We consider again $t_{E,I,R} < \infty$) (case {\bf (i)}).
The evolution equations (\ref{SCEIRS-D-model}) then read
\beq
 \label{SCEIRS-D-model_zero-mortality}
 \begin{array}{lll}
 \displaystyle \frac{d S(t)}{dt} = &\ds  - {\cal A}(t) + \Big\langle {\cal A}(t-t_E-t_I-t_R) \Big\rangle +  e_0 \Big\langle \delta(t-t_E -t_I-t_R)\Big\rangle +j_0 \Big\langle \delta(t-t_I-t_R)\Big\rangle +  r_0\Big\langle \delta(t-t_R) \Big\rangle &  \\[3ex]
 \displaystyle \frac{d E(t)}{dt} = &\ds  {\cal A}(t) - \Big\langle {\cal A}(t-t_E) \Big\rangle -e_0\Big\langle \delta(t-t_E)\Big\rangle &\\[3ex] 
\displaystyle \frac{d J(t)}{dt}  = &\ds \Big\langle {\cal A}(t-t_E) \Big\rangle  - \Big\langle {\cal A}(t-t_E-t_I)\Big\rangle  - j_0 \Big\langle \delta(t-t_I) \Big\rangle + e_0\Big\langle \left\{\delta(t-t_E) -\delta(t-t_E-t_I) \right\}\Big\rangle    & \\[3ex]
 \displaystyle \frac{d R(t)}{dt} = & \ds  \Big\langle {\cal A}(t-t_E-t_I) \Big\rangle  - \Big\langle{\cal A}(t-t_E-t_I-t_R) \Big\rangle     - r_0\Big\langle \delta(t-t_R) \Big\rangle    &\\[3ex] 
 & \ds +j_0 \Big\langle \left\{\delta(t-t_I)-\delta(t-t_I-t_R) \right\}\Big\rangle 
 + e_0\Big\langle \left\{\delta(t-t_E-t_I) -\delta(t-t_E-t_I-t_R) \right\}\Big\rangle   &
 \end{array}
\eeq
in which $S(t)+E(t)+J(t)+R(t) =1$ and where $s_0=S(0), e_0=E(0), j_0=J(0), r_0 = R(0)$ are the initial condition
where we allow non-zero values $s_0, e_0, j_0, r_0 >0$. It is straightforward to retrieve
the `memory representation' of (\ref{SCEIRS-D-model_zero-mortality}) by setting $m_d(t)=0$ ($b_d(t)=0$) and $M_r(t)= M_{IM}(t) = M_I(t)$ in (\ref{convol_memory_SEIRDS}).

Now, our goal is to derive the endemic equilibrium. To that end it is 
convenient to consider Eqs. (\ref{SCEIRS-D-model_zero-mortality}) in Laplace space 
\beq
 \label{SCEIRS-D-model_zero-mortality_laplace}
 \begin{array}{lll}
 \ds \lambda {\widehat S}(\lambda) = & \ds s_0 + e_0  {\widehat K}_{E}(\lambda) {\widehat K}_{I}(\lambda){\widehat K}_{R}(\lambda) + j_0 {\widehat K}_{I}(\lambda){\widehat K}_{R}(\lambda) +  r_0{\widehat K}_{R}(\lambda)  - 
 {\widehat {\cal A}}(\lambda)\left(1- {\widehat K}_{E}(\lambda){\widehat K}_{I}(\lambda){\widehat K}_{R}(\lambda)\right)
 &  \\[3ex]
 \ds \lambda {\widehat E}(\lambda) = & \ds   
 \left[{\widehat {\cal A}}(\lambda) +e_0\right] \left(1- {\widehat K}_{E}(\lambda)\right)  &\\[3ex] 
\ds \lambda {\widehat J}(\lambda)  = & \ds 
\left\{ \left[e_0+{\widehat {\cal A}}(\lambda)\right] {\widehat K}_{E}(\lambda) 
 + j_0 \right\}\left(1-{\widehat K}_{I}(\lambda)\right)   & \\[3ex]
 \ds \lambda {\widehat R}(\lambda) = & \ds 
 \Big\{ \Big(  \left[e_0+{\widehat {\cal A}}(\lambda)\right] {\widehat K}_{E}(\lambda) 
 + j_0 \Big){\widehat K}_{I}(\lambda) + r_0 \Big\}\left( 1-{\widehat K}_{R}(\lambda)\right) .  & 
 \end{array}
\eeq
Then by using the limit value theorem (\ref{limit_value}) and assuming that the compartment waiting times have finite means
\beq
\label{limit_laplace}
\Big\langle t_{E,I,R} \Big\rangle  =\lim_{\lambda \to 0}  \frac{1-{\widehat K}_{E,I,R}(\lambda)}{\lambda} = -\frac{d}{d\lambda} {\widehat K}_{E,I,R}(\lambda)\Big|_{\lambda=0}, 
\eeq
and a constant asymptotic value for the infection rate ${\cal A}(\infty) = A_e$ with ${\widehat {\cal A}}(\lambda) \sim A_e/\lambda $ as $\lambda \to 0$, leading us in (\ref{SCEIRS-D-model_zero-mortality_laplace}) to limiting expressions, see for instance the second equation 
$$\ds \lim_{\lambda\to 0} A_e\frac{1-{\widehat K}_E(\lambda)}{\lambda} = 
-A_e \frac{d}{d\lambda}{\widehat K}_E(\lambda)\Big|_{\lambda=0} = A_e \big\langle t_E \big\rangle.$$
The infinite time asymptotics being stationary solutions of (\ref{SCEIRS-D-model_zero-mortality}),
($S(\infty) =S_e, E(\infty)=E_e, J(\infty)=J_e, R(\infty)= R_e$) then read
\beq
 \label{SCEIRS-D-model_zero-mortality_inifinite_time}
 \begin{array}{lll}
 \ds S_e = & \ds 1 - A_e\left( \Big\langle t_E \Big\rangle + \Big\langle t_I \Big\rangle +\Big\langle t_R \Big\rangle  \right) 
 \\[2ex]
 \ds E_e = & \ds A_e \Big\langle t_E \Big\rangle  &\\[1ex] 
\ds J_e  = & \ds A_e \Big\langle t_I \Big\rangle  & \\[2ex]
 \ds R_e = & \ds  A_e \Big\langle t_R \Big\rangle  . & 
 \end{array}
\eeq
We point that system (\ref{SCEIRS-D-model_zero-mortality_inifinite_time}) is independent of the
initial conditions $s_0,e_0,j_0,r_0$ since ${\widehat K}_{E,I,R}(\lambda)\Big|_{\lambda=0} =1$ (they are LTs of properly normalized complete PDFs) wiping out their contributions in the second, third, and fourth equation of (\ref{SCEIRS-D-model_zero-mortality_laplace}), and entering with $s_0+e_0+j_0+r_0=1$ into the first equation. We emphasize that, in order to achieve independence of the endemic equilibrium of the initial conditions, it is crucial to account for their transition dynamics. 
To see this directly, for instance integrate the second equation in
(\ref{SCEIRS-D-model_zero-mortality}) which yields
\beq
\label{integrate_second}
\int_0^{\infty} \frac{d E(t)}{dt}{\rm d}t = E_e-e_0=
\lim_{t\to\infty} \Big\langle \int_{t-t_E}^t {\cal A}(\tau){\rm d}\tau  \Big\rangle 
-e_0 \Big\langle \int_0^{\infty} \delta(t-t_E) {\rm d}t \Big\rangle = A_e \big\langle t_E\big\rangle -e_0
\eeq
since ${\cal A}(\tau) \to A_e$ in the integral when $t\to \infty$. In the same way one can retrieve the complete system (\ref{SCEIRS-D-model_zero-mortality_inifinite_time}) by integrating 
(\ref{SCEIRS-D-model_zero-mortality}) directly.
From (\ref{infection_law}) we connect the endemic values of the compartments $S_e, E_e, J_e, R_e$
and the infection rate $A_e$ as
\beq
\label{infection_endemic}
A_e = \beta S_e \Lambda_e  , \hspace{1cm} \Lambda_e =\mu_E E_e  + J_e .
\eeq
Plugging (\ref{SCEIRS-D-model_zero-mortality_inifinite_time}) into this relation yields 
\beq
\label{self-consist}
1 = S_e \beta \left(\mu_E \Big\langle t_E \Big\rangle + \Big\langle t_I \Big\rangle \right)
\eeq
from which one has $S_e = 1/{\cal R}_0$ with
\beq
\label{basic_reproduction_number}
{\cal R}_0= \beta \left( \mu_E \Big\langle t_E \Big\rangle + \Big\langle t_I \Big\rangle \right) =\beta \Big\langle \tau_{eff} \Big\rangle .
\eeq
We interpret (\ref{basic_reproduction_number}) 
as the {\it basic reproduction number} (without mortality), 
i.e. the number of secondary infections generated by a single infectious agent (being either in E or I) in an entirely heathy population of individuals in S. 
${\cal R}_0$ is an upper bound of the basic reproduction number 
with mortality ${\cal R}_M \leq {\cal R}_0$ (see Eq. (\ref{repro_num}) for $s_0=1$ corresponding to the disease-free equilibrium of case {\bf (i)}).
(\ref{basic_reproduction_number}) generalizes the result obtained in reference \cite{Granger-et-al2023}
to two predator compartments E and I.
The constant values of the endemic equilibrium yield
\beq
\label{endem}
\begin{array}{clr} 
\displaystyle  S_e & = \displaystyle  \frac{1}{{\cal R}_0} , &   \\[2ex]
\displaystyle   E_e & = \displaystyle  \frac{{\cal R}_0-1}{{\cal R}_0} \frac{\langle t_E \rangle }{\langle T \rangle}  & \\[2ex]
\displaystyle  J_e & = \displaystyle  \frac{{\cal R}_0-1}{{\cal R}_0} \frac{\langle t_I \rangle}{\langle T \rangle}  &\\[2ex]
\displaystyle  R_e & =  \displaystyle  \frac{{\cal R}_0-1}{{\cal R}_0} \frac{\langle t_R \rangle}{\langle T \rangle} , & 
\end{array} \hspace{2cm} \displaystyle  
\langle T \rangle = \langle t_E \rangle +\langle t_I\rangle +\langle t_R \rangle .
\eeq
Clearly, the existence of the endemic equilibrium requires that the spreading condition ${\cal R}_0>1$ is fulfilled.
We point out that (\ref{endem}) also covers 
fat-tailed distributed compartmental sojourn times for which
$\big\langle t_{E,I,R}\big\rangle \to \infty$, however $t_{E,I,R} < \infty$ corresponding to case {\bf (i)} (recall Sect. \ref{fractional_dyn}). 

Let us briefly discuss the behavior of the endemic equilibrium
in such `fractional settings'. Consider first the situation (a) where
one of the compartmental sojourn times, say $t_E$ is fat-tailed and all others have finite means. Then ${\cal R}_0 =\infty$ and $\Big\langle T \Big\rangle = \infty$, where compartments with finite mean sojourn times and compartment S become empty $S_e=J_e=R_e=0$ where all agents are eventually trapped in the compartment with infinite mean with $E_e=1$.
More delicate becomes this question when (b) two compartments have infinite mean sojourn times, say
E and I. Then again as ${\cal R}_0=\infty$ and $\Big\langle T \Big\rangle = \infty$ thus
compartments S and R become empty $S_e=R_e=0$ and from $(\big\langle t_E\big\rangle +\big\langle t_I\big\rangle)/\big\langle T\big\rangle =1$, follows that $E_e+J_e=1$ and $E_e,J_e \in [0,1]$ may take any value on this 1D manifold. 
Similarly, when (c) three compartments 
E, I and R have fat-tailed waiting times, then their endemic values may take any values on the 2D manifold $E_e+J_e+R_e=1$ with $E_e,J_e, R_e \in [0,1]$.

\begin{figure}[H]
\centerline{
\includegraphics[width=0.55\textwidth]{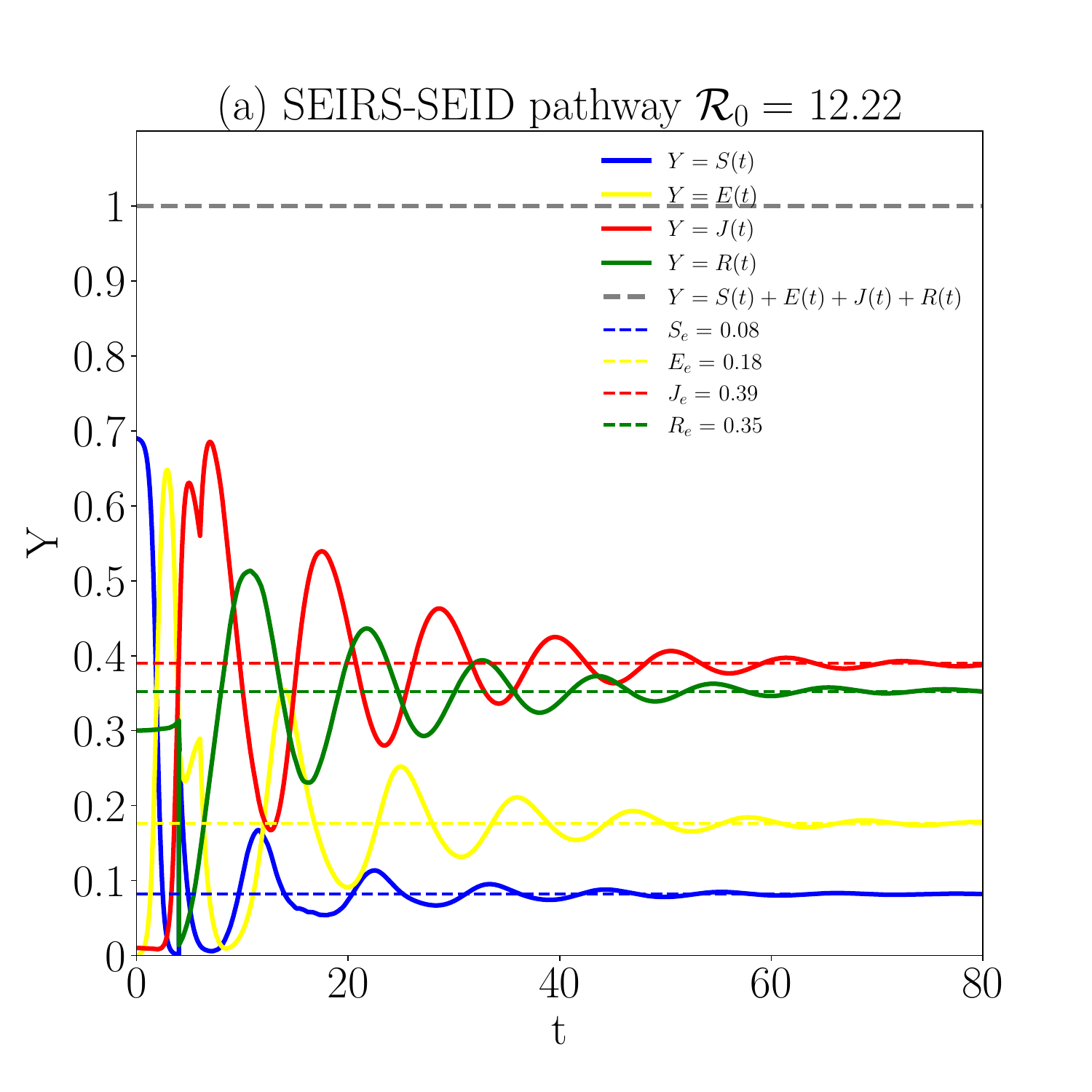}
\includegraphics[width=0.55\textwidth]{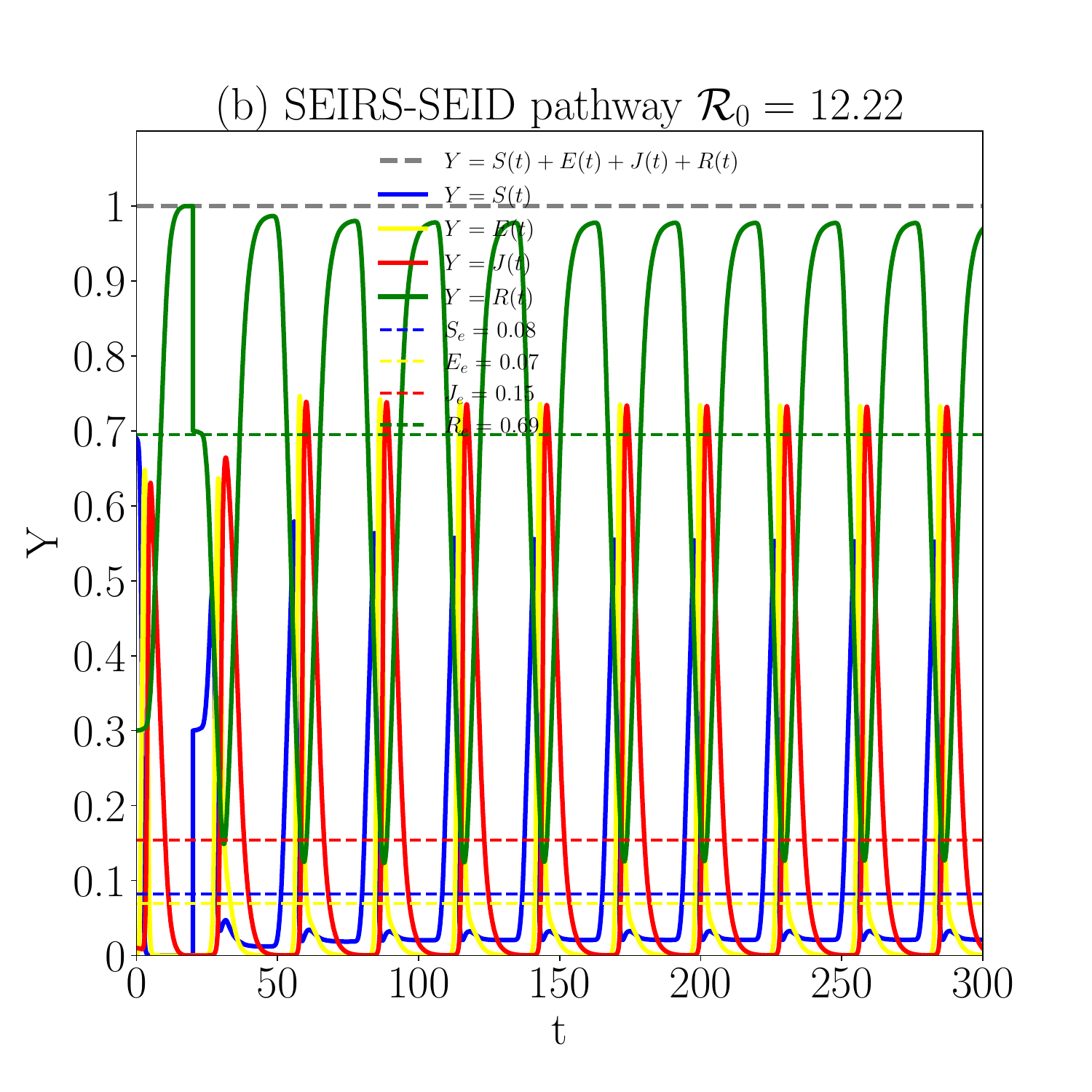}
}
\caption{
Cyclic SEIRS dynamics without mortality above the spreading threshold ${\cal R}_0>1$. (a) Emergence of the stable endemic equilibrium (\ref{endem}) in the presence of presymptomatic infections. The parameters are $\beta=0.5$, $\mu_E=10$, with a sharp $t_E=2$, $t_R=4$, and Weibull distributed $t_I$ with parameters $t_0=5$ and $\alpha=2$. (b) Same setting and color codes as in frame (a), except for the extended immunity time $t_R=20$, resulting in a Hopf instability with persistent oscillations.
} 
\label{endemic_evol}
\end{figure}

\paragraph{Linear stability analysis of the endemic equilibrium}
Let us briefly analyze stability of the endemic equilibrium (\ref{endem}).
To that end we set
\beq
\label{linear_ansatz}
S(t) = S_e + S_1 e^{\xi t} , \hspace{2cm} \Lambda(t) = \mu_E E(t)+J(t) = \Lambda_e + \Lambda_1 e^{\xi t}
\eeq
where $S_1,\Lambda_1 \ll 1 $  thus ${\cal A}(t) = \beta S(t)\Lambda(t) \approx  A_e + A_1 e^{\xi t}$
with $\Lambda_e=\mu_E E_e+J_e =
({\cal R}_0-1)/(\beta \langle T \rangle)$ and $A_e = \beta S_e \Lambda_e$, 
$A_1= \beta (S_e \Lambda_1 + S_1 \Lambda_e)$. Plugging (\ref{linear_ansatz}) into (\ref{SCEIRS-D-model_zero-mortality}) leads to the system
\beq
\label{lin_system}
\left(\begin{array}{ll}\ds  1 + 
\beta\Lambda_e h(\xi) \hspace{0.1cm} ;& 
\ds  \beta S_e h(\xi)   \\[3ex] 
\ds \beta \Lambda_e g(\xi)\hspace{0.1cm} ;& \ds -1 + \beta S_e g(\xi)\ds 
  \end{array}\right)\cdot 
\left(\begin{array}{l} S_1 \\[3ex] \Lambda_1  \end{array}\right) = \left(\begin{array}{l} 0 \\[3ex] 0  \end{array}\right) .
\eeq
with $h(\xi) =\frac{1- {\widehat K}_E(\xi) {\widehat K}_I(\xi){\widehat K}_R(\xi)}{\xi}$ 
and $g(\xi)= \mu_E {\widehat \Phi}_{E}(\xi) + {\widehat K}_E(\xi) {\widehat \Phi}_I(\xi)$ in which ${\widehat \Phi}_{E,I}(\xi)$ are the LTs of
the respective persistence probabilities
(recall (\ref{peristent_probas})). Setting the determinant 
of above matrix to zero leads to
\beq
\label{solvability_cond}
{\widehat G}(\xi) - 1 = 0 , \hspace{1cm} {\widehat G}(\xi) = \beta \left\{ S_e g(\xi) - \Lambda_e h(\xi) \right\} .
\eeq
The endemic state is stable if there is an eigenvalue $\xi>0$ solving (\ref{solvability_cond}), otherwise unstable. With similar arguments as 
in Sect. \ref{spreading_cond}, one has 
$\lim_{\xi \to \infty} [g(\xi), h(\xi)] = [0,0]$ and on the other hand
\beq
\label{xi_to_0_limit}
\lim_{\xi \to 0} \left(\begin{array}{l} g(\xi) \\[3ex] h(\xi)  \end{array}\right) = \left(\begin{array}{l} \mu_E \Big\langle t_E \Big\rangle + \Big\langle t_I \Big\rangle \\[3ex]    \Big\langle t_E \Big\rangle + \Big\langle t_I \Big\rangle + \Big\langle t_R \Big\rangle = \Big\langle T \Big\rangle  \end{array}\right) .
\eeq
Plugging (\ref{xi_to_0_limit}) into (\ref{solvability_cond}) for $\xi \to 0$ yields
\beq
\label{zero_cond_value}
{\widehat G}(0) = \beta S_e g(0) -\beta \Lambda_e h(0) = 2-{\cal R}_0  < 1 
\eeq
as ${\cal R}_0 >1$. Since 
$|{\widehat G}(\xi)|$ is monotonously decreasing with $\xi$
where $\lim_{\xi \to \infty} {\widehat G}(\xi) \to 0$, 
the inequality ${\widehat G}(\xi) <1$ is maintained $\forall \xi \geq 0$
proving stability of the endemic equilibrium, unless an oscillatory (Hopf) instability
occurs.
Let us briefly explore this issue. 
To this end we set $\xi = \pm i \omega$ where (\ref{solvability_cond}) yields the transcendental equations
\beq
\label{real_imainary}
\begin{array}{clr}
\Re\{ \beta S_e g(i\omega) - \beta \Lambda_e h(i\omega) \}  -1 &= 0 \\[2ex]
\Im \{ \beta S_e g(i\omega) - \beta \Lambda_e h(i\omega) \} & = 0
\end{array}
\eeq
to be simultaneously fulfilled for some Hopf frequencies $\omega=\omega_H$ allowing
persistent oscillations (Hopf instability of the endemic equilibrium).
To explore whether a Hopf instability might exist, we observe that the 
first term in (\ref{real_imainary})
${\widehat K}_g(\xi)=\beta S_e g(\xi) = \frac{g(\xi)}{g(0)}$ ($g(0) = 
\mu_E\langle t_E\rangle + \langle t_I\rangle$) is the LT of a
non-negative, properly normalized function (given that $g(0) < \infty$) which can be hence conceived as a PDF, namely
\beq
\label{construct_pdf}
K_g(t) = \frac{1}{\mu_E\langle t_E\rangle + \langle t_I\rangle} [\mu_E\Phi_E(t) + (K_E\star \Phi_I)(t)] .
\eeq
Indeed, the second term can be traced back on a PDF as well, namely
taking into account that ${\widehat \Phi}_T(\xi) = h(\xi)=
[1-{\widehat K}_E(\xi){\widehat K}_I(\xi){\widehat K}_R(\xi)]/\xi$ is the LT  of the persistence probability
$\Phi_T(t) = Prob(T=t_E+t_I+t_R >t)$ which is a non-negative function 
from which can be constructed a (normalized) PDF $K_T(t) = \Phi_T(t)/\big\langle T \big\rangle$
given $\big\langle T \big\rangle < \infty$. Hence one has
\beq
\label{second_term}
\begin{array}{clr}
\beta \Lambda_e h(\xi) & = ({\cal R}_0-1)\frac{{\widehat \Phi}_T(\xi)}{\big\langle T \big\rangle} & \\[2ex]
  & = ({\cal R}_0-1) {\widehat K}_T(\xi) &
\end{array}
\eeq
Therefore we can conceive ${\widehat G}(\xi)$ in (\ref{solvability_cond}) as the LT of the following
function
\beq
\label{G_t}
G(t)= K_g(t) - ({\cal R}_0-1) K_T(t) .
\eeq
Since $K_g(t),K_T(t)$ are normalized PDFs it follows that
\beq
\label{Re_Im_ineq}
-1 < \Re\{{\widehat K}_{g, T} (i\omega)\} = 
\int_0^{\infty} K_{g,T}(t)\cos{\omega t} \, {\rm d} t \leq 1
\eeq
(equality with the r.h.s. only for $\omega =0$) and therefore
\beq
\label{G_hat_real}
-{\cal R}_0 < \Re\{{\widehat G}(i\omega)\} < {\cal R}_0 , \hspace{1cm} \omega \neq 0
\eeq
where the same inequality holds for 
$\Im\{{\widehat G}(i\omega)\}$.
These inequalities tell us that there may exist situations for some $\omega=\omega_H$ for which the system (\ref{real_imainary}) is simultaneously fulfilled. Hence, we infer that oscillatory (Hopf) instabilities of the endemic equilibrium with persistent oscillations may exist. Indeed, we give hereafter numerical evidence for the emergence of persistent oscillations due to Hopf instability (see frame Fig. \ref{endemic_evol}(b)).
We refer to the references
\cite{Granger-et-al2023,besmi_periodic2025} for further evidence of occurrence of Hopf instabilities in a model without presymptomatic infections, and see also the discussion in the reference \cite{entropy_paper}. 

In Fig. \ref{endemic_evol} we depict a cyclic SEIRS evolution without mortality. In frame (a) the endemic equilibrium (\ref{endem}) is stable and eventually approached. In frame (b) we increase the sharp immunity time, leaving all other parameters identical as in frame (a). The basic reproduction number ${\cal R}_0$ is in both frames identical as it does not depend on 
immunity parameters, however the endemic equilibrium does apart from $S_e$ (see (\ref{basic_reproduction_number}), (\ref{endem})). The results in this figure suggest
that the endemic equilibrium becomes oscillatory (Hopf) unstable when in this case $t_R$ is increased. 

\subsection{Late time limit of the non-cyclic SEIR dynamics without mortality}
\label{SEIRD}
Here we consider the late time asymptotics of the non-cyclic SEID pathway with permanent immunity where it turns out that an endemic equilibrium does not exist.
In this setting clearly $S(t)$ is non-increasing and $R(t)$ non-decreasing, both approaching finite values with eventually empty predator compartments.
The asymptotic state is then given by 
(recall (\ref{infinite_time_limit_everlasting_imm}) for $B_r=1$)
\beq
\label{infinite_time_limit_everlasting_imm_no_mort}
\begin{array}{clr}
\ds S(\infty) & = \ds s_0 - {\widehat {\cal A}}(0) &\\[3ex]
\ds E(\infty) & =\ds  J(\infty) = 0 &\\[3ex] 
\ds R(\infty) & = \ds e_0 + j_0 + r_0 + {\widehat {\cal A}}(0)  &
\end{array}
\eeq
and depends on the initial conditions and the infection rate history.
Since $t_{E,I}$ are finite, the initial fractions $e_0,j_0$ and all ever infected agents ${\widehat {\cal A}}(0) = 
\int_0^{\infty}{\rm d}t {\cal A}(t)$ eventually accumulate in compartment R. 
We corroborate asymptotic relation (\ref{infinite_time_limit_everlasting_imm_no_mort}) hereafter numerically.

\section{Some pertinent variants: SIS, SIR, SIR-SID, SEIR-SEID, SEIRS, SEIRS-SEID}
\label{pertinent_special}
This section is devoted to shed light on stochastic aspects of some common compartment
models which are contained in (\ref{SCEIRS-D-model}), some of them without mortality
such as cyclic SIS and the non-cyclic SIR, and others include mortality such as SIR-SID 
and SEIR-SEID pathways.  

\subsection{Cyclic SIS pathway}
\label{SIS}
An especially instructive case is the simple cyclic SIS pathway without mortality \cite{Hethcode1995}. 
It is retrieved from system (\ref{SCEIRS-D-model}) for $t_E=t_R= 0, t_M=\infty$ ($E(t)=R(t)=D(t)=0$) and reads
\beq
\label{SIS_simple}
\begin{array}{clr}
  \ds \frac{d}{dt}J(t) & = \ds - j_0\Big\langle \delta(t-t_I)\Big\rangle +  {\cal A}(t) - \Big\langle {\cal A}(t-t_I) \Big\rangle  & \\[3ex]
 & =  \ds - j_0 K_I(t) - \int_0^t {\cal A}(\tau)K_I(t-\tau){\rm d}\tau + {\cal A}(t)
 \end{array}
\eeq
where here ${\cal A}(t)= \beta S(t)J(t)$ and $S(t)+J(t)=1$.
We consider 
initial conditions $s_0=1-j_0$. Integrating (\ref{SIS_simple}) yields 
\beq
\label{SIS_integral}
J(t) = j_0 \Phi_I(t) + \int_0^t {\cal A}(\tau)\Phi_I(t-\tau){\rm d}\tau .
\eeq 
(\ref{SIS_simple}) is the non-Markovian generalization of the common SIS model \cite{KermackMcKendrick1927,Lajmanoviv_Yorke1976} which is retrieved for exponentially distributed $t_I$ representing the memoryless (Markovian) case. With ${\widehat M}_I(\lambda) = {\widehat K}_I(\lambda)/{\widehat \Phi}_I(\lambda)$ the memory representation of (\ref{SIS_simple})  reads
\beq
\label{memory_evoleq}
\frac{d}{dt}J(t) = - \int_0^t M_I(t-\tau) J(\tau) + {\cal A}(t) .
\eeq
Of special interest is the case
of long $t_I$ for which $K_I(t)$ is fat-tailed where (\ref{SIS_simple}) involves fractional derivatives, which we considered already in the general setting above (recall (\ref{convol_memory_SEIRDS_gen_der})-(\ref{convol_memory_SEIRDS_fractional RL_der})).

For exponentially distributed $t_I$ one has $\Phi_I(t) = e^{-\xi_I t}$ and $K_I(t) = \xi_I e^{-\xi_I t} = \xi_I \Phi_I(t)$
with
${\widehat K}_I(\lambda) =\frac{\xi_I}{\xi_I+\lambda}$ corresponding to
${\widehat M}_I(\lambda) = \xi_I$ 
corresponding to the Dirac memory-kernel $M_I(t) = \xi_I \delta(t)$
representing the memoryless feature.
Plugging this kernel into (\ref{memory_evoleq}) yields the memoryless 
standard SIS equation
\beq
\label{exponential_SIS}
\frac{d}{dt} J(t) = -\xi_I J(t) + \beta J(t) (1-J(t)) ,
\eeq
This equation exhibits two fixed points for which the solution is stationary. The first one is the disease-free state $J=0$ ($S=1$) and the second one is the endemic equilibrium $J=J_e= 1-S_e$ which exists only for $S_e =\frac{\xi_I}{\beta} =\frac {1}{{\cal R}_0}< 1$.
(\ref{exponential_SIS}) with initial condition $J(0)=j_0$ 
has the analytical solution 
\beq
\label{exp_SIS}
\ds J(t) =   \frac{\ds  e^{({\cal R}_0-1)\xi_I t} }{\ds  \frac{1}{j_0} + \frac{1}{1-\frac{1}{{\cal R}_0}}(e^{({\cal R}_0-1)\xi_I t} -1) } .
\eeq
The asymptotic state is either the disease-free state $J(\infty) = 0$ if ${\cal R}_0< 1$ or the endemic state $J(\infty)=J_e= 1-\frac{1}{{\cal R}_0}$ for ${\cal R}_0>1$.

\begin{figure}[H]
\centerline{
\includegraphics[width=0.55\textwidth]{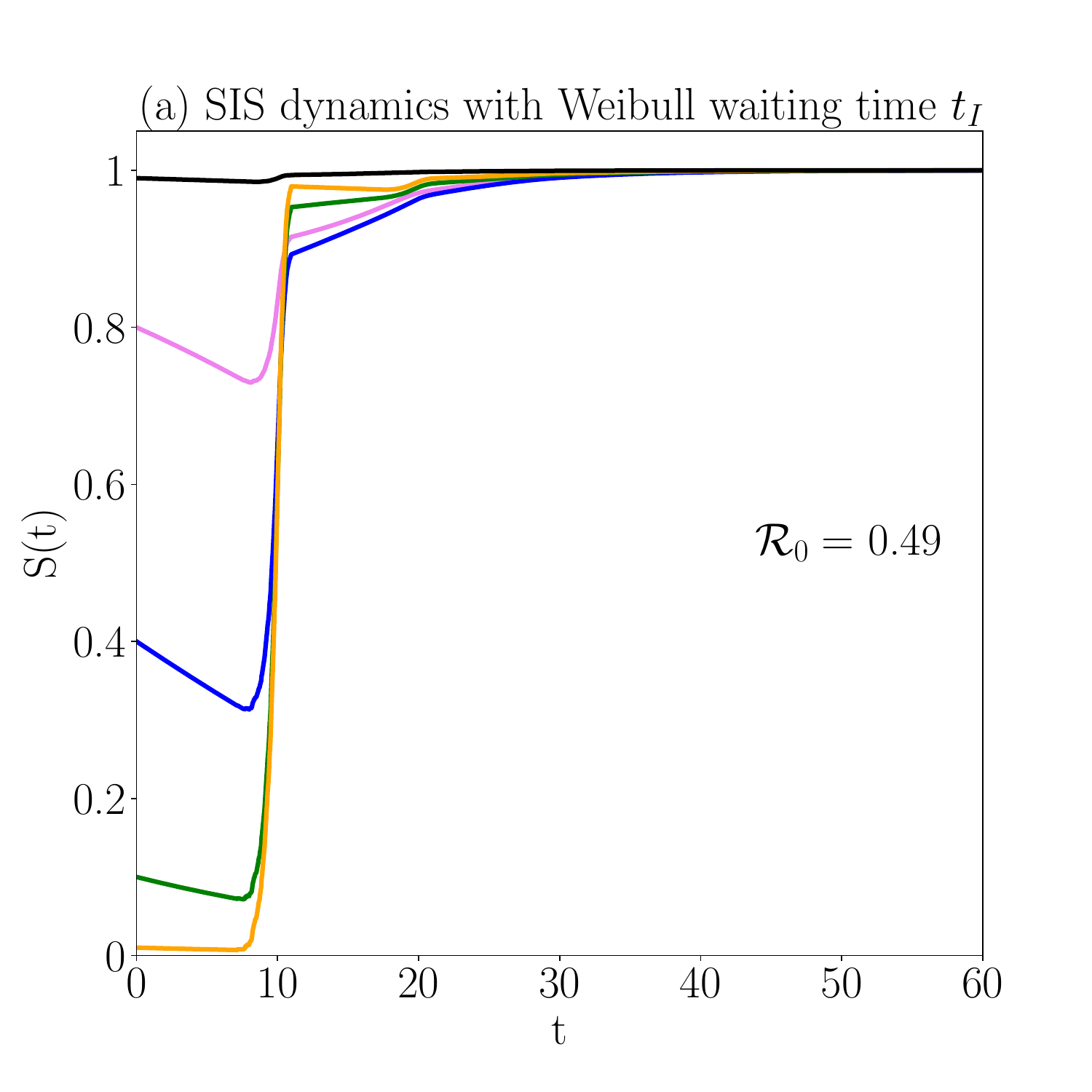}
\includegraphics[width=0.55\textwidth]{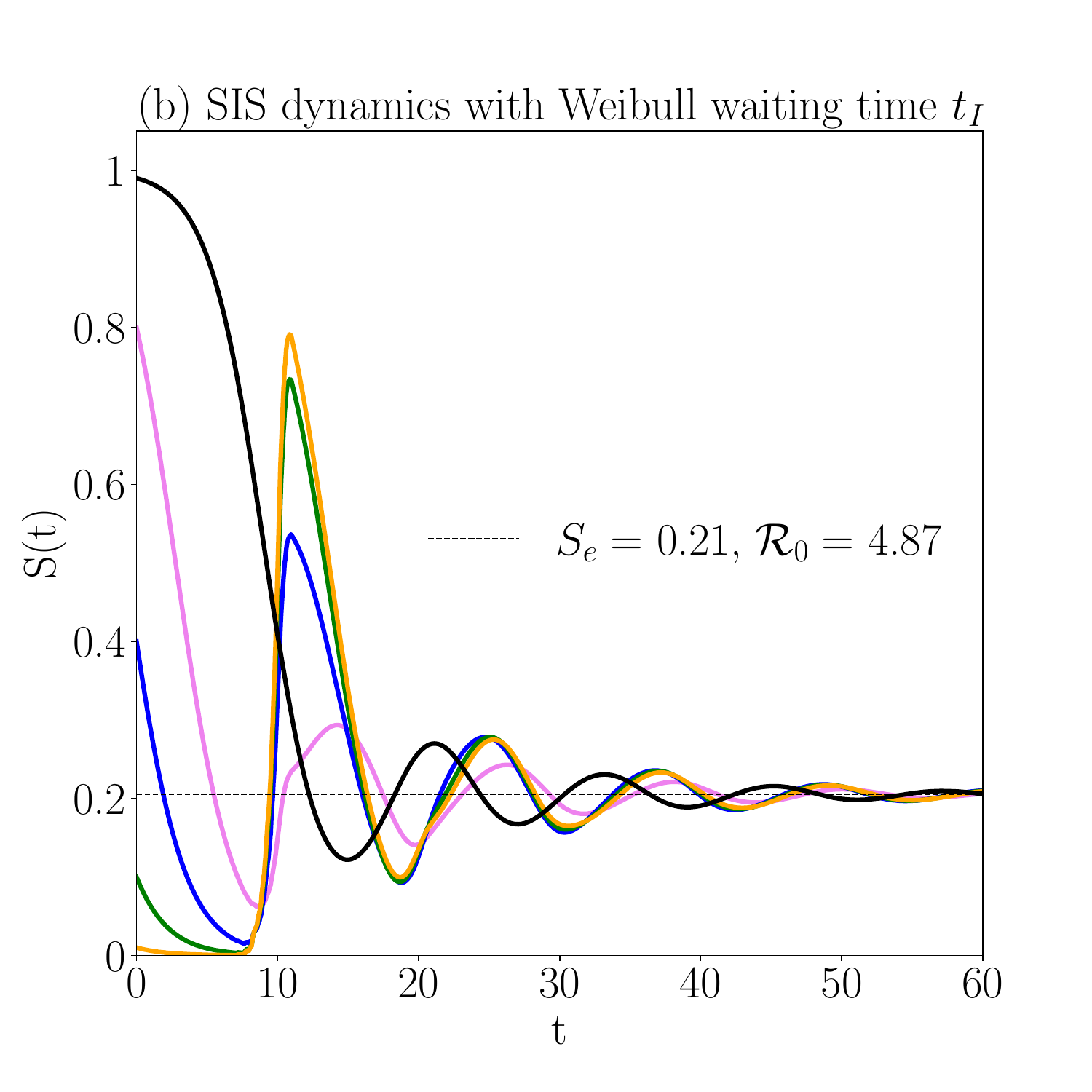}
}
\caption{
SIS dynamics for several initial conditions with Weibull distributed $t_I$ (Weibull parameters $t_0=10$ and $\alpha=20$), corresponding to a narrow unimodal PDF $K_I(t)$ concentrated around its mean 
$\Big\langle t_I\Big\rangle \approx t_0=10$ (see (\ref{Weibull_mean})). Both frames consider the same setting, except for the parameter $\beta$, which determines the basic reproduction number ${\cal R}_0$. (a) $\beta=0.05$, with basic reproduction number ${\cal R}_0<1$, below the spreading threshold: no spreading occurs, and the system relaxes to the stable disease-free equilibrium $S=1$. (b) $\beta=0.5$, with ${\cal R}_0>1$: the disease spreads and a stable endemic equilibrium $S_e=1/{\cal R}_0$ emerges.
}
\label{FIG_SIS} 
\end{figure}

Considering a case with memory in which $t_I$ is drawn from a Weibull PDF (\ref{Weibull_proba_funct})) (see Fig. \ref{weibull_pdf})), we depict in Fig. \ref{FIG_SIS} the SIS dynamics $S(t)=1-J(t)$ by integrating 
(\ref{SIS_simple}) in a combined Monte Carlo - Runge-Kutta 4 (RK4) scheme for several initial conditions 
with respectively ${\cal R}_0<1$ (frame (a)) and ${\cal R}_0>1$ (frame (b)) where we use the stochastic formulations of the equations and average over $1000$ realizations of Weibull $t_I$. 
One can see in frame \ref{FIG_SIS}(b) that for all chosen initial conditions the disease is spreading, leading to the same
asymptotically stable endemic state $S_e = 1/{\cal R}_0$. For ${\cal R}_0<1$ the system relaxes for all considered initial conditions to the disease-free equilibrium state $S=1$.
The effect of the memory (distributed delay of $t_I$) generates a damped oscillatory behavior, a feature which was observed previously in delayed systems \cite{Bes_Mi_Bif_Chaos2023}.

\begin{figure}[H]
\centerline{
\includegraphics[width=0.55\textwidth]{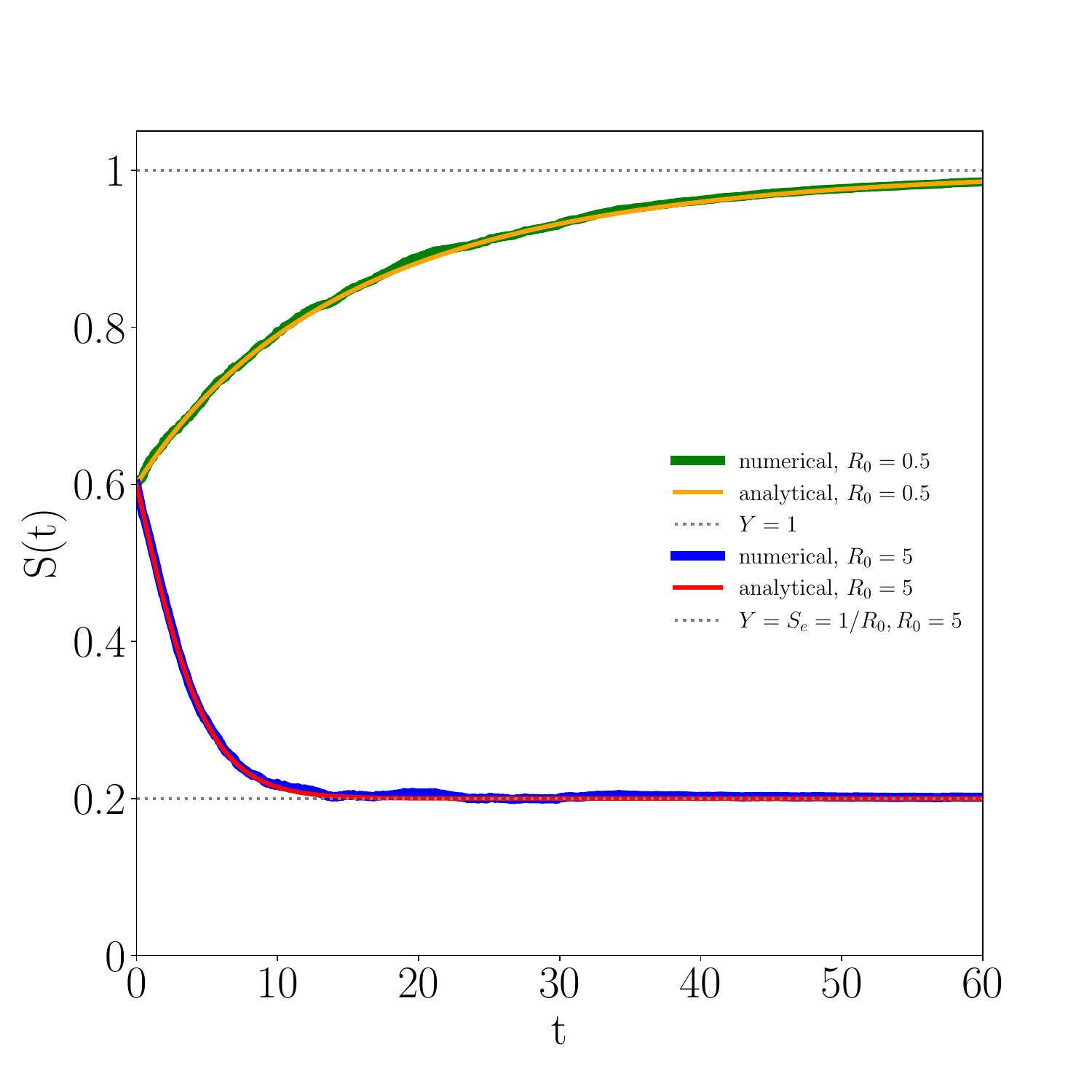}
}
\caption{
Comparison of the numerical solution for $S(t)=1-J(t)$ with the analytical solution (\ref{exp_SIS}) for two cases: ${\cal R}_0=5$ ($\beta=0.5$), for which the disease spreads, and ${\cal R}_0=0.5$ ($\beta=0.05$), where the disease-free state is stable (${\cal R}_0=\beta t_0$). Both cases correspond to Weibull parameters $\alpha=1$ (exponential) and $t_0=10$. The numerical solutions (blue and green curves) are obtained by Monte Carlo averaging over $1000$ realizations of $t_I$. The analytical and numerical solutions are in excellent agreement, validating our numerical Monte Carlo--RK4 scheme.
}
\label{FIG_SIS_validation} 
\end{figure}

In Fig. \ref{FIG_SIS_validation} we consider the memoryless case (\ref{exp_SIS}). The memoryless effect is such that no oscillations occur. 
We compare in this figure the numerical solutions of $S(t)$ obtained from integrating (\ref{exponential_SIS}) with analytical solutions (\ref{exp_SIS}). We focus on two situations below and above the spreading threshold. Below the threshold (${\cal R}_0< 1$) the asymptotic state is the disease-free stable fixed point $S=1$. Above the spreading threshold (${\cal R}_0> 1$) the asymptotically stable endemic state $S_e=1/{\cal R}_0$ is approached.
Both analytical and numerical solutions are in perfect accordance validating our combined Monte Carlo -- RK4 numerical scheme which we utilize throughout this paper.
\subsection{Non-cyclic SIR pathway}
\label{SIR}

An interesting variant is the non-cyclic SIR pathway with everlasting immunity $t_R=\infty$ (case {\bf (ii)}) and zero mortality.
System (\ref{SCEIRS-D-model}) without compartment E (i.e. $t_E=0$) boils then down to 
\beq
\label{SIR_simple_memory}
\begin{array}{clr}
  \ds \frac{d}{dt} S(t) & = \ds - {\cal A}(t)  & \\[3ex]
 \ds   \frac{d}{dt} J(t) & =  \ds - j_0 K_I(t) - \int_0^t {\cal A}(\tau)K_I(t-\tau){\rm d}\tau + {\cal A}(t) &\\[3ex]
\ds   \frac{d}{dt} R(t) & =\ds  j_0 K_I(t) + \int_0^t {\cal A}(\tau)K_I(t-\tau){\rm d}\tau 
 \end{array}
\eeq
where here ${\cal A}(t) = \beta S(t) J(t)$ and with the disease-free fixed point manifold $S+R=1$. 
So it makes sense to consider the spreading condition close this manifold and the effective reproduction number with presence of vaccinated agents (recall Sect. \ref{spreading_cond}).
System (\ref{SIR_simple_memory}) is the non-Markov generalization of the 
classical SIR model of Kermack and McKendrick \cite{KermackMcKendrick1927},
which is recovered when $t_I$ is exponentially distributed where one has $\Phi_I(t) =e^{-\xi_I t}$ 
and $K_I(t) =\xi_I \Phi_I(t)$ ($\xi_I >0$). Then (\ref{SIR_simple_memory}) boils down to
\beq
\label{SIR_simple_no_memory}
\begin{array}{clr}
  \ds \frac{d}{dt} S(t) & = \ds - {\cal A}(t)   & \\[3ex]
 \ds   \frac{d}{dt} J(t) & =  \ds -\xi_I J(t)+ {\cal A}(t) &\\[3ex]
\ds   \frac{d}{dt} R(t) & =\ds \xi_I J(t) .
 \end{array}
\eeq
The late time limit of this system is given by
(\ref{infinite_time_limit_everlasting_imm_no_mort}) for $e_0=0$, and depends on the infection history captured in ${\widehat {\cal A}}(0)$ and on the initial conditions. 

\begin{figure}[H]
\centerline{
\includegraphics[width=0.55\textwidth]{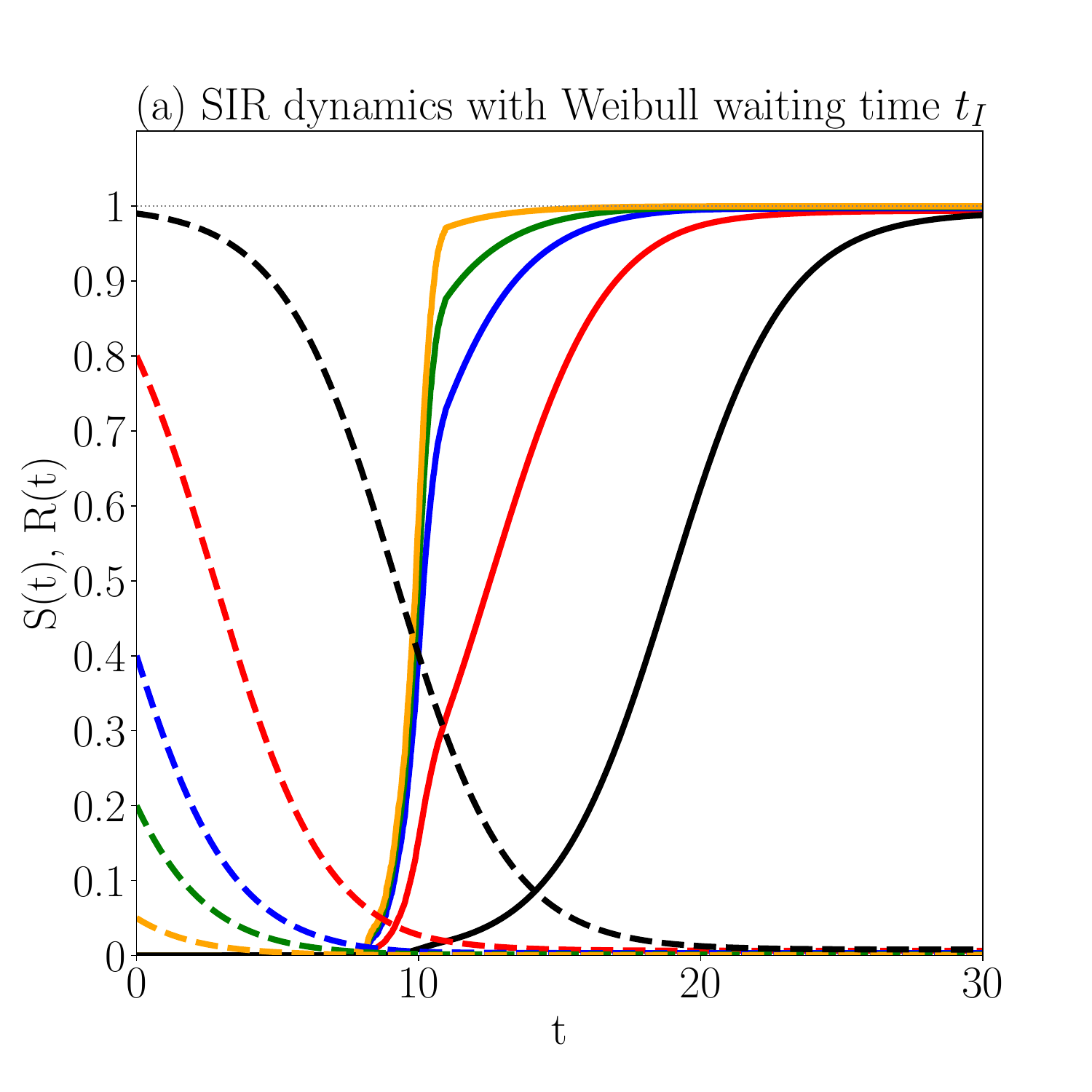} 
\includegraphics[width=0.55\textwidth]{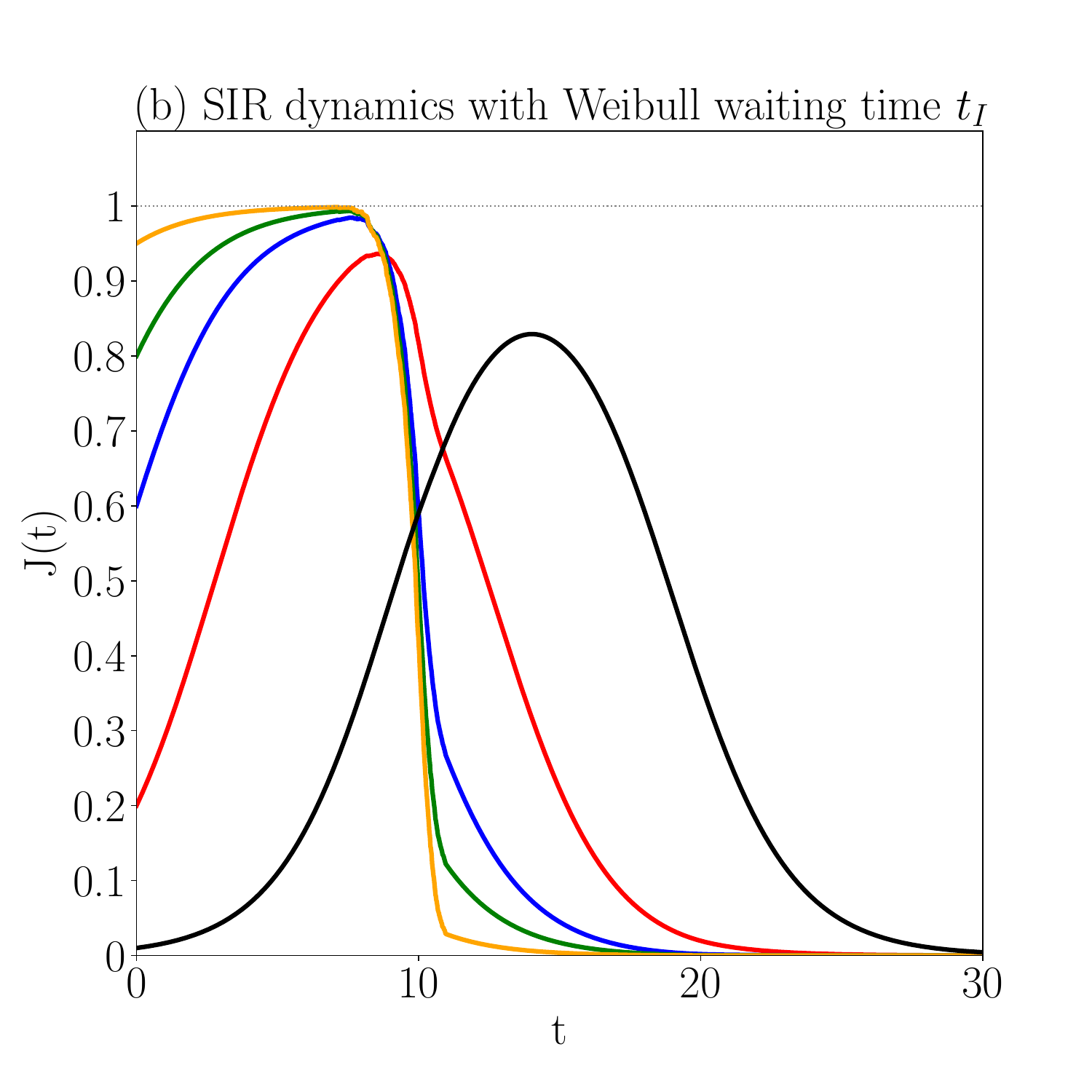} 
}
\caption{
SIR dynamics without vaccination of the compartmental fractions in Eq. (\ref{SIR_simple_memory}) with $\beta=0.5$ and Weibull parameters $t_0=10$ and $\alpha=20$ for $t_I$. The $S(t)$, $J(t)$, and $R(t)$ curves corresponding to each case are shown using the same color. The basic reproduction number ${\cal R}_0=\beta\big\langle t_I\big\rangle=4.87>1$ is meaningful here only close to the fixed point $s_0\to1$ (black curves). We consider the initial conditions $[S(0),J(0),R(0)]=[1-j_0,j_0,0]$, corresponding to an initially susceptible and infected population with no immune agents. Colors indicate identical initial conditions.
}
\label{FIG_SIR_4} 
\end{figure}

\begin{figure}[H]
\centerline{
\includegraphics[width=0.55\textwidth]{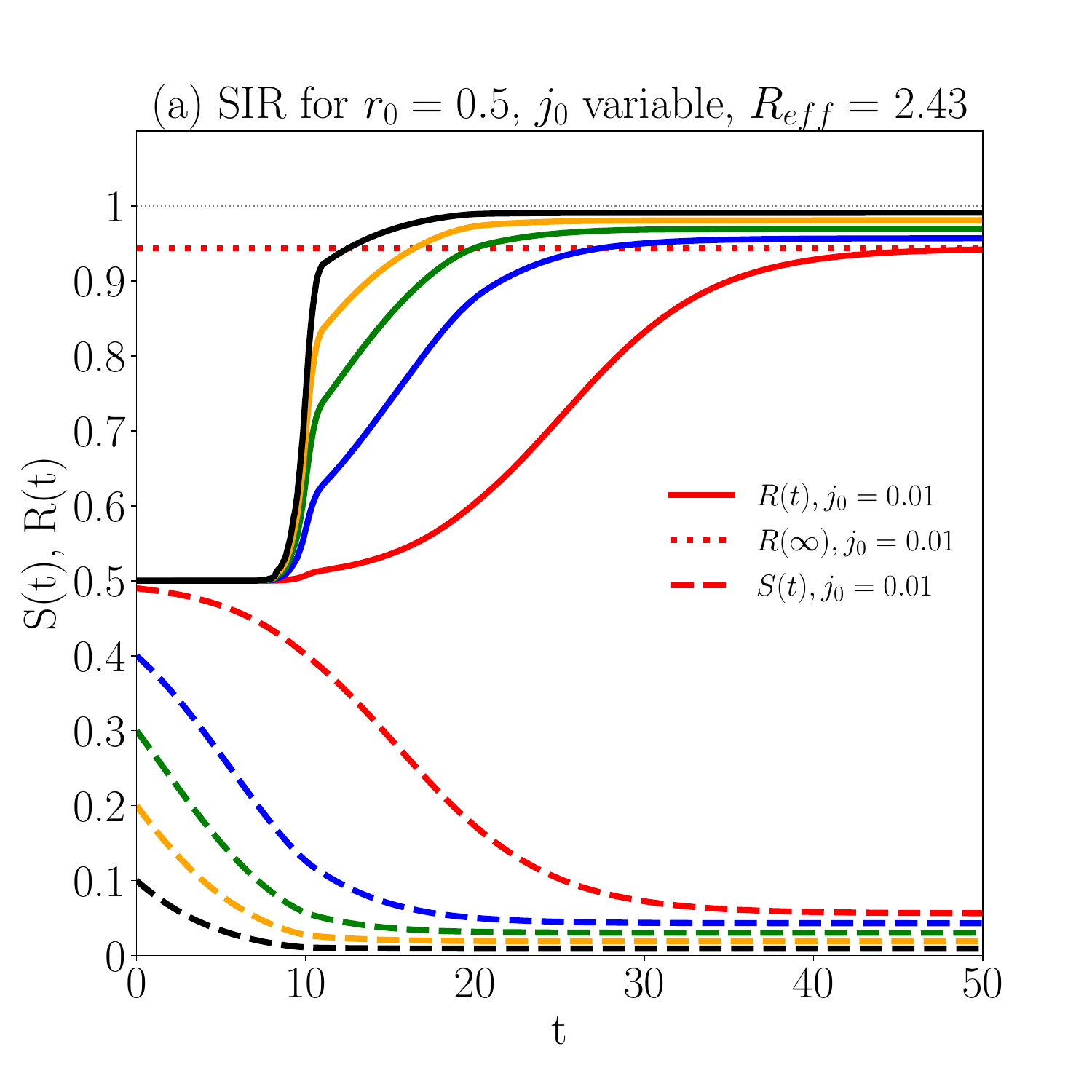} 
\includegraphics[width=0.55\textwidth]{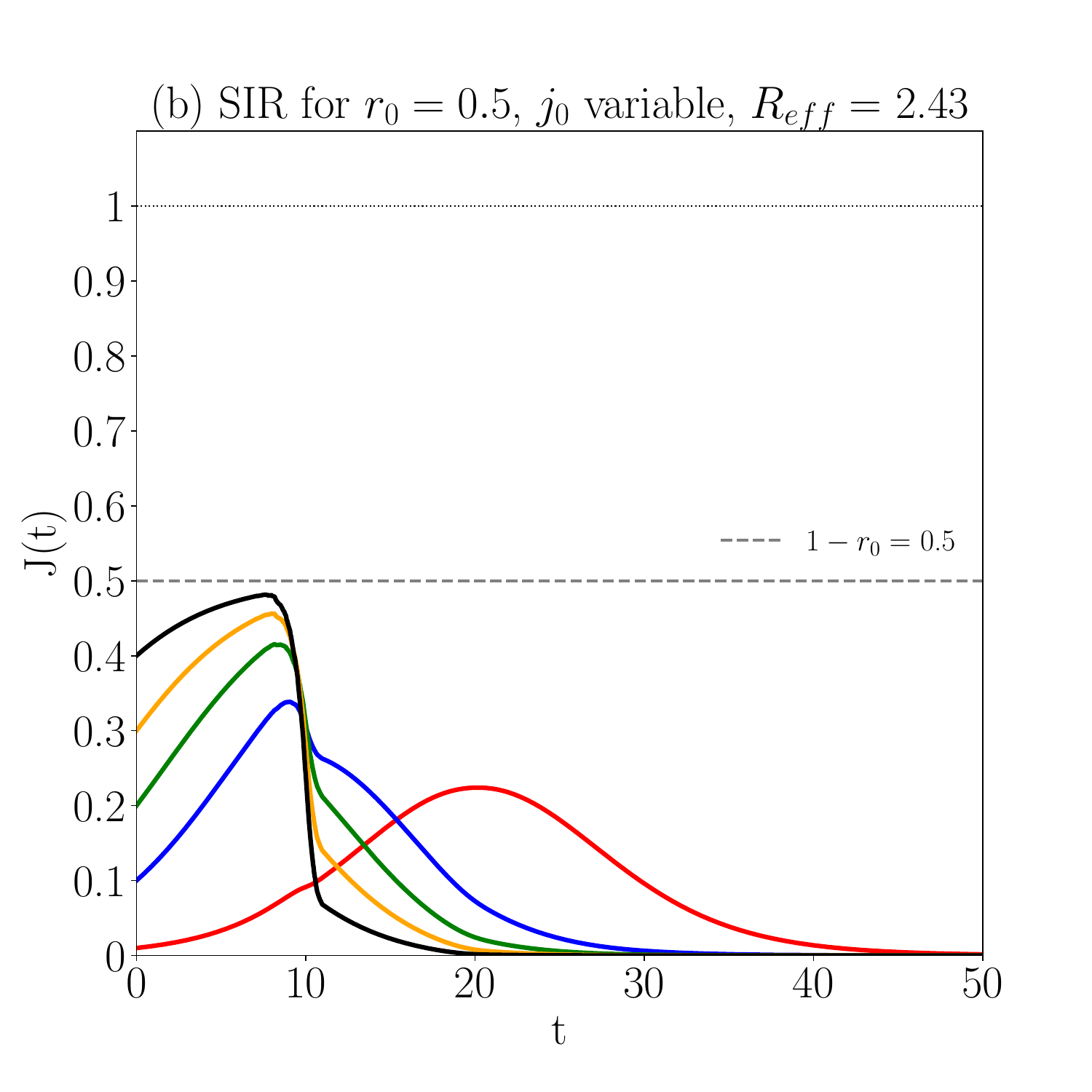} 
}
\caption{
SIR dynamics with vaccination for initial conditions $[S(0),J(0),R(0)]=[1-r_0-j_0,j_0,r_0]$, with fixed $r_0=0.5$ and variable $j_0$, using the same Weibull parameters for $t_I$ as in Fig. \ref{FIG_SIR_4}. The fixed point $[1-r_0=0.5,0,r_0=0.5]$ is unstable, with $R_{eff}=2.43$. The red dotted horizontal line in frame (a) indicates $R(\infty)=r_0+j_0+{\widehat {\cal A}}(0)$ for the case $j_0=0.01$, obtained from (\ref{infinite_time_limit_everlasting_imm_no_mort}) by numerical integration. The colors correspond to the same initial conditions.
}
\label{FIG_SIR_5} 
\end{figure}

\begin{figure}[H]
\centerline{
\includegraphics[width=0.55\textwidth]{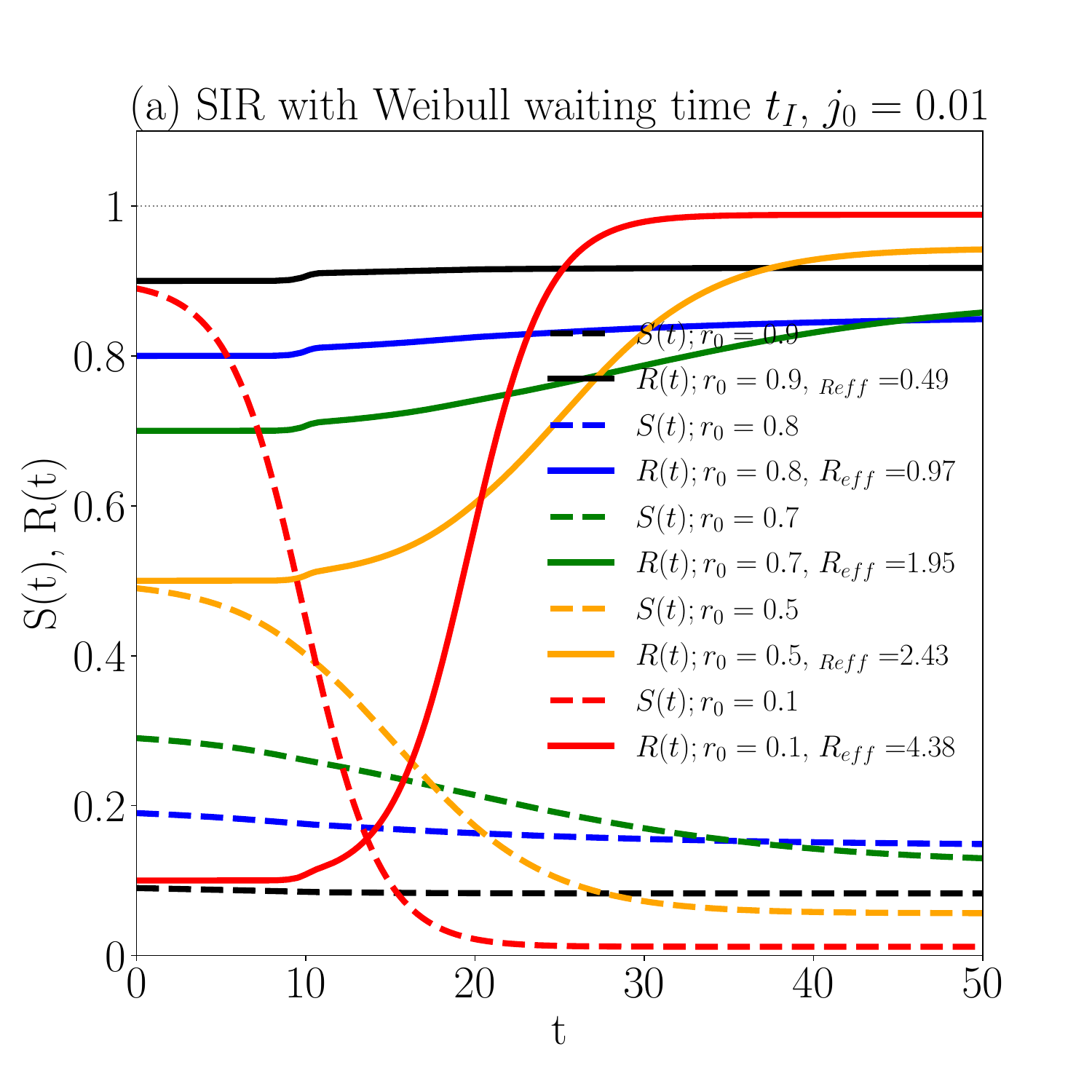} 
\includegraphics[width=0.55\textwidth]{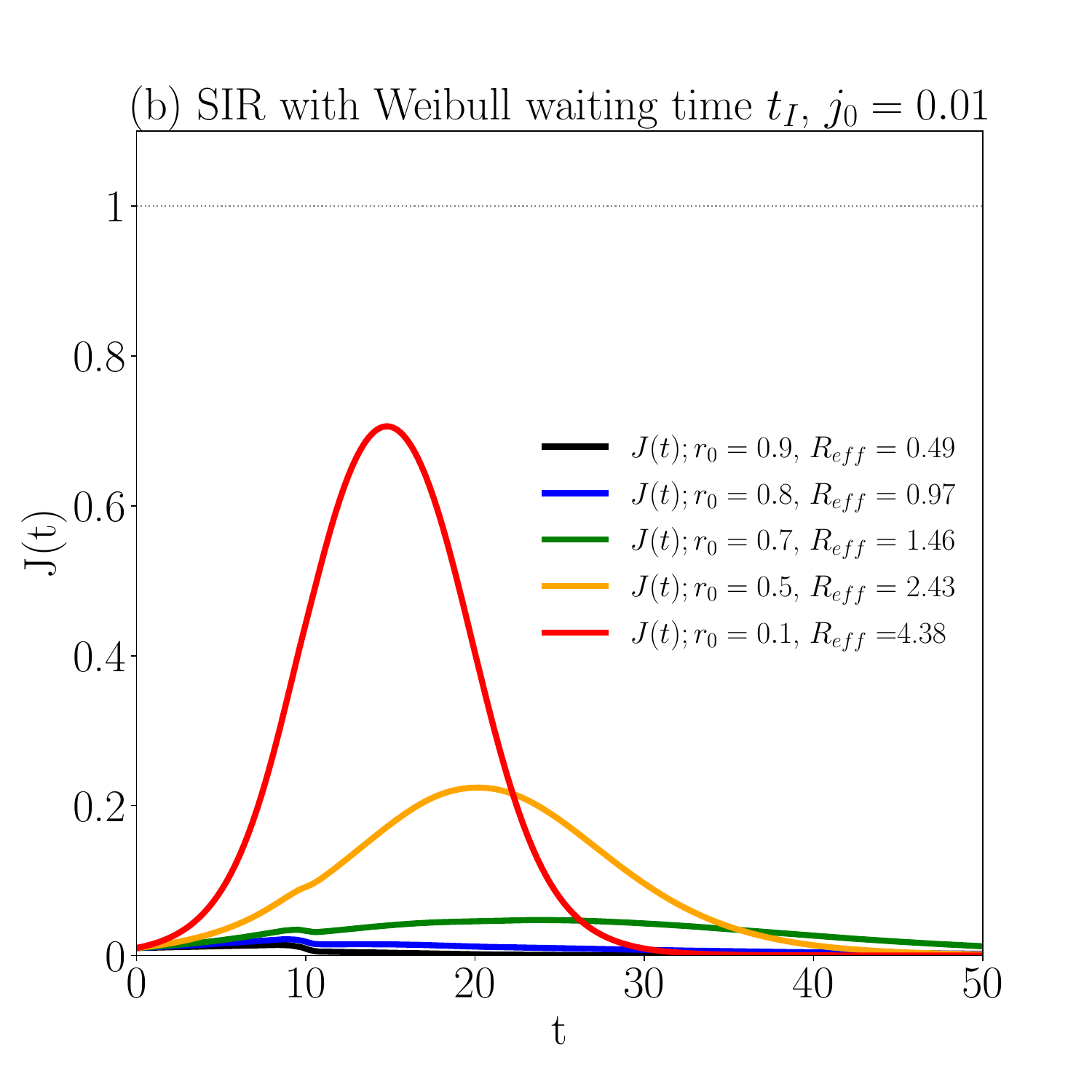} 
}
\caption{
Compartmental fractions for fixed small $j_0=0.01$ and variable $r_0$, exploring the effect of initially immune (``vaccinated'') agents. The colors correspond to the same initial conditions. We use the same Weibull parameters for $t_I$ as in Figs. \ref{FIG_SIR_4} and \ref{FIG_SIR_5}.
}
\label{FIG_SIR_6} 
\end{figure}

We integrate system (\ref{SIR_simple_memory}) with our Monte Carlo -- RK4 scheme and obtain the numerical solutions, which are plotted in Figs. \ref{FIG_SIR_4} -\ref{FIG_SIR_6} for several initial conditions and identical 
Weibull parameters $t_0=10$ and $\alpha=20$ for the distribution of $t_I$ corresponding to a narrow unimodal Weibull PDF.
In Fig. \ref{FIG_SIR_4} we consider ${\cal R}_0 >1$ thus the disease-free state is unstable and the disease for all initial conditions the disease is spreading where unimodal infection waves 
emerge. It takes a delay of around 
$\big\langle t_I\big\rangle \approx 10$ until immunity occurs in the part of 
population which has suffered the disease.
For late times the complete population has suffered an infection and is eventually reaching immunity.
Since there are no transitions from R back to S, the compartment fraction $S(t)$ is a non-increasing
and $R(t)$ is non-decreasing function.

Fig. \ref{FIG_SIR_5} shows the SIR dynamics with presence of $50\%$ of vaccinated agents ($r_0=0.5$) for variable $j_0$ for a setting with 
${\cal R}_{eff} =(1-r_0){\cal R}_0  >1$ (only meaningful for small $j_0 = 0.01$). The disease-free equilibrium $s_0=1-r_0 =0.5$ is unstable and therefore for any $j_0$ an infection $J$-wave evolves (see frame (b)).
The maximum amplitude of this wave is the larger, the larger is $j_0$ and the larger also becomes $R(\infty)$ (see frame (a)). For the largest $j_0$-value the amplitude of the $J$-wave is reaching 
almost $1-r_0 \approx 0.5$. Without vaccination this effect can be seen in Fig. \ref{FIG_SIR_4}(b)
where the entire population suffers infection.

In Fig. \ref{FIG_SIR_6} we consider very small fixed $j_0=0.01$ and variable $r_0$ 
in order to explore the effect of initially immune ('vaccinated') agents. 
The initial conditions are very close to the fixed points $s_0=1-r_0$ thus the effective
reproduction number
${\cal R}_{eff} =(1-r_0) \beta \big\langle t_I\big\rangle$  ($s_0=1-r_0$) is meaningful, see
expression (\ref{repro_num}) where here $t_{IM} = t_I$ (zero mortality). One can see (frame (b)) that the larger the fraction 
$r_0$ of vaccinated agents (i.e. the smaller ${\cal R}_{eff}$) the smaller the amplitude of the unimodal infection wave becomes. 
For sufficiently large fraction $r_0 \approx 0.8$ of vaccinated 
agents the effective reproduction number falls of the spreading limit $R_{eff} \to 1$ thus herd immunity is reached, and no infection wave emerges anymore (blue $J$-curve, right frame). These results clearly demonstrate efficiency of vaccination in order to reduce the impact of a disease.
\subsection{Non-cyclic SIR - SID pathway}
\label{SIRD}
\begin{figure}[H]
\centerline{
\includegraphics[width=0.55\textwidth]{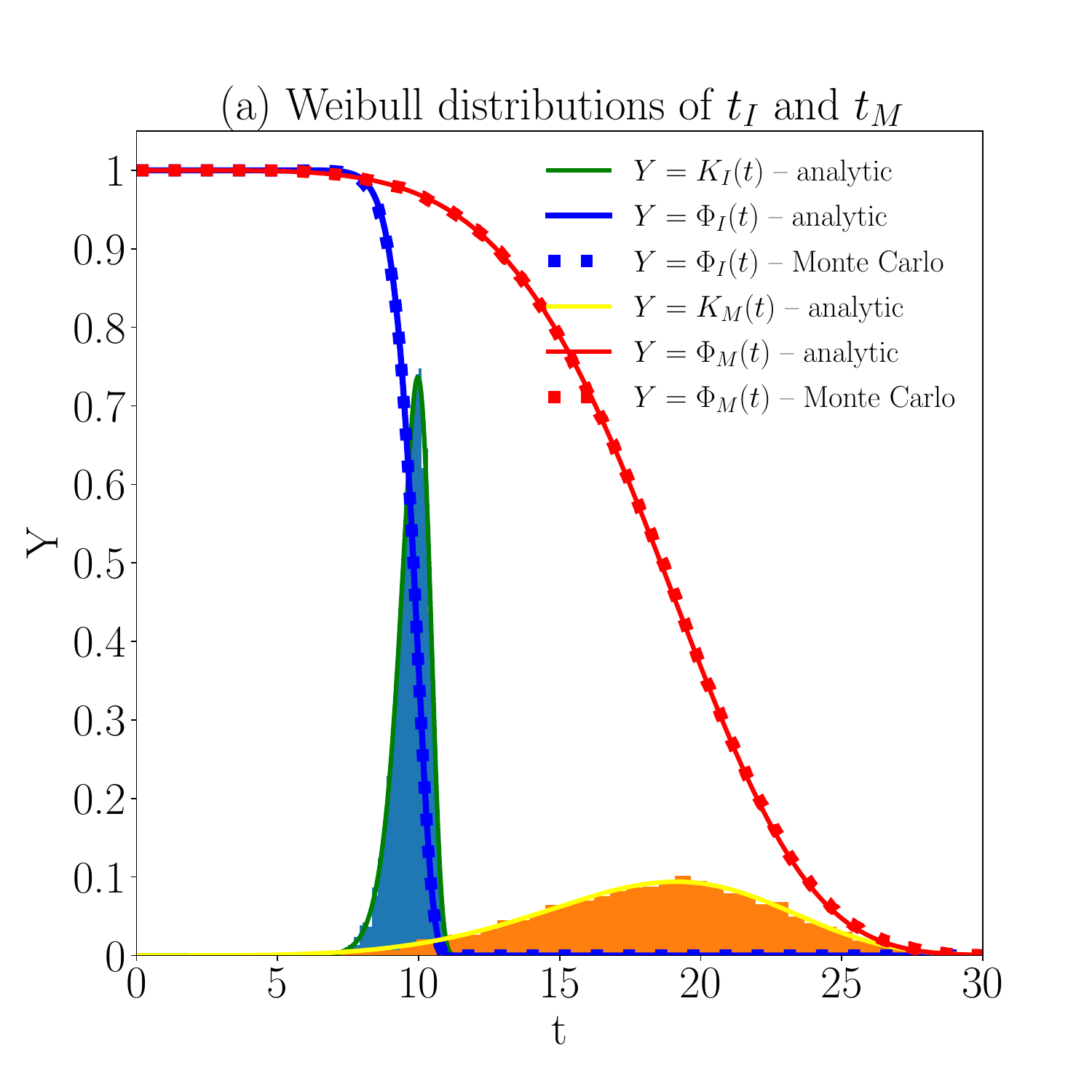}
\href{https://drive.google.com/file/d/1RuF0O8nLnut4V4kO_66G3ZWia46osvJ-/view}{ 
\includegraphics[width=0.55\textwidth]{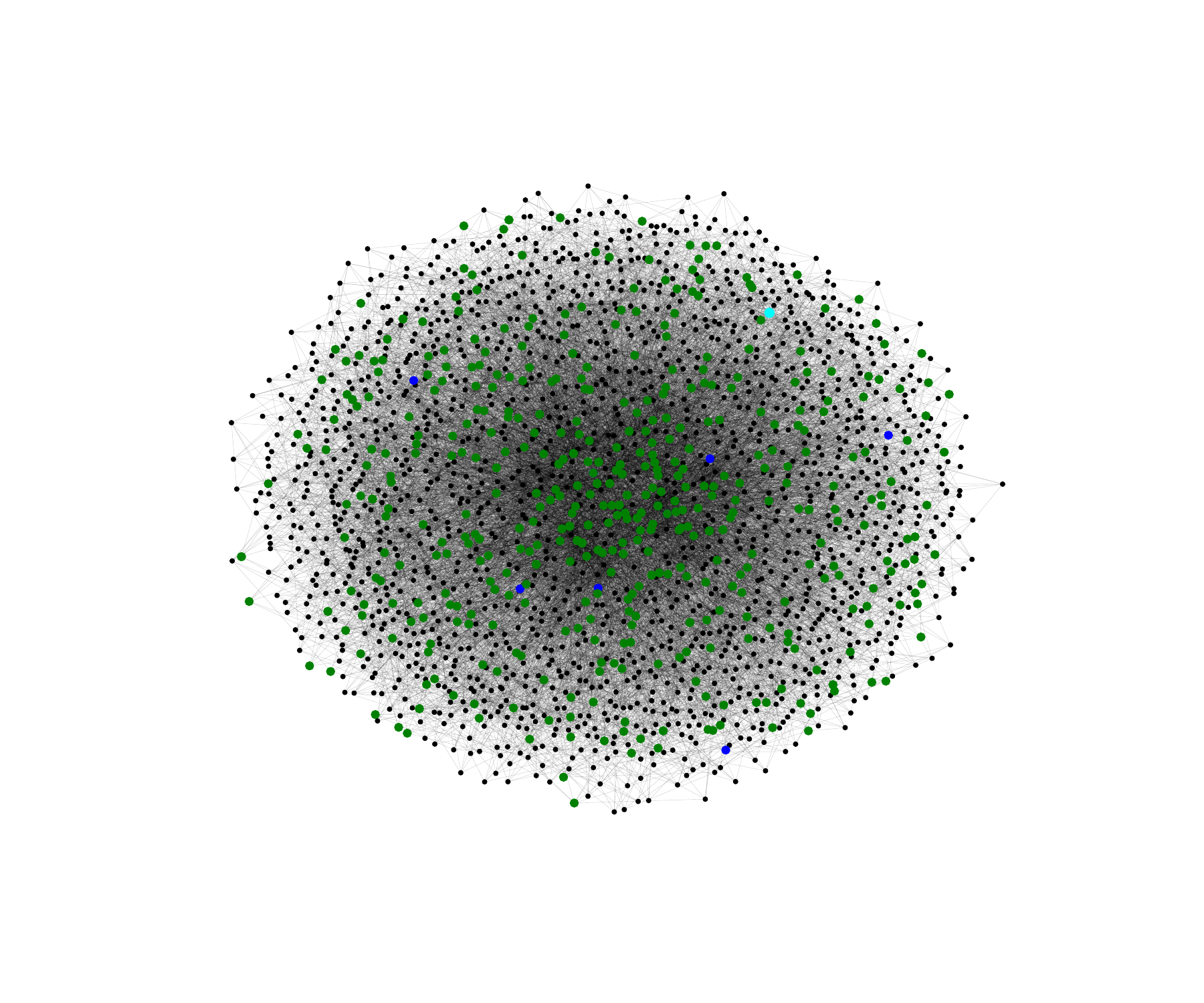}}
}
\centerline{
\includegraphics[width=0.55\textwidth]{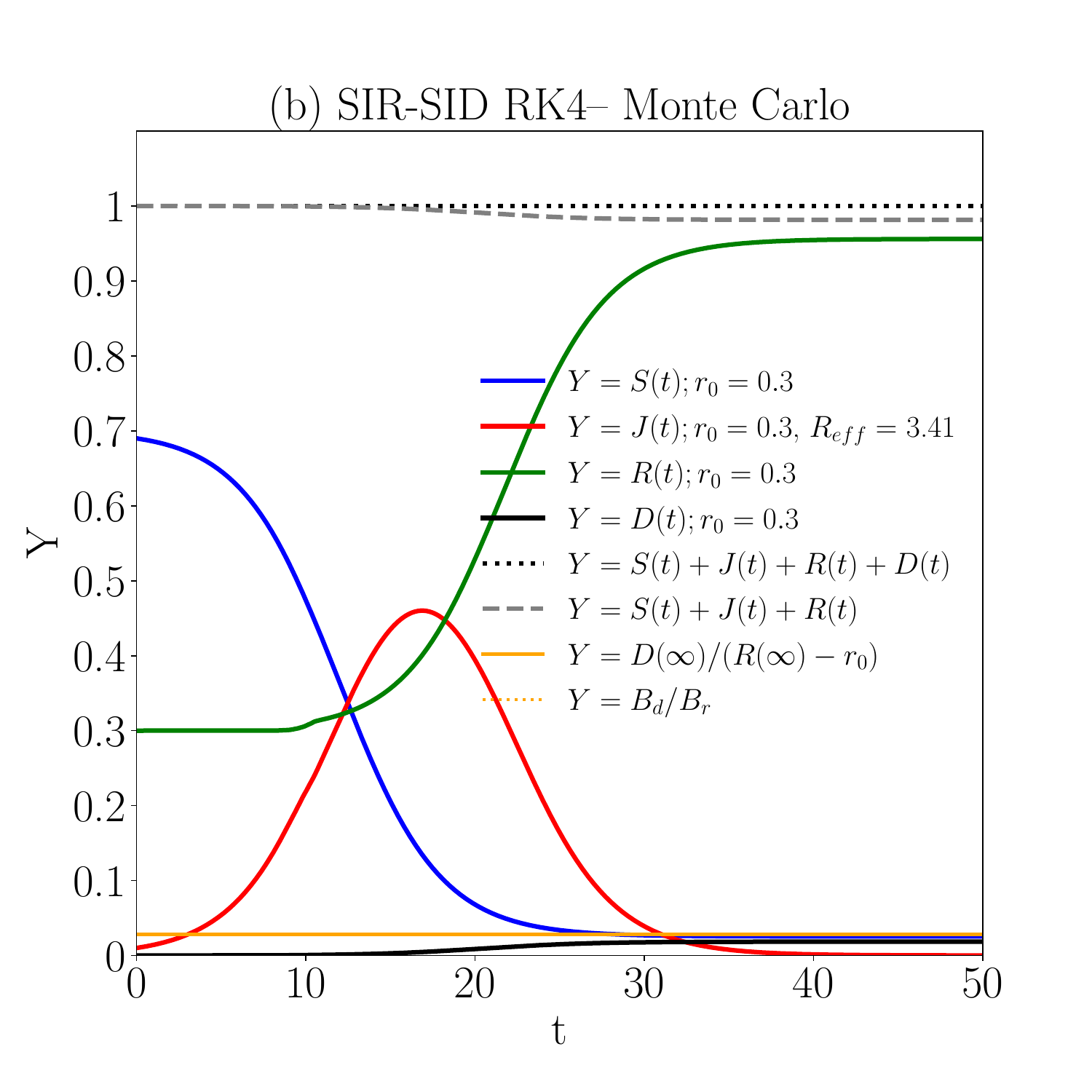}
\includegraphics[width=0.55\textwidth]{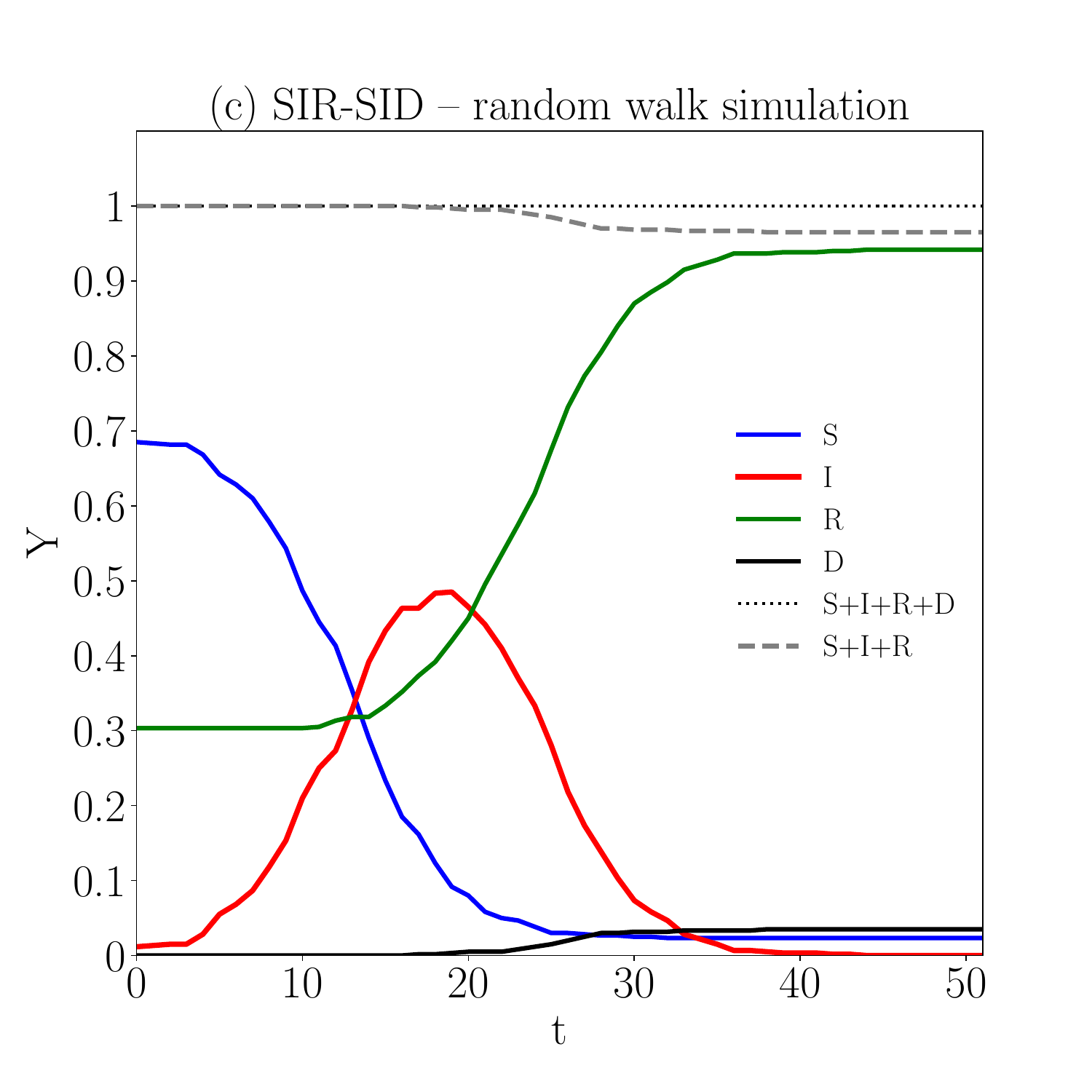}
}
\caption{
(a) Weibull waiting time distributions of $t_M$ and $t_I$ with Weibull parameters $t_0=10$ and $\alpha=20$ for $t_I$, and $t_0=20$ and $\alpha=5$ for $t_M$, used in the evolutions shown in frames (b) and (c). These parameters result in a small mortality $B_d\approx0.03$ and an effective reproduction number ${\cal R}_M=3.41$ ($\beta=0.5$). Initial conditions in frames (b) and (c) are $s_0=1-r_0-j_0$, with small $j_0=0.01$, and $r_0=0.3$, corresponding to a state ``close'' to the fixed point manifold $R+S=1$ of case {\bf (ii)} with everlasting immunity. (b) Numerical solution. (c) Random walk simulation of $600$ random walkers (agents) navigating independently on a Barab\'asi--Albert network of $2000$ nodes (upper right frame) for the same setting. An animated simulation of (c) is available \href{https://drive.google.com/file/d/1RuF0O8nLnut4V4kO_66G3ZWia46osvJ-/view}{online here or by clicking on the upper right frame}. The animation uses the same color codes to indicate the compartments of the walkers as those used in the curves in frames (b) and (c).
}
\label{FIG_SIRD_7} 
\end{figure}

Now we consider again case {\bf (ii)} with everlasting immunity and retrieve from (\ref{SCEIRS-D-model}) 
above SIR and include mortality.
We refer to \cite{Tome_Oliveira2020} for a related model.
We observe that two possible non-cyclic transition pathways SIR and SID emerge. The evolution equations
then take the form
\beq
\label{SIR_SID}
\begin{array}{clr}
  \ds \frac{d}{dt} S(t) & = \ds - {\cal A}(t)  & \\[3ex]
 \ds   \frac{d}{dt} J(t) & =  \ds - j_0 K_{IM}(t) - \int_0^t {\cal A}(\tau)K_{IM}(t-\tau){\rm d}\tau + {\cal A}(t) &\\[3ex]
\ds   \frac{d}{dt} R(t) & =\ds  j_0 b_r(t) + \int_0^t {\cal A}(\tau)b_r(t-\tau){\rm d}\tau 
 \end{array}
\eeq
with $D(t)=1-S(t)-J(t)-R(t)$.
The late time limit is contained in (\ref{infinite_time_limit_everlasting_imm}) and reads
\beq
\label{SIR_SID_late_time}
\begin{array}{clr}
  \ds S(\infty) & = \ds  s_0 - {\hat {\cal A}}(0) & \\[3ex]
 \ds   J(\infty) & = \ds 0  &\\[3ex]
\ds    R(\infty) & =\ds  r_0 +  B_r\left(j_0+{\hat {\cal A}}(0)\right) &\\[3ex]
\ds D(\infty) & = \ds  B_d\left(j_0+{\hat {\cal A}}(0)\right) .  &
 \end{array}
\eeq
As already mentioned the ratio
\beq
\label{kappa}
\kappa = \frac{B_d}{B_r} = 
\frac{D(\infty)}{R(\infty)-r_0}
\eeq
depends only on the distributions of $t_M$ and $t_I$ and
is independent of the infection history and specific form of the infection rate ${\cal A}(t)$,
recall (\ref{b_r_def}), (\ref{b_d_def}) and (\ref{BRD_relations}).
We depict in Fig.~\ref{FIG_SIRD_7} the time evolution of the compartmental fractions governed by the non-cyclic SIR-SID pathway for small mortality.
The small mortality can be confirmed from the shapes of the PDFs of $K_M(t)$ and $K_I(t)$ in Fig.~\ref{FIG_SIRD_7}(a): $K_M(t)$ is concentrated at larger values of $t$ than $K_I(t)$. Hence, in most realizations the survival condition $t_M >t_I$ is fulfilled. 
For late times the non-decreasing $R(t)$ and $D(t)$ saturate to finite values, whereas $J(t) \to 0$ reflecting finiteness of $t_I$,
and $S(t)$ is monotonously decreasing and converging to a finite value as predicted in (\ref{SIR_SID_late_time}). 
In this setting the effective reproduction number is
${\cal R}_M = 3.41$ thus
the manifold of disease-free fixed points $S + R = 1$ is unstable, and the disease is spreading where an unimodal $J$-wave emerges.
We verified validity of relation (\ref{kappa}) numerically which is found to be fulfilled with high accuracy (orange horizontal lines in frame (b)).
In order to obtain this accordance and stable numerical solutions, we averaged over $200$ different realizations of $t_I$ and $t_M$, respectively in our Monte Carlo scheme.
In Fig.~\ref{FIG_SIRD_7}(c) we depict the outcome of a random walk simulation in a connected
Barabási–Albert (BA) graph with the same setting as in frame (b) (i.e. same initial condition and Weibull distributions of $t_I$ and $t_M$ as depicted in frame (a)).
In this simulation the SIRD agents are represented
by random walkers navigating independently on the BA network in normal random walks 
\cite{BesMi-etal2021,BesMiRias2022,Granger-et-al2023,RiascosSanders2021}.
Consult \cite{Barabasi2016,BarabasiAlbert1999,Newman2010} for a general account of related network science.
We assume that the infection is transmitted (with probability one) if an S walker meets an I walker on the same node. 
By fitting the `density of walkers' ($600$ walkers on a network of $2000$ nodes), we achieved impressive quantitative agreement of the numerical solution of frame (b) and the outcome of the random walk simulation of frame (c), strongly corroborating our analysis.

\subsection{Non-cylic SEIR-SEID pathway: presymptomatic transmission and vaccination}
\label{RW_simulations_SEIRDS}
Here our goal is to investigate the interplay of the enhancing effects of presymptomatic transmission (from agents in E) on the spreading and immune ('vaccinated') agents to limit the propagation of the disease. We consider the non-cyclic SEIR-SEID pathway (case {\bf (ii)}) for everlasting immunity and presence of mortality with infection rate (\ref{infection_law})
${\cal A}(t) = \beta S(t)[\mu_E E(t)+ J(t)]$ 
and $\mu_E >1$ corresponding to `large' infectiousness of agents in E. 
Indeed, it has been observed in various diseases such as COVID-19 (SARS-CoV-2), Measles, 
Chickenpox (Varicella-Zoster) or Influenza that presymptomatic transmission is responsible for a substantial contribution of the total spread \cite{Qiu_etal2020}. 
The evolution equations read 
\beq
\label{SEIR_SEID_eqs}
\begin{array}{clr}
  \ds \frac{d}{dt}S(t) & = \ds - {\cal A}(t)  & \\[3ex]
  \ds  \frac{d}{dt}E(t) & = \ds -e_0K_E(t) - (K_E \star {\cal A})(t) + {\cal A}(t) &\\[3ex]
 \ds   \frac{d}{dt}J(t) & =  \ds e_0 K_E(t) +(K_E \star {\cal A})(t) - j_0 K_{IM}(t) 
 - (K_{IM} \star K_E \star {\cal A})(t) &\\[3ex]
\ds   \frac{d}{dt}R(t) & =\ds  j_0 b_r(t) + (b_r \star K_E \star {\cal A})(t) &\\[3ex]
\ds  \frac{d}{dt}D(t) & = \ds  j_0 b_d(t) + (b_d \star K_E \star {\cal A})(t) .&
 \end{array}
\eeq
The formulation of this system in terms of memory kernels (\ref{memory_kernels}) 
can be retrieved from (\ref{convol_memory_SEIRDS}) by setting there $M_R(t)=0$ (as $K_R(t)=0$).

\begin{figure}[H]
\centerline{
\includegraphics[width=0.55\textwidth]{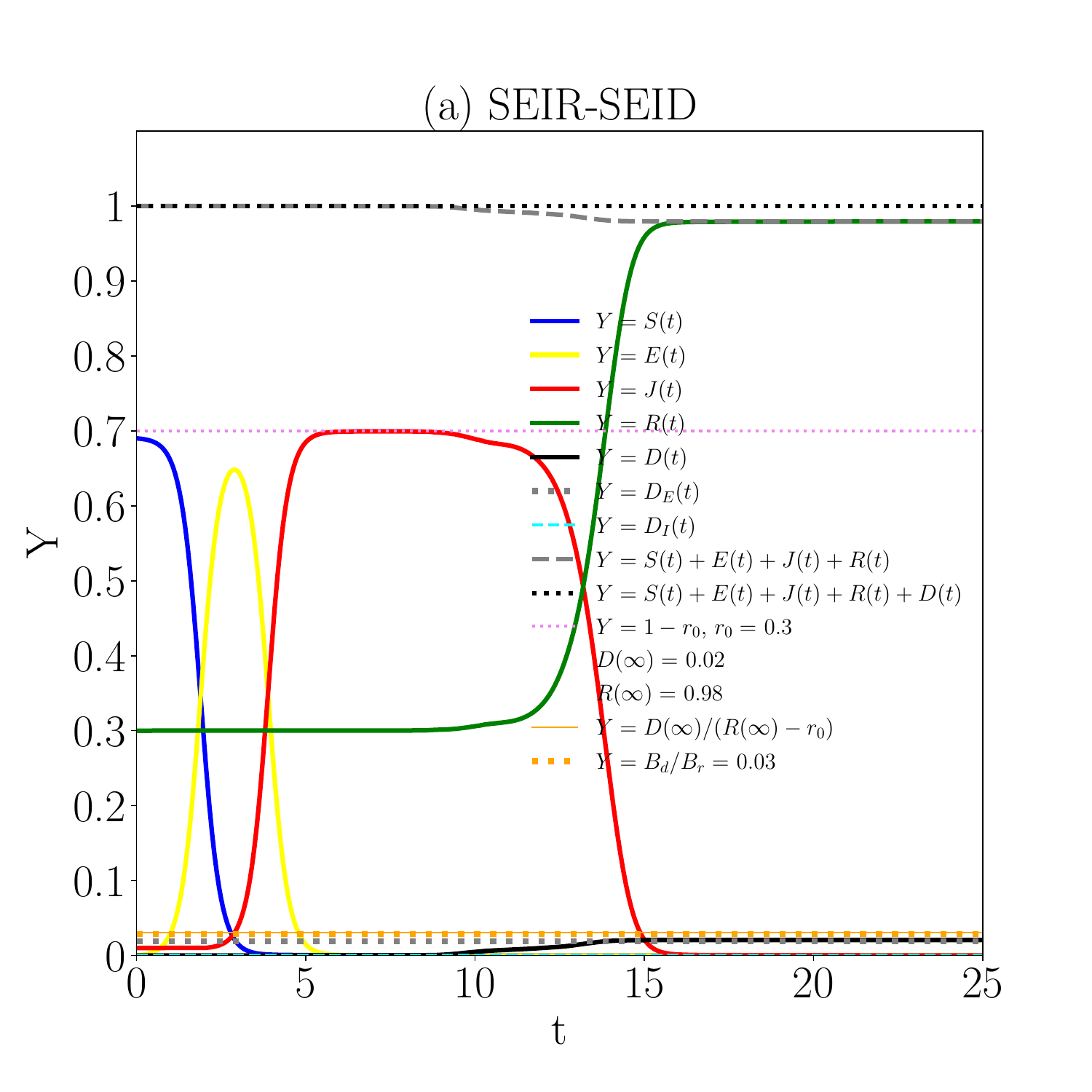}
}
\centerline{
\includegraphics[width=0.55\textwidth]{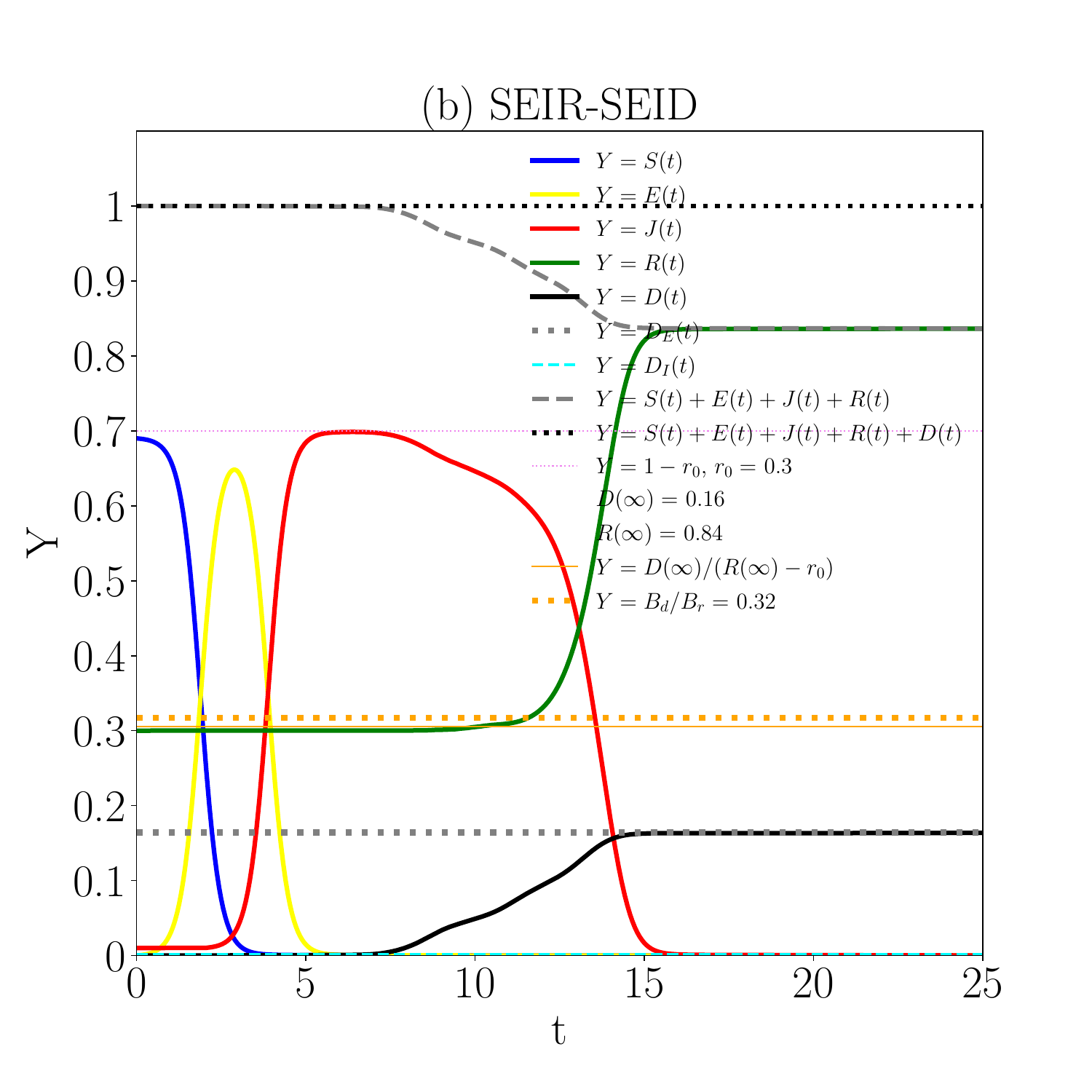}
\includegraphics[width=0.55\textwidth]{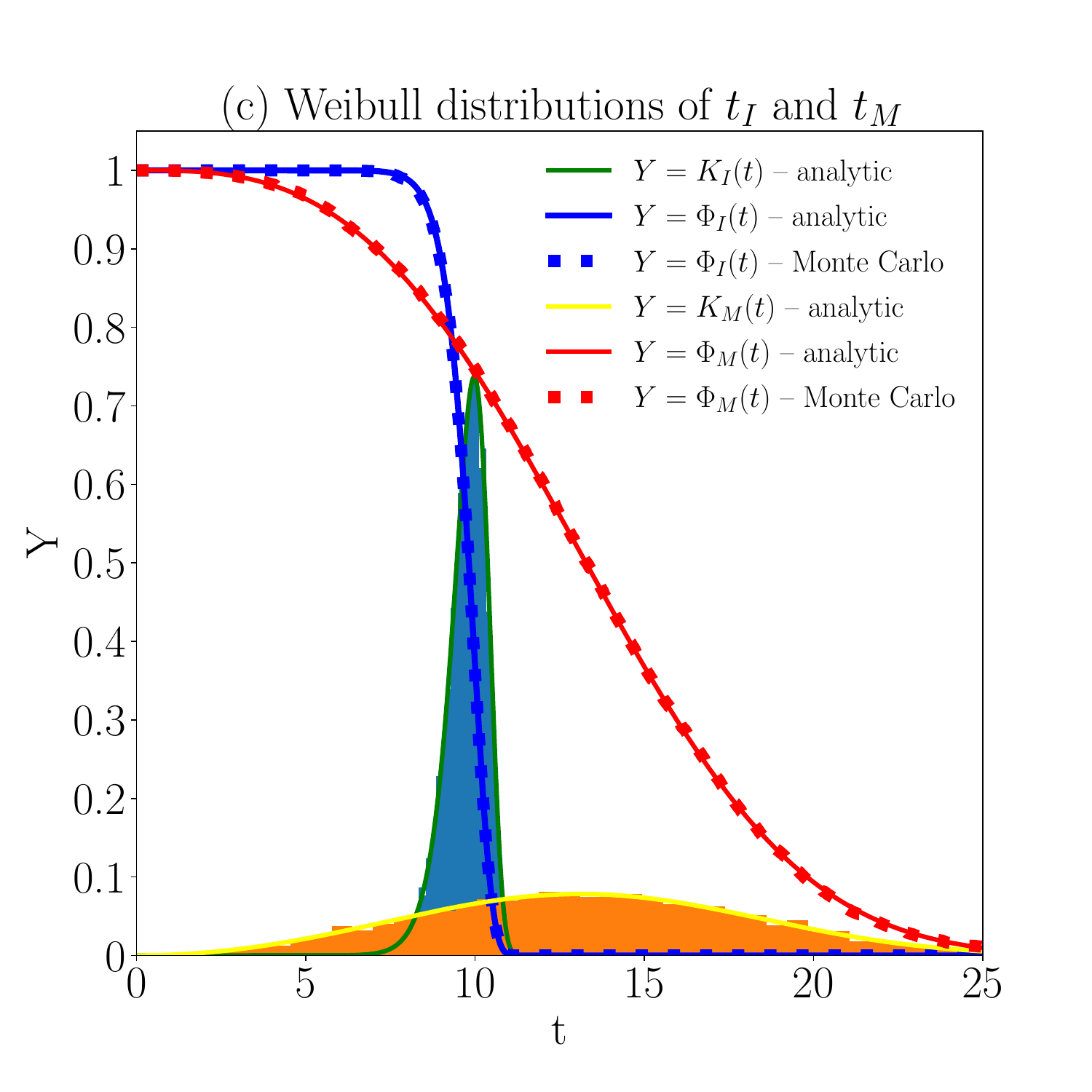}
}
\caption{
Non-cyclic SEIR-SEID pathway with presymptomatic transmission and infection rate parameters $\beta=0.5$ and $\mu_E=10$. Frames (a) and (b) use the same parameters $t_E=2$ and Weibull parameters for $t_I$: $t_0=10$, $\alpha=20$ and 
same initial conditions $[1-r_0-j_0,0,j_0,r_0]$ as in Fig. \ref{FIG_SIRD_7}, with $r_0=0.3$ fraction of `vaccinated' agents, $e_0=0$, and $j_0=0.01$, ``close'' to the fixed point manifold $S+R=1$ of case {\bf (ii)}. \\
Frame (a): Low-mortality case with Weibull parameters for $t_M$: $t_0=20$ and $\alpha=5$, using the same setting as in Fig. \ref{FIG_SIRD_7}, apart from the additional compartment E with sharp $t_E=2$. We determined numerically ${\cal R}_M\approx9.37$ using Eq. (\ref{repro_num}) accounting for the fraction $r_0=0.3$ of the vaccinated population.\\ Frame (b): Evolution with higher mortality than in frame (a). The Weibull parameters for $t_M$ are $t_0=15$ and $\alpha=3$, while all other parameters are the same as in (a). We determined numerically ${\cal R}_M\approx9.23$, which is smaller than in (a) due to the higher mortality. Frame (c): Weibull distributions of $t_I$ and $t_M$ used in (b).
}
\label{SEIRD_SEID_fig}
\end{figure}

We consider in Fig.~\ref{SEIRD_SEID_fig}(a) 
sharp latency time $t_E=2$ (i.e. $K_E(t)=\delta(t-2)$) and otherwise
the same setting as in Fig. \ref{FIG_SIRD_7} with identical parameters governing the distributions of $t_I$ and $t_M$ (small mortality, see Fig.~\ref{FIG_SIRD_7}(a)).
The presymptomatic infections lead to a more pronounced $J$-wave
in Fig.~\ref{SEIRD_SEID_fig}(a)
than without presymptomatic infections (due to absence of compartment E) in Figs. \ref{FIG_SIRD_7}(b,c). Fig.~\ref{SEIRD_SEID_fig}(a) shows that the amplitude of the $J$-wave is substantially enhanced by the contribution of presymptomatic infections, but remains strictly limited to the fraction $1-r_0$ of non-vaccinated agents. 
The same limiting effect is observed for higher mortality in
Fig.~\ref{SEIRD_SEID_fig}(b). The effect of the higher mortality is visible by the faster decay of the $J$-wave 
and larger value $D(\infty)$ in frame (b) than in (a).
The late time limit of system (\ref{SEIR_SEID_eqs}) was already obtained above (Eqs. (\ref{infinite_time_limit_everlasting_imm})).
The numerically obtained values of $\kappa \approx 0.03$ (see (\ref{kappa})) in Figs. \ref{FIG_SIRD_7}(b), \ref{SEIRD_SEID_fig}(a) coincide with high accuracy (see orange lines in both figures) reflecting independence of $\kappa$ of the different infection rates with and without presymptomatic infections, corroborating our analysis.
Of particular relevance is whether the contribution of presymptomatic infections increases the late time mortality $D(\infty)$. The answer is yes, since it increases ${\widehat A}(0)$, namely
\beq
\label{increased_mort}
{\widehat A}(0) = \beta \int_0^{\infty} S(t)[\mu_E E(t) + J(t)]{\rm d}t = {\widehat A}_E(0)+{\widehat A}_I(0) .
\eeq
From (\ref{infinite_time_limit_everlasting_imm}) one obtains for the mortality originating from presymptomatic infections 
\beq
\label{pre-symptomatic_morts}
D_E(\infty) = B_d {\widehat A}_E(0) = B_d \beta\mu_E \int_0^{\infty} S(t)E(t){\rm d}t ,
\eeq
while the mortality caused by infections from agents in I is $D_I(\infty) = B_d {\widehat A}_I(0)$.
The part $D_E(\infty)$ may be small for small overall mortality probability $B_d$, however it may become the dominating contribution for sufficiently large $\mu_E$. We consider such a case in Fig. \ref{SEIRD_SEID_fig}(b). 
Here almost the complete mortality is caused by presymptomatic infections. 
This behavior can be identified considering the evolution of $S(t)$ and $J(t)$ in the plot:
There is almost no time range where $S(t)J(t) \neq 0$ and therefore here $D_I(\infty) \ll D_E(\infty) =0.16 \approx D(\infty)$ (see dotted grey and cyan dashed lines).
The higher mortality in Fig. \ref{SEIRD_SEID_fig}(b)
comes along not only with a larger $D(\infty)= 0.16$ and $\kappa=0.32$, but also with a smaller value of ${\cal R}_M$ than in frame (a).

\subsection{Cyclic SEIRS-SEID dynamics}
\label{RW_simulations_SEIRDS_dynamics}
\begin{figure}[H]
\centerline{
\includegraphics[width=0.55\textwidth]{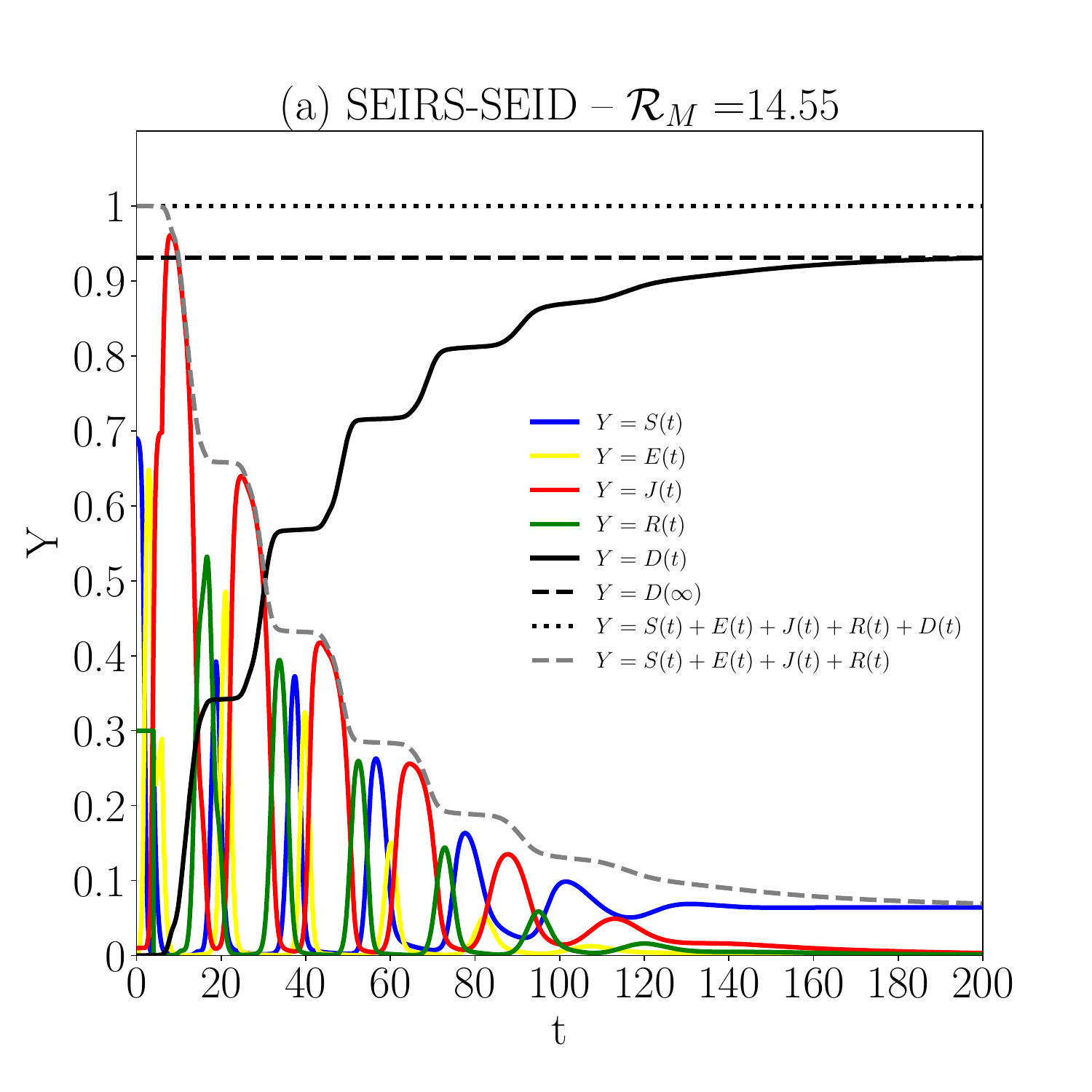}
\includegraphics[width=0.55\textwidth]{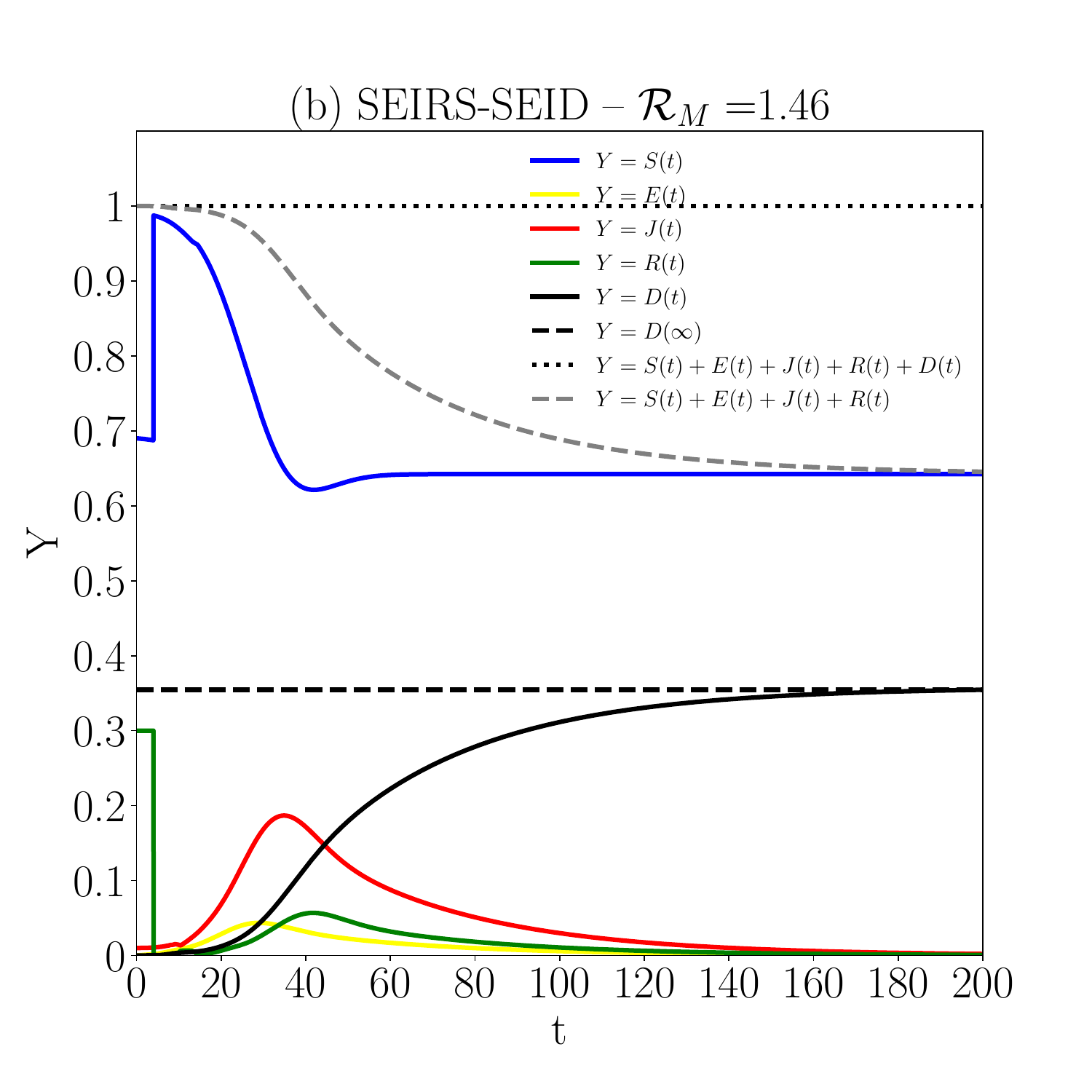}
}
\caption{
Cyclic SEIRS-SEID pathway with sharp immunity time $t_R=4$ and otherwise the same parameters and initial conditions as in frame (b) of Fig. \ref{SEIRD_SEID_fig}. The expected number of infections during an agent's lifetime is determined numerically as $\big\langle {\cal N}(\infty) \big\rangle = B_r/B_d = 3.16$ (with $B_d=0.24$ and $B_r=0.76$); recall relation (\ref{late_time}).\\ Frame (a): $\beta=0.5$ and $\mu=10$, with a relatively large ${\cal R}_M$, leading to an evanescent sequence of oscillations. \\ Frame (b): Same setting as in frame (a), but with reduced $\beta=0.05$ (mimicking confinement measures) where ${\cal R}_M$ still remaining slightly above the spreading threshold, resulting in a single infection wave.
}
\label{SEIRS_SEID_fig}
\end{figure}

Finally, we explore in Fig. \ref{SEIRS_SEID_fig} the complex interplay of 
presymptomatic infections, confinement measures and mortality
in a complete cyclic SEIRS-SEID transition pathway with sharp latency and immunity times, $t_E=2$ and $t_R=4$, respectively. In
Fig. \ref{SEIRS_SEID_fig}~(a) apart from finite immunity time, all parameters and initial condition are the same as in Fig. \ref{SEIRD_SEID_fig}~(b).
Although the probability to die in a single infection 
$B_d=0.24$ is identical in these plots, the effect of repeated infections in an agent's lifetime ($\big\langle {\cal N}(\infty) \big\rangle \approx 3$, 
recall (\ref{late_time})) due to cyclicity of the pathway
is dramatically increasing the overall mortality in 
Fig.~\ref{SEIRS_SEID_fig}~(a) with $D(\infty) > 0.9$
compared to the non-cyclic pathway of Fig.~\ref{SEIRD_SEID_fig}(b) 
with $D(\infty) < 0.2$. 
As an essential feature of cyclicity and high basic reproduction number in 
Fig.~\ref{SEIRS_SEID_fig}~(a) one observes a damped 
infection wave train with eventual extinction of the disease as predicted in relation (\ref{infinite_time_limit_with_mort}). 

In frame \ref{SEIRS_SEID_fig}~(b) we mimic `confinement measures' by reducing the
infection rate constant $\beta$ by a factor $10$ ($\beta=0.05$, reducing ${\cal R}_M$ correspondingly, still remaining above the spreading threshold), whereas leaving $\mu_E$ and all other parameters the same as in frame (a). 
This leads to reduced late time mortality $D(\infty) \approx 0.35$ where only a single
uni-modal infection wave occurs before extinction of the disease.

These results support the conclusion that consequently applied confinement measures reduce mortality and prevent emergence of multiple infection waves.

\section{Conclusions}
\label{conclusions}
We studied a stochastic formulation of cyclic SEIRS-SEID and non-cyclic SEIR-SEID transition pathways. We accounted for the mortality of agents during their illness period and investigated the effects of presymptomatic infections, vaccination, confinement measures and mortality on spreading dynamics. The cyclic SEIRS-SEID and non-cyclic SEIR-SEID dynamics are, apart from the memoryless cases with exponentially distributed waiting times, non-Markovian stochastic processes that retain a memory of the infection history. 

Starting from the stochastic formulation of the evolution equations (\ref{SCEIRS-D-model}), we derived a system of `memory equations' (\ref{convol_memory_SEIRDS}) with connections to general fractional calculus (Eqs. (\ref{convol_memory_SEIRDS_gen_der})). In particular, we analyzed the connection between compartment waiting time distributions and the emergence of fractional operators in the evolution equations (recall Sect. \ref{fractional_dyn}). For fat-tailed waiting time distributions, fractional (Riemann-Liouville) derivatives arise naturally (see (\ref{convol_memory_SEIRDS_fractional RL_der})). In contrast, for exponentially distributed sojourn times, the memory kernels reduce to Dirac kernels, and the standard memoryless evolution equations commonly considered in the literature are recovered.

From a linear stability analysis of the disease-free equilibrium, we obtained the basic reproduction number for cyclic pathways and the effective reproduction number for non-cyclic pathways with everlasting immunity, both in the presence of mortality and presymptomatic infections (Eq. (\ref{repro_num}). Non-cyclic pathways also allow us to obtain the spreading condition in the presence of vaccinated agents (Sect. \ref{spreading_cond}). For zero mortality and finite compartmental sojourn times, we obtained, above the spreading threshold, an endemic equilibrium that is either the asymptotically stable late-time limit solution, independent of the initial conditions, or Hopf unstable (Eq. (\ref{endem}) and recall Sect. \ref{endemic_equilibrium}). For zero mortality in non-cyclic infection pathways with everlasting immunity, an endemic equilibrium does not exist. Instead, a stationary asymptotic disease-free state emerges (see (\ref{infinite_time_limit_everlasting_imm_no_mort})). In non-cylic pathways with mortality we identified the characteristic quantity
$\kappa= D(\infty)/(R(\infty)-r_0) = B_d/B_r$ (recall (\ref{infinite_time_limit_everlasting_imm}))
as being independent of the infection rate and uniquely determined by the characteristics of the distributions of $t_I$ and $t_M$. In contrast, for cyclic infection pathways, the quantity $B_r/B_d$ has the interpretation of the expected number of infection cycles an agent experiences during its lifetime (recall (\ref{late_time})). We explored the effect of presymptomatic infections on overall mortality (recall Sect. \ref{RW_simulations_SEIRDS}). We found that presymptomatic infections may, under certain conditions, substantially increase overall mortality and may even become its dominant contribution (see Fig. \ref{SEIRD_SEID_fig}(b)).

Our model has considerable flexibility and can be adapted to spreading processes across a wide range of real-world problems. It can also be extended in various directions. For instance, additional intermediate compartments can be included, some of which may be infectious (predator). Moreover, accounting for post-symptomatic infections, in which agents in R may become infected with positive probability due to incomplete immunity, may also be of interest. Another interesting extension could involve cyclic SIRI pathways, in which agents in R may relapse to compartment I \cite{Bhat_SIRI}.

The inclusion of demographic population changes could provide a further interesting extension, allowing diseases evolving ``slowly'' to be described on timescales over which such changes become relevant \cite{Anderson_May1991}. Moreover, variants could be considered in which the infection law (\ref{infection_law}) is modified beyond a simple mass action law \cite{besmi_periodic2025}. 
Finally, the class of vector-borne transmission constitutes a challenging field \cite{Granger_et_al_Springerbook2026}.

The cross-disciplinary nature of our model allows applications beyond epidemiology, including the description of a wide class of spreading phenomena such as the propagation of pollutants, political opinions, wildfires, financial crises with ``contagion,'' and many others.
\section*{Acknowledgments} 
T.\,M.\,M. and A.\,D.\,B. gratefully acknowledge valuable comments of
Reza Jafari, Physics department, Shahid Beheshti University.
M.\,B. is indebted to the Institut Jean le Rond 
d'Alembert for the hospitality in the framework of the present study.

\appendix
\setcounter{equation}{0}
\renewcommand{\theequation}{A\arabic{equation}}
\section{Appendix}
\subsection{Explicit representation of evolution equations (\ref{SCEIRS-D-model}) with mortality}
\label{explicit_evol_eqs}
In order to evaluate (\ref{SCEIRS-D-model}), we need the persistence probability that an individual is in compartment I and alive.
This state is governed by the r.v. $t_{IM} = min(t_I,t_M)$. An indicator function being one if an agent is in this state is 
$\Theta(t_{IM}-t) = \Theta(t_M-t)\Theta(t_I-t)$ 
thus the persistence probability of this state (in compartment I and alive) is
\beq
\label{in_I_and_alive}
\Phi_{IM}(t) = \Big\langle \Theta(t_M-t)\Theta(t_I-t) \Big\rangle = \Big\langle \Theta(t_M-t) \Big\rangle \Big\langle \Theta(t_I-t) \Big\rangle 
= \Phi_M(t)\Phi_I(t)
\eeq
with 
\beq
\label{complete_pdf}
\begin{array}{clr}
\ds \frac{d}{dt} \Phi_{IM}(t) =-K_{IM}(t) & = \ds  - \Big\langle\delta(t-t_I) \Big\rangle\Big\langle\Theta(t_M-t) \Big\rangle
- \Big\langle \delta(t-t_M) \Big\rangle \Big\langle \Theta(t_I-t) \Big\rangle &\\[2ex]
& = \ds  -K_M(t) \Phi_I(t) - K_I(t) \Phi_M(t) &\\[2ex]
 & = \ds   -b_d(t) -b_r(t) &
\end{array}
\eeq
where the PDF $K_{IM}(t)$ captures all transitions out of compartment I.
We emphasize that $K_{IM}(t)$ is a complete properly normalized PDF, whereas $b_r(t), b_d(t)$ are defective (not well normalized PDFs), recall (\ref{b_r_def}), (\ref{b_d_def}) and (\ref{BRD_relations}).
Now consider the second equation in (\ref{SCEIRS-D-model})
$\frac{d}{dt}E(t) = rate_E^{in}(t) -rate_E^{out}(t)$ where $rate_E^{in} ={\cal A}(t)$ are the transitions into E (new infections), and
$rate_E^{out}= rate_I^{in}= e_0 K_E(t)+ (K_E \star {\cal A})(t) $ are the transitions out from E to I (the two negative terms in this equation), i.e. the rate of incoming individuals into I.
Then it is straight-forward to obtain the following representations for the compartment populations
\beq
\label{comp_E}
E(t) = e_0\Phi_E(t) + \int_0^t \Phi_E(t-\tau){\cal A}(\tau){\rm d}\tau
\eeq
and
\beq
\label{repres}
J(t)= j_0 \Phi_{IM}(t)  + \int_0^t \Phi_{IM}(t-\tau) rate_I^{in}(\tau){\rm d}\tau = 
j_0 \Phi_{IM}(t)  + \int_0^t \Phi_{IM}(t-\tau)\left[e_0 K_E(\tau)+ (K_E \star {\cal A})(\tau)\right]{\rm d}\tau .
\eeq
The first term describes the decay of 
the proportion of the agents $j_0$ initially in I. In the convolution
the quantity $\Phi_{IM}(t-\tau) rate_I^{in}(\tau){\rm d}\tau$ is the contribution of
the agents that came at time $\tau$ from E into I
and still are at time $t$ (with probability $\Phi_{IM}(t-\tau)$). An analog 
interpretation applies to (\ref{comp_E}).
Further consider the mean of the term
\beq
\label{mean_example_A}
\Big\langle {\cal A}(t-t_E-t_I-t_R)\Theta(t_M-t_I) \Big\rangle = \int_0^t {\rm d}\tau {\cal A}(\tau) \Big\langle \delta(t-\tau-t_E-t_I-t_R)\Theta(t_M-t_I) \Big\rangle 
\eeq
which contains the defective PDF $b_r(t)$, which takes us to
\beq
\label{defective_density}
\Big\langle \delta(t-t_E-t_I-t_R)\Theta(t_M-t_I) \Big\rangle = \Big\langle \delta(t-t_E-t_I-t_R)\Phi_M(t_I) \Big\rangle = (K_E \star (\Phi_M K_I) \star K_R)(t)
\eeq
thus (\ref{mean_example_A}) yields
\beq
\label{averaging_example_A}
\Big\langle {\cal A}(t-t_E-t_I-t_R)\Theta(t_M-t_I) \Big\rangle = \left({\cal A} \star K_E 
\star b_r \star K_R\right)(t) .
\eeq
Keep in mind that all these convolutions commute. To this end, recall that $(1 \star {\cal A})(t) = \int_0^t {\cal A}(\tau){\rm d}\tau$ 
and $(K_{E,I,R} \star 1)(t) =\Big\langle \Theta(t-t_{E,I,R})\Big\rangle=\int_0^t K_{I,I,R}(\tau) {\rm d}\tau  =1-\Phi_{I,E,R}(t)$, see  (\ref{peristent_probas}).
With these considerations, we can integrate system (\ref{SCEIRS-D-model}) to obtain the self-consistent convolutional representation
\beq
\label{convol_integraytion_appendix}
\begin{array}{clr}
\ds S(t) = & \ds s_0 - \left(\left[1- 1 \star K_E \star  b_r \star K_R\right]\star {\cal A} \right)(t) + e_0(1\star K_E \star b_r\star K_R)(t) + j_0(1 \star b_r\star K_R)(t)+r_0 (1\star K_R)(t)   &\\[3ex]
\ds E(t)  = & \ds e_0\Phi_E(t) + \left({\cal A} \star \Phi_E\right)(t)   & \\[3ex]
\ds J(t) = & \ds j_0 \Phi_{IM}(t) + e_0 \left(\Phi_{IM} \star K_E\right)(t) +\left(\Phi_{IM}\star K_E \star {\cal A}\right)(t)  &\\[3ex]
\ds R(t) = & \ds  r_0\Phi_R(t) + j_0 \left(\Phi_R \star b_r\right)(t)+e_0 \left(\Phi_R \star K_E \star b_r\right)(t) + \left(\Phi_R \star  K_E 
\star b_r \star {\cal A}\right)(t)  &\\[3ex]
\ds D(t) = & j_0 \left(1\star b_d\right)(t) +e_0\left(1\star K_E \star b_d\right)(t) + \left(1\star K_E \star b_d \star {\cal A}\right)(t)    &
\end{array}
\eeq
where one easily confirms that $D(t)+S(t)+E(t)+J(t)+R(t)=1$.
We point out that system (\ref{convol_integraytion_appendix}) is self-consistent and is not an explicit solution, since ${\cal A}(t) = {\cal A}(S(t),\Lambda(t))$ implicitly depend on the unknown compartmental fractions $S(t),E(t),J(t)$. 
Applying the time derivative on (\ref{convol_integraytion_appendix}) straightforwardly
yields the convolutional representation of the averaged system (\ref{SCEIRS-D-model}). In these equations we assume finite waiting times $t_{E,I,R,M}$ (i.e., case {\bf (i)}).
In the limit of zero mortality one retrieves the convolutional representations of Eqs. (\ref{SCEIRS-D-model_zero-mortality}) (set $t_M \to \infty, \Phi_M(t)=1, b_r(t) \to K_I(t), b_d(t) \to 0$).

An instructive representation of system (\ref{convol_integraytion_appendix}) is obtained when
using the memory kernels (\ref{memory_kernels}).
Differentiating (\ref{convol_integraytion_appendix}) with respect to time [Use $\frac{d}{dt}\left(\Phi \star f \right)(t) = - (K \star f)(t) + f(t)$] and
plugging in (\ref{memory_kernels}) takes us to an evocative convolutional representation of (\ref{SCEIRS-D-model})
involving above memory kernels, namely
\beq
\label{convol_memory_SEIRDS_appendix}
\begin{array}{clr}
\ds \frac{dS(t)}{dt} = & \ds - {\cal A}(t) + (M_R \star R)(t) &\\[3ex]
\ds \frac{dE(t)}{dt}  = & \ds - (M_E \star E)(t) + {\cal A}(t)  & \\[3ex]
\ds \frac{dJ(t)}{dt} = & \ds -(M_{IM} \star J)(t) + (M_E \star E)(t) &\\[3ex]
\ds \frac{dR(t)}{dt} = & \ds  -(M_R \star R)(t) + (m_r \star J)(t)  & \\[3ex]
\ds \frac{dD(t)}{dt} = & \ds (m_d \star J)(t) &
\end{array}
\eeq
where the last equation is redundant. It is straightforward to retrieve from these equations 
the limit of zero mortality by setting $m_d(t)=0$ and $m_r(t)=M_{IM}(t) = M_I(t)$.
\paragraph{Everlasting immunity} 
Finally we consider everlasting immunity (case {\bf (ii)}) $t_R \to \infty$ ($K_R(t)=0$,
$\Phi_R(t)=1$) for which (\ref{convol_integraytion_appendix}) boils down to
\beq
\label{convol_integraytion_appendix_immunity_forever}
\begin{array}{clr}
\ds S(t) = & \ds s_0 - \left(1\star {\cal A}\right)(t)  &\\[3ex]
\ds E(t)  = & \ds e_0\Phi_E(t) + \left({\cal A} \star \Phi_E\right)(t)   & \\[3ex]
\ds J(t) = & \ds j_0 \Phi_{IM}(t) + e_0 \left(\Phi_{IM} \star K_E\right)(t) +\left(\Phi_{IM}\star K_E \star {\cal A}\right)(t)  &\\[3ex]
\ds R(t) = & \ds  r_0 + j_0 \left(1 \star b_r\right)(t)+ e_0 \left(1\star K_E \star b_r\right)(t) + \left(1\star K_E \star b_r \star {\cal A}\right)(t)  &\\[3ex]
\ds D(t) = & \ds j_0 \left(1\star b_d\right)(t) +e_0\left(1\star K_E \star b_d\right)(t) + \left(1\star K_E \star b_d \star {\cal A}\right)(t)    &
\end{array}
\eeq
where $S(t)+E(t)+J(t)+E(t)+ D(t) =1$ (recall $b_r+b_d=K_{IM}$ and $\Phi(t) + (1 \star K)(t) = 1$). Since both compartments R and D have infinite sojourn times, the quantities
$R(t)-r_0, D(t)$ exhibit exchange symmetry with respect to $b_r \leftrightarrow b_d$.
Straightforwardly one also retrieves from (\ref{convol_memory_SEIRDS_appendix})
the memory representation for everlasting immunity by $M_R(t)=0$.

In all above relations, one obtains late time limits directly from the correspondence
$\lim_{t\to \infty}(1 \star f)(t) = \int_0^{\infty} f(t){\rm d}t = {\hat f}(0) = \lim_{\lambda \to 0} \lambda ({\hat f}(\lambda)/\lambda)$.

\end{document}